\documentclass[]{silvia}
\usepackage{graphicx}
\usepackage{natbib}
\usepackage{lscape}         
\usepackage[normalem]{ulem}
\def\new#1 {{\bf #1} }
\def\cut#1 {\sout{#1}}
\def\new#1 { #1 }
\def\cut#1 {}

\bibpunct{(}{)}{;}{a}{}{,}
\usepackage{subfigure}
\usepackage{txfonts}
\bibpunct{(}{)}{;}{a}{}{,}
\begin{document}

\title{Physical characteristics 
 of  bright Class\,I methanol masers}  
\author{S. Leurini\inst{1} \and  K. M. Menten\inst{1} \and  C. M. Walmsley\inst{2,3}}
\institute{Max-Planck-Institut f\"ur Radioastronomie, Auf Dem H\"ugel 69, D-53121, Bonn\\ email sleurini@mpifr-bonn.mpg.de 
\and INAF – Osservatorio Astrofisico di Arcetri, Largo E. Fermi 5, 50125 Firenze, Italy
\and Dublin Institute of Advanced Studies, 31 Fitzwilliam Place, 2 Dublin, Ireland}
\offprints{S. Leurini} \date{\today}
 \abstract
     {Class\,I methanol masers are thought to be tracers of interstellar shock
       waves. However, they have received relatively little attention mostly as a consequence of their low luminosities compared to other maser transitions. This situation has changed  recently and Class\,I
 methanol masers are now  routinely used as signposts of outflow activity especially in high extinction regions. The recent detection of polarisation
in Class\,I lines now makes it possible to obtain direct observational information about magnetic fields in interstellar shocks.}
     {We make use of newly calculated collisional rate coefficients for methanol to investigate
       the excitation of Class\,I methanol masers and to reconcile the observed Class\,I methanol maser
       properties with model results.}
     {We performed large velocity gradient calculations with a plane-parallel slab geometry appropriate for shocks
       to compute the pump  and loss rates which regulate the interactions of the different maser systems with
       the maser reservoir. We study the dependence of the pump rate coefficient, the 
maser loss rate, and the inversion efficiency of the pumping scheme of several Class\,I masers on the physics of the emitting gas.}
   {We predict inversion in all transitions where maser emission is observed. Bright
     Class\,I methanol masers are mainly high-temperature ($>100$\,K) high-density ($n({\rm H_2})\sim10^7 - 10^8$\,cm$^{-3}$) structures with methanol maser emission measures, $\xi$, corresponding to high methanol abundances close to the limits set by collisional quenching. 
     Our model predictions reproduce reasonably well most of the observed properties of Class\,I methanol masers. Class\,I masers in the 25\,GHz series are the most sensitive to the density of the medium and mase at higher densities than other lines. Moreover, even at high density and high methanol abundances, their luminosity is predicted to be lower than that of the 44\,GHz and 36\,GHz masers. Our model predictions also reflect the observational result that
   the 44\,GHz line is almost always stronger than the 36\,GHz maser. By comparison between observed isotropic photon luminosities and our model predictions,  we infer maser beam solid angles of roughly $10^{-3}$ steradian.}
   {We find that the Class\,I masers can reasonably be separated into
 three families: the $(J+1)_{-1}-J_{0}$-$E$ type series, the 
 $(J+1)_0-J_{1}$-$A$ type, and the $J_{2}-J_{1}$-$E$ lines at 25\,GHz. The 25\,GHz lines behave in a different fashion from the other masers as they are only inverted at high densities
  above $10^6$\,cm$^{-3}$ in contrast to other Class\,I masers. Therefore, the detection of maser activity in all three families is a clear indication of high densities.}

\titlerunning{Bright Class\,I methanol masers}
 \authorrunning{Leurini et al.} \maketitle

%
 \section{Introduction}
As first realised by \citet{1987Natur.326...49B}, methanol masers come
in two varieties, termed Class\,I and II by \citet{1991aimn.conf..119M,1991ApJ...380L..75M}. 
Class\,II methanol masers 
are closely associated with high-mass  young stellar objects (YSOs) and arise from 
the hot molecular layers heated by the embedded YSOs before and after the formation of an ultracompact H{\sc ii} region. When they occur after the formation of   an ultracompact H{\sc ii} region, they frequently coexist with hydroxyl (OH) masers \citep{1992ApJ...401L..39M} and are, like the OH masers \citep{1991A&A...241..537C},
most likely pumped by far-infrared radiation in dense ($n \sim 10^7$\,cm$^{-3}$) warm
($T \sim$ 150 K) gas \citep{1994ApJ...433..719S,1994A&A...291..569S,1997A&A...324..211S}. 
In contrast, Class\,I masers  are collisionally pumped and their occurrence can be explained from
the basic properties of the CH$_3$OH molecule \citep{1973ApJ...184..763L}.

The presence of a Class\,II methanol maser associated with a star forming core provides unambiguous and sufficient proof of the presence of a high-mass protostellar or young stellar object \citep{2015MNRAS.446.3461U}, and more than a thousand of these sources have been found in Galaxy-wide surveys \citep{2009MNRAS.392..783G}. In contrast,  until recently 
relatively little attention was
 paid to Class\,I methanol masers.
However,  recent observational results have changed this situation.
\citet{2009ApJ...702.1615C}  show that these masers are associated with shocked material in protostellar outflows and thus trace very early stages of massive star formation.
It is 
also worth noting that Class\,I masers have also been detected 
from a few outflows associated with low-mass protostars
 \citep{2010MNRAS.405..613K}. Moreover, 
\citet{2010ApJ...710L.111S} and \citet{2014AJ....147...73P} report the detection of Class\,I methanol masers towards  supernova  
remnants \citep[see][for an overview]{2011MmSAI..82..703F}. 

Additionally, \citet{2009ApJ...705L.176S, 2011ApJ...730L...5S} report the discovery
of  circular polarisation  in the 36 and 44\,GHz Class\,I maser lines, which they interpret as a result of the Zeeman effect. These results suggest that Class\,I masers could be exploited to
  measure magnetic fields in interstellar shocks and could deliver inputs for magneto-hydrodynamic modelling of shock models.
From the point of view of modelling, recent progress in the calculation of collisional rate coefficients
has resulted in numbers that now allow meaningful modelling of 
methanol  excitation over a wide range of interstellar 
conditions \citep{2010MNRAS.406...95R}.

 The aim of our study is to investigate the properties of bright Class\,I methanol
 masers focussing mainly on masers with brightness temperatures higher than the critical
 value corresponding to the saturated intensity $J_{s}$ beyond which
 maser amplification becomes linear rather than exponential \citep[see][]{1992ASSL..170.....E,2013ApJ...773...70H}. These masers are usually associated with regions of massive star formation, while lower luminosity
 Class\,I masers  are reported in supernovae remnants \citep[e.g.,][]{2014AJ....147...73P}.
 We assume the
 masers to be pumped by collisions with the para form of H$_2$ and
 compute the pump and decay rates,  as well as the pump efficiency
 for the  most commonly observed Class\,I methanol maser transitions.
 An implicit assumption is that the escape probabilities of the
 relevant transitions populating the upper and lower maser levels are
 highly anisotropic allowing coherence along the line of sight, while
 a large velocity gradient in the plane of the sky allows relatively
 high escape probabilities in other directions.  This type of
 velocity field is expected to be produced by a shock propagating
 perpendicular to the line of sight;  moreover,  a shock of this kind is  expected to produce a maser volume of high aspect ratio
 oriented along the line of sight.  The shock  is also expected
 to cause a rise in the methanol abundance as is suggested by
 some of the observational data discussed in Sect.\,\ref{obs} of this article.

 In this context, we have concentrated on geometry independent
 maser characteristics and their dependence on physical properties such as
 density and temperature. We have thus computed the maser output per
 cubic centimetre and have estimated the densities and methanol abundances
 at which maser action is quenched. We do not, on the other hand, attempt to
 compute maser beaming angles and brightness temperatures, although we
 do focus on  size scales  (roughly 10--300\,AU) corresponding to the
 observed maser diameters.  Our hope is that our results in
 combination with future interferometric work will allow an increased
 understanding of the structure of interstellar shocks and the dissipation
 of mechanical energy in regions of star formation.
  In a follow-up study, we plan to implement geometry effects in our model to compute brightness 
 temperatures and beaming angles and properly model observations of Class\,I masers in massive star forming regions.

In Sect.\,\ref{obs} we give a review of the observational properties of 
Class\,I methanol masers, followed in Sect.\,\ref{models} by a description of previous studies on the pumping mechanism for Class\,I masers. In Sect.\,\ref{stateq} we present our model for the pumping of Class\,I methanol masers, and compute the photon production rates of high-gain masers as a function  of the physical parameters of the gas. In Sect.\,\ref{dis}, we compare our model results with the observed properties of Class\,I masers. In Sect.\,\ref{outlook} we conclude with possible observational and theoretical perspectives to improve our knowledge of Class\,I masers.

\section{Phenomenology of Class\,I methanol masers}\label{obs}

\begin{figure*}
\centering
\includegraphics[width=0.8\textwidth]{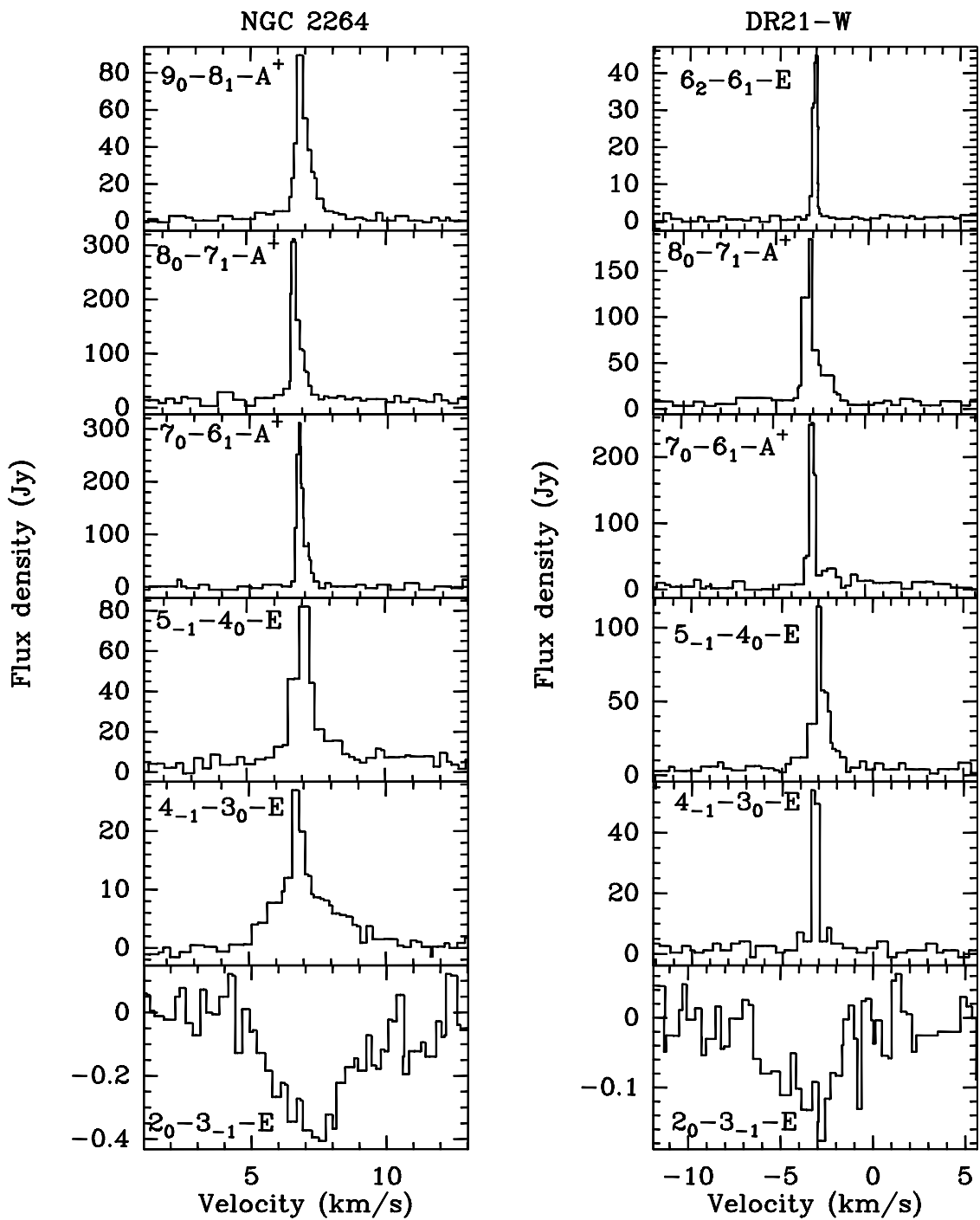}
\caption{ For both NGC\,2264 (left) and DR21-W (right) the lowest spectrum shows absorption in the
  CH$_3$OH $2_0\to 3_{-1}$-$E$ line, while the other
  spectra show various maser transitions taken with differing velocity resolutions and a number of telescopes. Adopted from \citet{1991aimn.conf..119M}. All spectra where taken between 1988 and 1990. The spectra of the $4_{-1}\to 3_0$-$E$ and  $7_0\to 6{_1}$-$A^+$ lines are from \citet{1990ApJ...354..556H} and those of the  $2_0\to 3_{-1}$-$E$ lines were taken with the Green Bank 140-foot telescope. The remaining spectra were taken with the IRAM 30\,m telescope.}\label{spectra}
\end{figure*}

After the first example of a Class\,I methanol maser was discovered, the 25 GHz $J_{k=2} -J_{k=1}~E$ transitions in Orion-KL \citep{1971ApJ...168L.101B}, it took 15 years before these lines were found in other 
regions \citep{1986A&A...157..318M}, and before the 44\,GHz $7_0\to 6_1$-$A^+$ and the 36\,GHz $4_{-1}\to 3_0$-$E$ methanol masers \citep{1985ApJ...288L..11M}  were discovered.
This paucity of detections of the 25 GHz lines was partially a result of the masers' relatively modest luminosities and because they are frequently 
found offset from traditional tracers of star formation such as strong infrared sources and ultracompact H{\sc ii} regions \citep{1986A&A...157..318M}. As discussed above,  an association between Class\,I methanol masers and protostellar outflows was established later on.

Table\,\ref{masertable} reports all known Class\,I methanol maser
transitions to date. For reference, examples of spectra of Class\,I
masers are given in Fig.\,\ref{spectra} towards DR21-W and NGC\,2264
where the $4_{-1}\to3_{0}E$ line also shows
 broader thermal emission. 
The 44\,GHz $7_0\to 6_1$-$A^+$ and 95\,GHz $8_0\to 7_1$-$A^+$ lines are
generally the most widespread and the strongest Class\,I transitions
observed \citep[e.g.,][]{1990ApJ...354..556H, 2000MNRAS.317..315V}, and therefore they have
traditionally been the target of searches for Class\,I methanol masers
in astronomical sources
\citep[e.g.,][]{1990A&A...240..116B,1994MNRAS.268..464S,2004ApJS..155..149K,2005MNRAS.359.1498E,2009ApJ...702.1615C,2011ApJS..196....9C}. The next most widespread Class\,I masers are the 36\,GHz $4_{-1}\to 3_0$-$E$ and 
84\,GHz $5_{-1}\to 4_0$-$E$ transitions 
\citep[e.g.,][]{1989ApJ...339..949H,2001ARep...45...26K,2014MNRAS.439.2584V}.
Recent observations at high angular resolution \citep{2007IAUS..242..182V,2012IAUS..287..282B} show
that even the 25\,GHz masers are more common than previously thought. The new capabilities of the upgraded VLA and ATCA interferometers allow the simultaneous observations of several transitions in the ${J}_{2}\to{J}_{1}E$ series thus allowing better investigation of these masers. The masers at 9.9\,GHz, 23.4\,GHz, and 104\,GHz are rare and they have been detected in only  a handful of objects \citep[see][]{2012IAUS..287..433V}. Interestingly, many of the remaining Class\,I masers fall within ALMA bands.

\begin{table*}
\begin{center}
\caption{Known interstellar Class\,I methanol maser transitions}\label{masertable}
\begin{tabular}{lrll}
 \hline  \hline
Transition    &Frequency (MHz)&Lab. Ref. & Astronomical Reference\\
           \hline
$9_{-1}\to8_{-2}$-$E$               &9\,936&  L2  &A28    \\
$10_1\to 9_2$-$A^-$                &23\,445& L5&A48\\
$3_{2}\to3_{1}$-$E$                 &24\,929&  L3  &A12, A20    \\
$4_{2}\to4_{1}$-$E$                 &24\,933&  L1  &A1, A6, A8, A12, A20     \\
$2_{2}\to2_{1}$-$E$                 &24\,934&  L1  &A7, A12, A17, A20    \\
$5_{2}\to5_{1}$-$E$                 &24\,959&  L4  &A1, A6, A12, A17, A20,
{\it A24}    \\
$6_{2}\to6_{1}$-$E$                 &25\,018&  L4  &A1, A2, A3, A17, A20,{ \bf A4*}, A5, A6, A7,{ \it A8}, { \bf A9}, A12,\\
                                 &               &      &{\bf A13}, A17, A20,{ \it A24},{ \bf A26*},{ \it A36},{ \bf A38*}\\
$7_{2}\to7_{1}$-$E$                   &25\,125&  L4  &A1, A2, A3, A17, A20    \\
$8_{2}\to8_{1}$-$E$                   &25\,294&  L4  &A1, A6, A17   \\
$9_{2}\to9_{1}$-$E$                   &25\,541& L4      &A12, A32         \\
$10_{2}\to10_{1}$-$E$                 &25\,878&L4&   A39    \\
$12_{2}\to12_{1}$-$E$                 &26\,847&L5&A32     \\
$13_{2}\to13_{1}$-$E$                 &27\,473&L5&A32      \\
$14_{2}\to14_{1}$-$E$                 &28\,169&L5&A32     \\
$15_{2}\to15_{1}$-$E$                 &28\,906&L5&A32     \\
$16_{2}\to16_{1}$-$E$                 &29\,637&L5&A32     \\
$17_{2}\to17_{1}$-$E$                 &30\,308&L5 &A32        \\
$4_{-1}\to3_{0}$-$E$                  &36\,169&L5&{\bf A4, A9}, A10,{ \bf A13}, A19, A25,{ \bf A26}, A27, A33, {\it A51}\\ 
$7_{0}\to6_{1}$-$A^{+}$               &44\,069&L5&{\bf A4, A9},A10, A21, A22,{ \bf A26}, A30, A34, A37, {\it A44}, {\it A46}, A47, {\it A51}\\        
$5_{-1}\to4_{0}$-$E$                 &84\,521&L3&{\it A14}, {\bf A13}, {\bf A26},A42         \\
$8_{0}\to7_{1}$-$A^{+}$              &95\,169&L3&A11, A18,{ \it A23},{ \bf A26}, A29, A31, A49, A50         \\
$11_{-1}\to10_{-2}$-$E$             &104\,300& L3 & A43\\ 
$6_{-1}\to5_{0}$-$E$                &132\,891&L5&{\bf A26}, A35, A40         \\
$9_{0}\to8_{1}$-$A^{+}$             &146\,618&L5& {\bf A26}         \\
$4_{2}\to3_{1}$-$E$                 &218\,440&L6& {\bf {\it A52}}\\
$8_{-1}\to7_{0}E$                   &229\,758& L6& A43\\
$9_{-1}\to8_{0}$-$E$                &278\,305&L6& {\bf {\it A53}}\\

\hline
\end{tabular}
\end{center}
{\scriptsize

NOTES -- References are in chronological order. 
Laboratory references  for higher accuracy frequencies are: L1: \citet{1974ApJ...191L..99G}; L2: \citet{1995ApJ...438..504B}; 
L3: \citet{2004A&A...428.1019M}, L4: \citet{tsunekawa}\footnote{http://www.sci.u-toyama.ac.jp/phys/4ken/atlas/}, L5: \cite{mehrotra}, L6: \citet{1984JMoSp.103..486S}.
Astronomical references presenting interferometric 
observations are given in {\it italics}, theoretical papers in
{\bf bold}face. Theoretical references discussing various of the 
$J_{2} \to J_{1}$-$E$ transitions are only listed for the $6_{2}\to6_{1}$-$E$  transition and are
marked by an asterisk. Astronomical references are:
A1: \citet{1971ApJ...168L.101B}; A2: \citet{1974ApJ...187L..19C}; A3:
\citet{1975A&A....39..149H}; {\bf A4*: \citet{1975MNRAS.172...41P}}; A5: {\it \citet{1976ApL....18...13B} (VLBI)}; 
A6: \citet{1975ApJ...198L.119B}; A7: \citet{1977AJ.....82..985B}; {\it A8:
\citet{1980ApJ...236..481M}}; {\bf A9: \citet{1983SvAL....9...12S}}; A10:
\citet{1985ApJ...288L..11M}; A11: \citet{1986PASJ...38..531N}; A12: \citet{1986A&A...157..318M};
{\bf A13 \citet{1987Ap&SS.132..263Z};} {\it A14: \citet{1988ApJ...329L.117B}}; 
A17: \citet{1988A&A...198..253M,1988A&A...198..267M}; {\it A18: \citet{1988ApJ...330L..61P}}; 
A19: \citet{1989ApJ...339..949H}; A20: \citet{1989ApJ...341..839M}; A21: \citet{1990ApJ...354..556H};
A22: \citet{1990A&A...240..116B}; {\it A23: \citet{1990ApJ...364..555P}};
{\it A24: \citet{1992ApJ...385..232J}}; {\it A25: \citet{1992AZh....69.1002K}};
{\bf A26* \citet{1992MNRAS.259..203C}}; A27: \citet{1989ApJ...339..949H}; 
A28: \citet{1993ApJ...413L.133S}; A29: \citet{1994A&AS..103..129K}; 
A30: \citet{1994MNRAS.268..464S}; A31: \citet{1995ARep...39...18V}
; A32: \citet{1996A&A...307..209W}; A33: \citet{1996A&A...314..615L}; {\it A34:\citet{1997ApJ...474..346M}}; 
A35: \citet{1997ApJ...478L..37S}; {\it A36: \citet{1997ApJ...490..758J}};
{\it A37: \citet{1998ApJ...497..800K}}; {\bf A38* \citet{1980ApJ...236..481M}};
{\bf A39: \citet{1998ApJ...498..763S}}; A40: \citet{1999ApJS..123..515S};
A41: \citet{2000MNRAS.317..315V}; A42: \citet{2001ARep...45...26K};  A43: \citet{2002ARep...46...49S}; A44: \citet{2004ApJS..155..149K}; A45: \citet{2005Ap&SS.295..217V}; A46 \citet{2009ApJ...702.1615C}; A47 \citet{2010A&A...517A..56F}; A48: \citet{2011MNRAS.413.2339V}; A49 \citet{2011ApJS..196....9C}; A50 \citet{2013ApJ...763....2G}; A51 \citet{2014MNRAS.439.2584V}; {\it A52 \citet{2014ApJ...788..187H}}; {\it A53 \citet{2014ApJ...794L..10Y}}}
\end{table*}

\subsection{Occurrence of Class\,I methanol masers}\label{online_a}

\begin{figure*}
\centering
\includegraphics[width=15cm, bb = 33 230 564 540,clip]{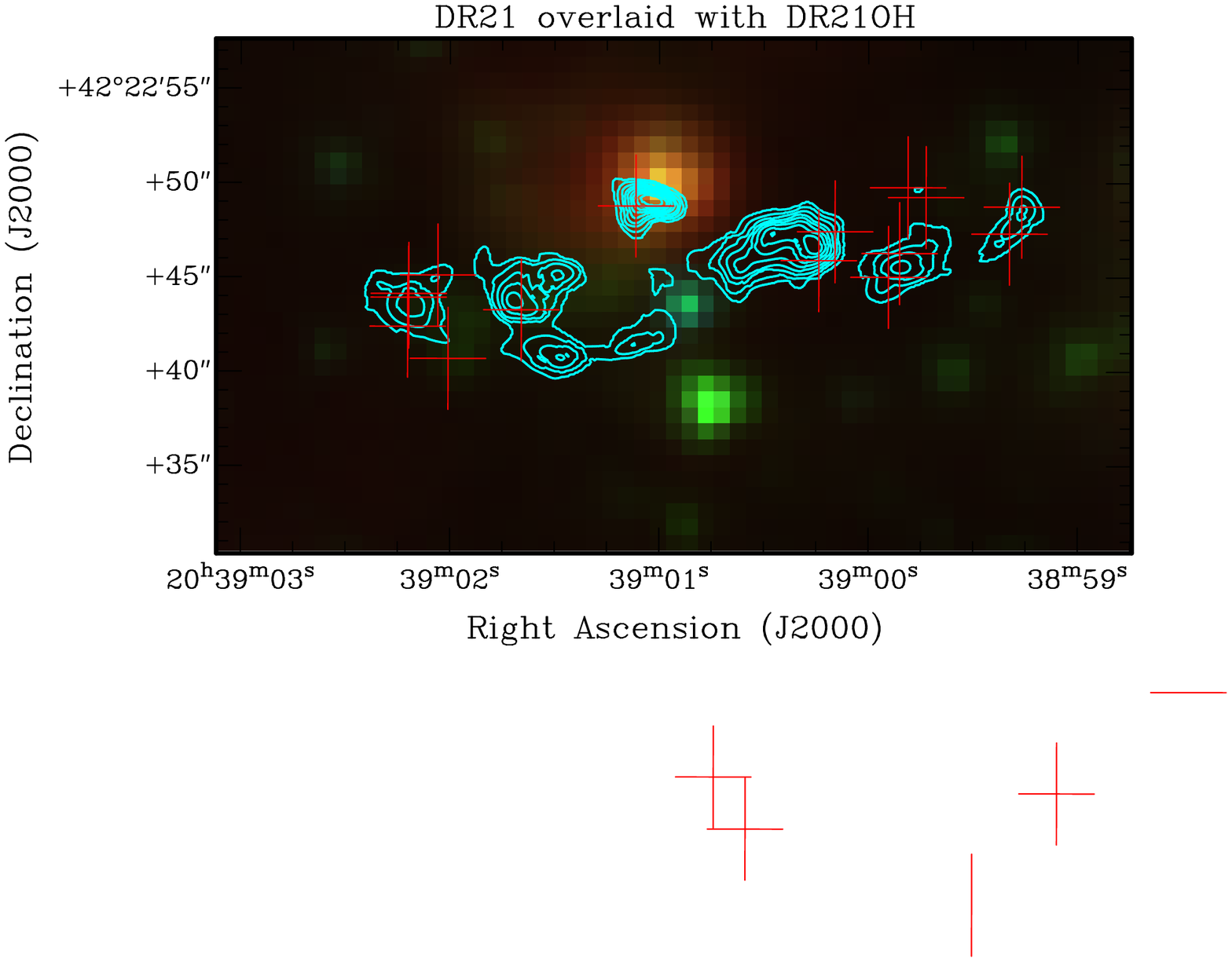}
\caption{Spitzer
IRAC image of the DR21(OH) region (3.6\,$\mu$m blue, 4.5\,$\mu$m green, 8.0\,$\mu$m red). 
The images were retrieved from the Spitzer archive.
The red crosses mark the positions of 36\,GHz masers from \citet{2011ApJ...729...14F}. Cyan contours represent the integrated intensity of the thermal $4_2\to3_1$-$E$ methanol  line \citep{2012ApJ...744...86Z} which traces the E-W outflow seen in the Class\,I masers.}\label{dr21}
\end{figure*}

In contrast to Class\,II masers,  which typically coincide  
with hot molecular cores, ultracompact H{\sc ii} regions,
OH masers, and near-IR sources \citep[e.g., ][]{1988ApJ...331L..41M,1988ApJ...333L..83M}, Class\,I methanol masers are frequently found in the general vicinity 
of intermediate- and  high-mass star formation, but often significantly offset (up to a parsec) from
prominent centres of activity, such as infrared sources or compact 
radio sources \citep{1986A&A...157..318M}. 
Based on large samples of sources, \citet{2004ApJS..155..149K} and \citet{2014MNRAS.439.2584V} concluded that the vast majority of Class\,I masers at 44 and 36\,GHz are within 1\,pc of 
a Class\,II methanol maser at 6.7\,GHz, taken as indicative of the position of the YSO. We note, incidentally, that the only
 two  cases reported by \citet{2004ApJS..155..149K} where Class\,I masers coincide with Class\,II features are observed with coarse angular resolution ($\sim0.1-0.2$\,pc).

\citet{1990ApJ...364..555P}  suggested that Class\,I methanol masers trace 
the swept-up interaction region of outflows
with ambient material.  This scenario is supported by an increasing number of 
observations that reveal 
a coincidence between Class\,I masers and other molecular shock tracers \citep[e.g.,][]{1992ApJ...385..232J,2004ApJS..155..149K,2006MNRAS.373..411V,2010MNRAS.408..133V}.
\citet{2009ApJ...702.1615C} surveyed a sample of massive young stellar objects 
with extended 4.5\,$\mu$m emission attributed to shocked H$_2$ in outflows and found a detection
rate  of 44\,GHz Class\,I methanol masers of $\sim89\%$.
In addition, Class\,I masers also seem  to be associated with molecular outflows from low-mass young stellar objects  \citep{2010MNRAS.405..613K,2010ARep...54..932K}.
It is important to note that Class\,I masers may be a unique tool for investigating outflow
  activity where other standard shock tracers are too weak to be detected. Figure\,\ref{dr21},
  for example, shows a zoom in DR21(OH) where several Class\,I maser
  lines clearly trace bow shocks in a bipolar outflow
  \citep{2009ApJ...698.1321A} not detected in Spitzer H$_2$ observations likely owing to extinction \citep{2007MNRAS.374...29D}.

A location in the swept-up interaction region of outflows
with ambient material may be a natural explanation for some
of the observed properties of Class\,I masers.
In C-shocks we expect  that the high relative velocities of charged
 grains and neutral gas cause methanol ice to be sputtered from the
 grain surface and released into the gas leading to the high methanol
 abundance and column density required for observable maser emission.
 Usually, no high-velocity Class\,I maser emission is observed 
and the  observed features are clustered within a
few km\,s$^{-1}$ around the ambient velocity. These features can be explained
by the fact that paths moving perpendicular to the 
line of sight have the longest velocity-coherent gain length
and have, naturally, velocities close to $\varv_{\rm LSR}$. 
However, it should be noted that \citet{2014MNRAS.439.2584V} recently reported  a small number of high-velocity Class\,I maser components, which are largely blue-shifted and detected more easily at 36\,GHz.

Class\,I masers, however, do not exclusively trace molecular outflows. 
 They have also been  detected in other astrophysical environments in the presence of shocked gas, for example
at the interaction of supernova remnants  with the molecular cloud 
\citep[e.g.,][]{1989IAUS..136..383S,1990ApJ...354..556H,2011ApJ...739L..21P}, in cloud-cloud collisions \citep[e.g.,][]{1992SvA....36..590S},
and  in layers where expanding H{\sc ii} regions interact with the ambient molecular environment \citep{2010MNRAS.405.2471V}.
Widespread Class\,I methanol masers have been  observed towards the  Galactic Centre \citep{2013ApJ...764L..19Y}. In this case, the authors have proposed  
that induced photodesorption by cosmic rays is responsible for the emission instead of large-scale shocks.
Recently,  \citet{2014ApJ...790L..28E} has  reported the first tentative detection of Class\,I maser emission at 36\,GHz in the nearby galaxy NGC\,253. This  detection would be intriguing since so far only the two Class\,II methanol maser lines that are by far the most  prominent in our Galaxy \citep[the 6.7 GHz and 12.2 GHz lines, e.g.,][]{1991ApJ...380L..75M} have been detected in a few nearby galaxies  \citep{2008MNRAS.385..948G,2010ApJ...724L.158S,2010MNRAS.404..779E}.

\subsection{Class\,I methanol maser spot sizes}\label{size}

Estimates of spot sizes for Class\,I masers have in general been  based upon VLA  measurements  and  VLBI non-detections  \citep[e.g.,][]{1976ApL....18...13B,1998AAS...193.7101L}.
For example, \citet{1997ApJ...490..758J}  used VLA observations of the  $6_2-6_1$-$E$ line at 25.018\,GHz to
 conclude that spot sizes in Orion-KL were in the range 22--56\,AU. \citet{2009ARep...53..519S}
 have also reported that the 44\,GHz maser has average spot sizes of 50\,AU in OMC-2 and in NGC\,2264, and have concluded that
Class\,I maser spots are much larger than Class\,II maser spots, which have typical sizes of a few AU 
\citep{1999ApJ...519..244M,2010A&A...517A..71S}, implying brightness temperatures of the order of  $10^{12}~$K.

Recently, \citet{2014ApJ...789L...1M} detected the 44\,GHz maser in the massive star forming region IRAS\,18151-1208 for the first 
time with a VLBI experiment. They found that  most of the 
maser emission is not due to a compact point source,
 but  originates from a  spatially extended structure which accounts for  about $73\%$  of the total flux of the maser. 
The strongest  maser feature has a  
linear size of $\sim15$\,AU$\times$9\,AU  and a brightness temperature of  $\sim10^{10}$\,K,  but this should be regarded as a lower limit to the real spot size given the severe filtering of extended emission that affects the data.  We note that several masers may have a compact high brightness temperature core surrounded by a larger envelope (hundreds of AU)  of lower brightness.

The  sizes and brightness temperatures measured for the other features are listed in  Table\,\ref{luminosity} together with
typical isotropic luminosities for different Class\,I masers from the literature. The scatter in the most commonly observed lines at 36\,GHz and 44\,GHz can be larger than reported as large samples were observed in these transitions. For the 25\,GHz series, values reported for the brightness temperature and the spot size are for the $6_2-6_1$ line only and are measured in Orion-KL.
For all lines the reported brightness temperatures and spot sizes are based on interferometric observations and are in most of the cases based on only one source. 
 In the following discussion, we will assume spot sizes of 50\,AU and 100\,AU.

\begin{table*}
\begin{center}
\caption{Summary of typical Class\,I maser luminosities, $L_{\rm{iso}}$ , brightness temperatures, $T_b$, and spot sizes.
\label{luminosity}}
\begin{tabular}{lllcc}             
\hline\hline
Line                         &$L_{\rm{iso}}$  & $T_b$&Spot sizes&Source\\
                             &photons\,s$^{-1}$  &     K                 &AU\\
\hline                       
9.9\,GHz             &$10^{43}\,^{a}$&$10^{4}$\,$^{a}$\\
25\,GHz series     &$10^{41}-10^{43}$\,$^{ a,b,c,d}$&$2\,10^6-3\,10^7$\,$^b$           & 22--56\,$^b$& Orion-KL$^b$\\
36\,GHz              &$10^{43}-10^{44}\,^{ e}$ &              &\\
44\,GHz             &$10^{42}-10^{45}\,^{ f,g}$&$4\,10^7-10^{10}$\,$^{g,h}$         & 6--50\,$^{g,h}$&IRAS\,18151--1208$^g$\\
84\,GHz              &$3\,10^{42}-2\,10^{43}\,^{a}$&$6\,10^3-5\,10^5\,^{a}$             &\\
95\,GHz         &$5\,10^{42}-10^{45}\,^{a,i,j,k}$  &$2\,10^4-10^5\,^{a}$              &\\
104\,GHz        &$2\,10^{43}\,^{a}$&$2\,10^4\,^a$\\
\hline
\end{tabular}
\end{center}
\tablefoot{Maser luminosities are taken from interferometric studies. Spot sizes
are from the VLA observations of  \citet{1997ApJ...490..758J} at 25\,GHz, and from the VLBI
detection of  \citet{2014ApJ...789L...1M} and are to be considered as upper and lower limits, respectively.
The measurements reported by \citet{2006MNRAS.373..411V} are based on linear resolutions of 4000--6000\,AU.
\tablefoottext{a}{\citet{2006MNRAS.373..411V}.}
\tablefoottext{b}{\citet{1997ApJ...490..758J} for the $6_2-6_1$ line.}
\tablefoottext{c}{\citet{1986A&A...157..318M}.}
\tablefoottext{d}{\citet{1992ApJ...385..232J}.}
\tablefoottext{e}{\citet{2011ApJ...729...14F}.}
\tablefoottext{f}{\citet{2004ApJS..155..149K}.}
\tablefoottext{g}{\citet{2014ApJ...789L...1M}.}
\tablefoottext{h}{\citet{2009ARep...53..519S}.}
\tablefoottext{i}{\citet{2007A&A...472..867M}.}
\tablefoottext{j}{\citet{1990ApJ...364..555P}.}
\tablefoottext{k}{\citet{2011ApJS..196....9C}.}
}
\end{table*}

\subsection{Are Class\,I masers saturated?}\label{saturation}

  Observed brightness temperatures
are upwards of $10^6$K when observed with high linear resolution and linear
sizes in the range
10-100\,AU (see Table\,\ref{luminosity}), whereas the expected limiting brightness temperature $T_{s}$ for
 an unsaturated maser (see Sect.\,\ref{sat}) is typically $10^5(4\pi/\Omega_{\rm b})$\,K, where $\Omega_{\rm b}$ is the maser beam solid angle.
 Moreover,  saturated
 masers are expected to show little or no evidence of variation.
Indeed,
saturated masers undergo linear amplification, whereas unsaturated
 masers should amplify exponentially. This suggests qualitatively
 that unsaturated masers should vary on much shorter time scales
 than their saturated counterparts. 
      A reasonable guess for the variation time scale of a saturated
      maser would be $l_{spot}/\varv_{sh}$, where $l_{spot}$ is the
      maser spot size and $\varv_{sh}$ is the shock velocity. Taking
      $l_{spot}$ =100\,AU and a shock velocity of 30\,km\,s$^{-1}$, we
      find a time scale of $\sim 15$ years.  This is consistent
      with most observations \citep{1988A&A...198..267M,
        2004ApJS..155..149K} which show that Class\,I masers have little time variation.  Dedicated observations of Class\,I masers to  detect variability over time scales of several years are  lacking, however,  and variability of Class\,I masers is basically still an uncharted territory.

 We also note  that theory suggests that with increasing amplification,
 unsaturated masers are expected first to narrow and then to
 re-broaden until the line width approaches the intrinsic value
 determined by local turbulent and thermal broadening. However,
 observationally,  the line profiles of different Class
 I masers are found to be very similar
 \citep[e.g.,][]{1988ApJ...329L.117B,1990ApJ...364..555P,2014MNRAS.439.2584V}.
 In particular for the DR21(OH) region, interferometric observations
 show that all 84\,GHz components have counterparts in the 95\,GHz
 line at identical velocities \citep[e.g.,][]{1990ApJ...364..555P}
 and that several transitions are spatially coincident \citep[from the
   36\,GHz line to the 229\,GHz maser,][their Fig.\,1 and references
   therein]{2011ApJ...729...14F}.  It is  worth noting, however, that
  real observed maser emission is likely a mixture of both saturated
  and unsaturated contributions. Taken all together, the observational properties seem to suggest that bright Class\,I masers are at least partially saturated.

\section{Previous excitation studies on Class\,I masers}\label{models}

\begin{figure*}
\centering
\subfigure[][]{\includegraphics[width=0.4\textwidth]{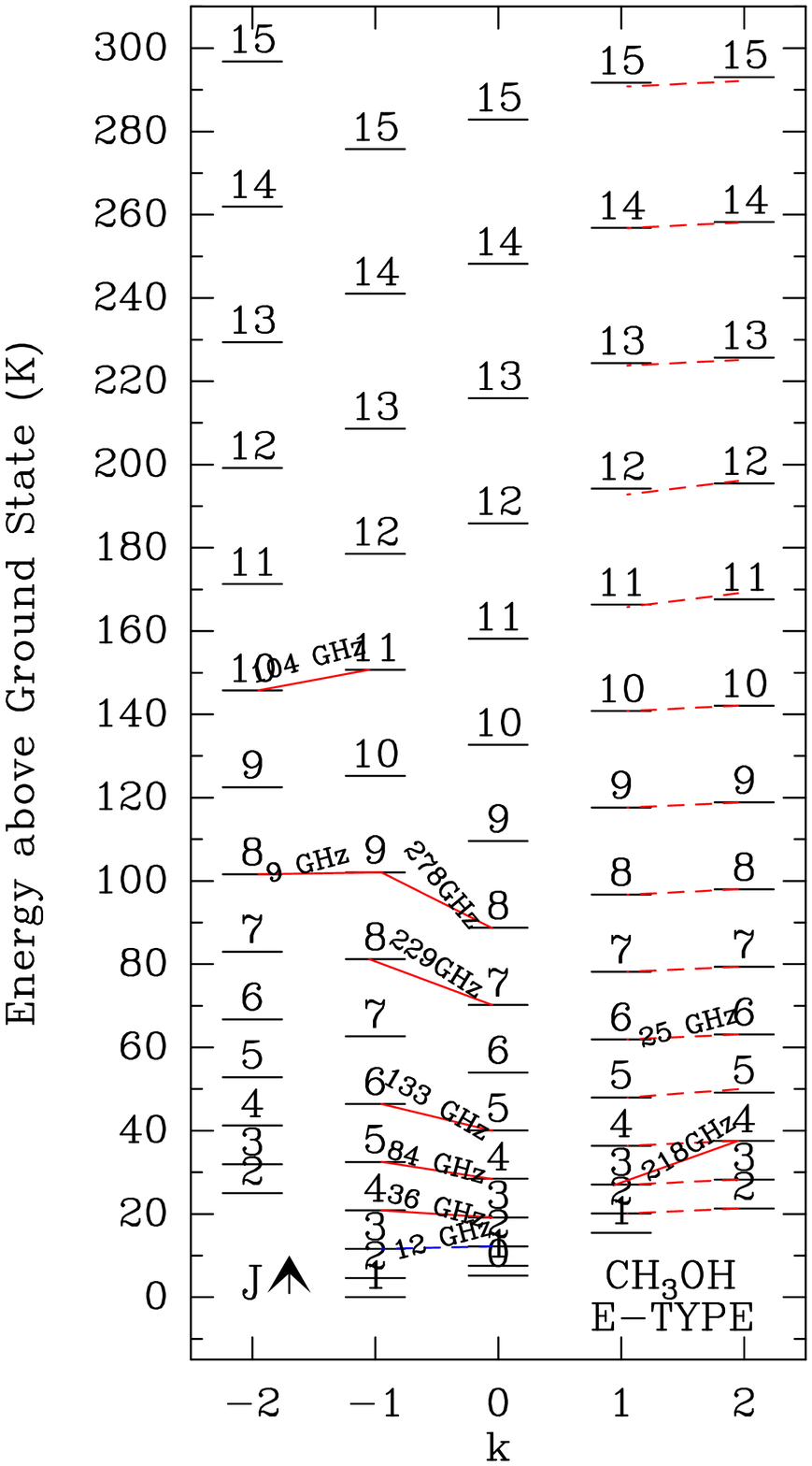}}
\subfigure[][]{\includegraphics[width=0.4\textwidth]{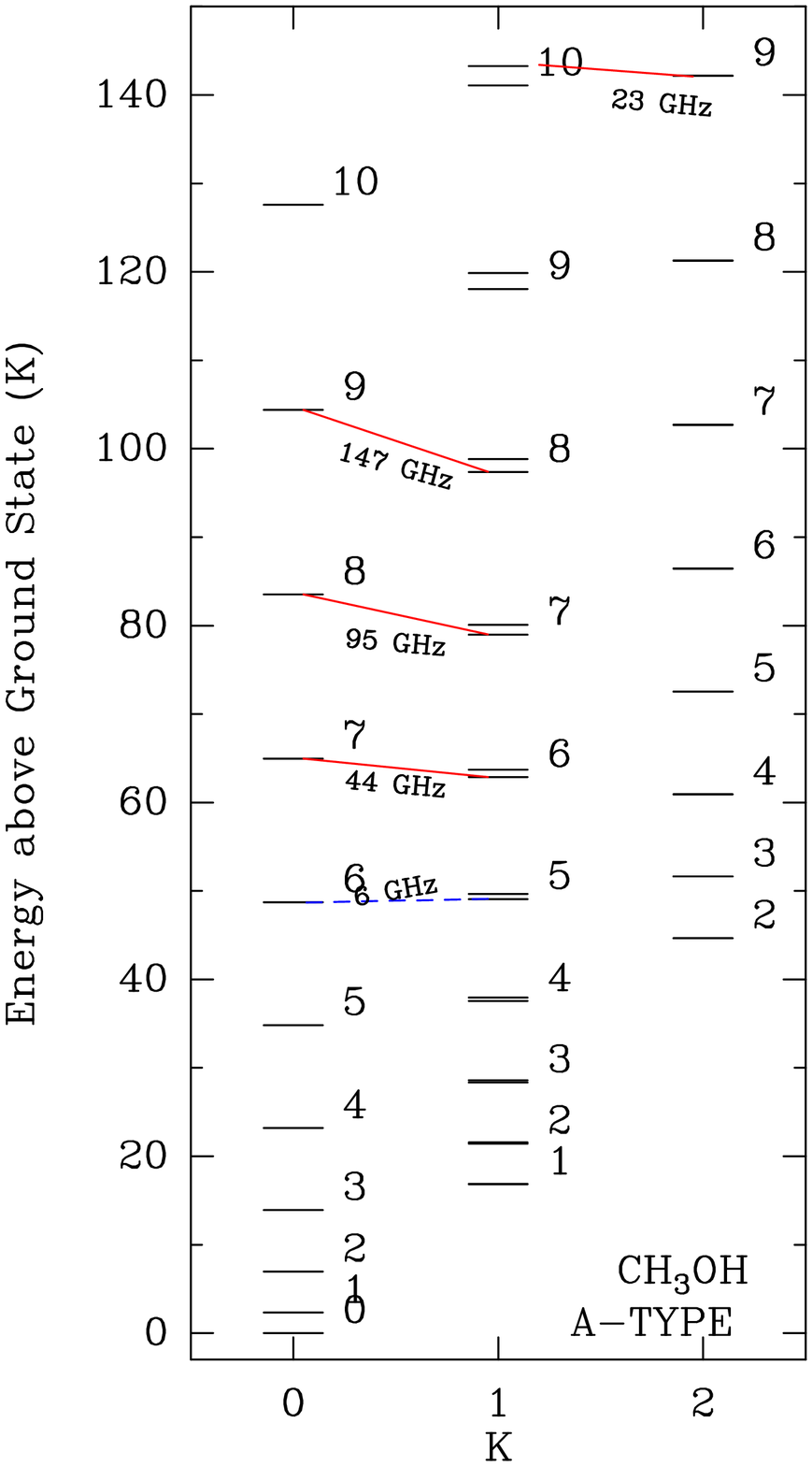}}
\caption{Partial rotational level diagram of $E$-type (left) and
$A$-type methanol (right). Known Class\,I maser transitions are connected
by red lines with frequencies indicated; blue dashed lines connect anti-inverted transitions.
 We note that all of the maser lines
originating in the  $J_2\to J_1E$ series near 25\,GHz are indicated by dashed red lines.}\label{levels}
\end{figure*}

Previous statistical equilibrium studies on CH$_3$OH Class\,I masers 
made use of collisional rates based on experimental results by \citet{lees1969} 
and \citet{lees1974}.
These rates show  a propensity for $\Delta k=0$ collisions and a dependence upon  $\Delta J$ 
as 1/$\Delta J$.
The propensity for
$\Delta k=0$ over $\Delta k = 1$ collisions causes molecules to preferentially
de-excite down a $k$ stack (see Fig.\,\ref{levels}), leading  to a situation 
where the lower levels in the $k=-1$ stack are  overpopulated relative to  those
in the $k=0$ stack (for CH$_3$OH-$E$), resulting in maser action in the  $J_{-1}\to (J-1)_{0}$-$E$ series
and enhanced absorption in 
the $2_{0}\to3_{-1}E$ line, the second most prominent 
Class\,II  maser line.  Similar mechanisms lead to inversion in the $9_{-1}\to8_{-2}$ and $11_{-1}\to10_{-2}$-$E$ lines.
Analogously, the $K=0$ ladder of CH$_3$OH-$A$ is overpopulated, leading to the 
$J_{0}\to(J-1)_{1}$-$A^{+}$  masers and to the anti-inversion of the $5_1\to 6_0A$ Class\,II transition,  the strongest Class\,II maser line.
 Indeed, observations showed that the $2_{0}\to3_{-1}E$ and $5_1\to 6_0A^+$ Class\,II lines are detected in absorption 
toward a number of Class\,I sources \citep[e.g.,][]{1987Natur.326...49B,1991ApJ...380L..75M} and also 
in dark clouds \citep{1988A&A...197..271W}.

However, while
the previously used collisional rates can reproduce the 
anti-inversion at 12.18\,GHz and 6.7\,GHz and several  Class\,I masers (4$_{-1}\to3_0E$ at 36\,GHz, 
5$_{-1}\to4_0E$ at 84\,GHz, 7$_{0}\to6_1A^+$ at 44\,GHz, 8$_{0}\to7_1A^+$ at 95\,GHz, and
9$_{0}\to8_1A^+$ 146\,GHz; see Fig.\,\ref{levels}), they do not account for  
the Class\,I maser series $J_{2}\to J_1E$ at 25\,GHz. For this reason, 
\citet{1992ApJ...385..232J} suggested an ad hoc
additional preference for the $\Delta k=3$ collisions and the assumption that 
$\Delta k =2$ collisions are not allowed to be able to collisionally pump the $J_{2}\to J_1E$ Class\,I masers. 
Alternatively, \citet{1992MNRAS.259..203C} modelled the pumping of interstellar methanol masers using the  hard sphere
model of \citet{1974ApJ...189..441G}, which represents the extreme case of completely unselective collisions.
In this case, all Class\,I masers, including the  $J_{2}\to J_1E$ series, appear for densities in the range $10^2-10^5$\,cm$^{-3}$.

The general conclusion of  previous models of Class\,I masers is that these
lines are collisionally excited and that they mase in different
density ranges.
These general results have recently been confirmed by \citet{2014ApJ...793..133M} and \citet{2016MNRAS.455.3978N}.

\section{Statistical equilibrium results}\label{stateq}
  
We performed large velocity gradient (LVG) calculations using a modified version of the code presented by \citet{2004A&A...422..573L}  with a plane-parallel slab geometry appropriate for shocks. Following a standard formalism \citep{1992ASSL..170.....E}, the maser levels are  described by pump ($P_m$) and loss ($\Gamma_m$) rates that regulate the interactions of the different maser systems with all other levels (which we refer to as the  maser reservoir)

\begin{equation}\label{pump}
P_{m}= \sum_{i\neq m} n_i\times(A_{im}\times\beta_{im}+C_{im}\times n)
\end{equation}

\begin{equation}\label{loss}
\Gamma_{m}= \sum_{i\neq m} (A_{mi}\times \beta_{mi}+C_{mi}\times n)
,\end{equation}where $m=1,2$ are the two levels of the maser system, $n_i$  the
CH$_3$OH density in level $i$, $n$  the molecular hydrogen density,
$A_{im}$  the Einstein coefficients for spontaneous
emission (with $A_{im}=0$ for $i<m$),  $C_{im}$ the temperature dependent collisional rates,
$\beta_{im}$ the escape probability ($\beta_{im}=\frac{(1-exp(-3\tau_{im}))}{3\tau_{im}}$ for a slab geometry,  and $\tau_{im}$  the optical depth (perpendicular
 to the line of sight) given by Eq.\,2 of \citet{2004A&A...422..573L}.  
The pump and loss rates  are obtained by solving the level populations for the full system (maser and reservoir) in the absence of maser radiation. We also note 
 that we neglect in these calculations the continuum dust opacity when
 computing $\beta_{im}$ and thus assume that the opacity at wavelengths of
 transitions feeding maser levels is small.

We updated the code with the more recent
collisional rates for the torsional ground states of $A$- and $E$-type methanol from \citet{2010MNRAS.406...95R}. 
The rates were computed for temperatures in the range from 10 to 200 K including rotational states up to $J=15$ for collisions with para-H$_2$. 
We assume in this study that  collisions with
 ortho-H$_2$ can be neglected or, equivalently, that the abundance ratio of ortho- to
 para-H$_2$ is low. This  needs to be investigated
 in future work, but we note that if the ratio of ortho- to para-H$_2$
 is low in the pre-shock gas, this may also hold
 in the post-shock layer  because the time taken to traverse the
 shock may be too short to thermalise the ratio. Indeed, \citet{2000A&A...356.1010W} found that significant conversion from para- to ortho-H$_2$ starts at 700\,K in C-type shocks (and at $T>1000$\,K in J-type). 
 
We ran models for densities between $10^4$ and $10^9$\,cm$^{-3}$  
and temperatures, $T$, of 20--400\,K. Since collisional de-excitation rates from \citet{2010MNRAS.406...95R} are computed for temperatures up to 200\,K, we use the rates computed at 200\,K for higher temperatures.
We used specific column densities ($N/\Delta\varv$) of CH$_3$OH-$A$ and -$E$ between $10^{14}$ and $10^{18}$\,cm$^{-2}$\,km$^{-1}$\,s.
In the  absence of any external radiation field except the cosmic background,
all known Class\,I maser transitions should be reproduced and the  
Class\,II 2$_0\to3_{-1}E$ and 5$_1\to6_0A^+$ masers should be  anti-inverted.
All Class\,I transitions reported in Table\,\ref{obs} show inversion in the analysed range of physical conditions.  In addition, the
2$_0\to3_{-1}E$ transition and the  5$_1\to6_0A^+$ line present anti-inversion 
as expected. In particular, the range of densities for which the lines are anti-inverted depends on the temperature and column density. For the same value of $T$, the 2$_0\to3_{-1}E$ line is anti-inverted at densities lower than the 5$_1\to6_0A$ transition. The absorption temperature increases with methanol column density  and kinetic temperature.

In the following discussion, we  present the pumping mechanisms that lead to maser action in the Class\,I lines and 
 the dependency of the masers on the physics of the gas. We
should note that although the general behaviour is similar to the
results recently published by \citet{2014ApJ...793..133M}, our 
calculations produce inversion in the  36\,GHz line at lower densities ($10^4-10^5$\,cm$^{-3}$)  
than found by \citet{2014ApJ...793..133M} ($10^6$\,cm$^{-3}$). The main difference between our models and those by
\citet{2014ApJ...793..133M} is that the latter use a 30\,K dust radiation field, while our models include only the cosmic radiation 
field.  This assumption is justified by the fact that Class\,I masers are usually found far from continuum sources
  \citep[e.g.,][]{2014MNRAS.439.2584V}.
To test our results, we benchmarked our code with the RADEX
LVG code \citep{2007A&A...468..627V} with a uniform slab geometry. The results of the two codes are consistent  up
to a specific column density of $10^{17}$\,cm$^{-2}$\,km$^{-1}$\,s.

\subsection{Pump and loss rates}

In this paper, we  confine ourselves to calculating the maser
 efficiency and other characteristics of saturated masers
  following the formalism used  by \citet{2013ApJ...773...70H} for water masers.
   We  define the methanol maser emission measure, $\xi$, 
 which is  a measure of the rate at  which pump photons are produced.
  It is defined as the ratio of the product of the
 molecular hydrogen and methanol number densities over the velocity
 gradient in the plane of the sky:

\begin{equation}\label{csi}
\xi = \frac{X_{6}({\rm{CH_3OH}})\times n_6^2}{(dV/dr)}  
.\end{equation}Here,  $X_{6}(\rm{CH_3OH})$ is the methanol abundance
($n({\rm{CH_3OH}})/n({\rm{H_2}})$)
 in units of $10^{-6}$ (we assume equal abundances for A- and E-type
 methanol), $n_6$  the H$_2$ number density in units of $10^6$\,cm$^{-3}$, and
 $dV/dr$  the velocity gradient in units of km\,s$^{-1}$\,pc$^{-1}$ 
 (for rough estimates, we assume a shock velocity of 30\,km\,s$^{-1}$ and a shock width of $10^{16}$\,cm; see \citealt{1997A&A...321..293S,2008A&A...482..809G}).
 We note that  $\xi$ can also be expressed  in terms of the
 ratio of methanol column density (in the plane of the sky) to
 line width, $N\rm{(CH_3OH})/d\varv$:

\begin{equation}\label{ncol}
  \frac{N{\rm{(CH_3OH)}}}{\rm{d\varv}} = \frac{3.3\,10^{18} \times n(CH_3OH)}{dV/dr} 
\end{equation}

\begin{equation}\label{csi2}
\xi = \frac{{{n({\rm{H_2}})}\times\frac{N(\rm{CH_3OH)}}{d\varv}}}{3.3\,10^{24}} 
.\end{equation}

\begin{figure*}
\centering
\subfigure{\includegraphics[width=0.85\textwidth]{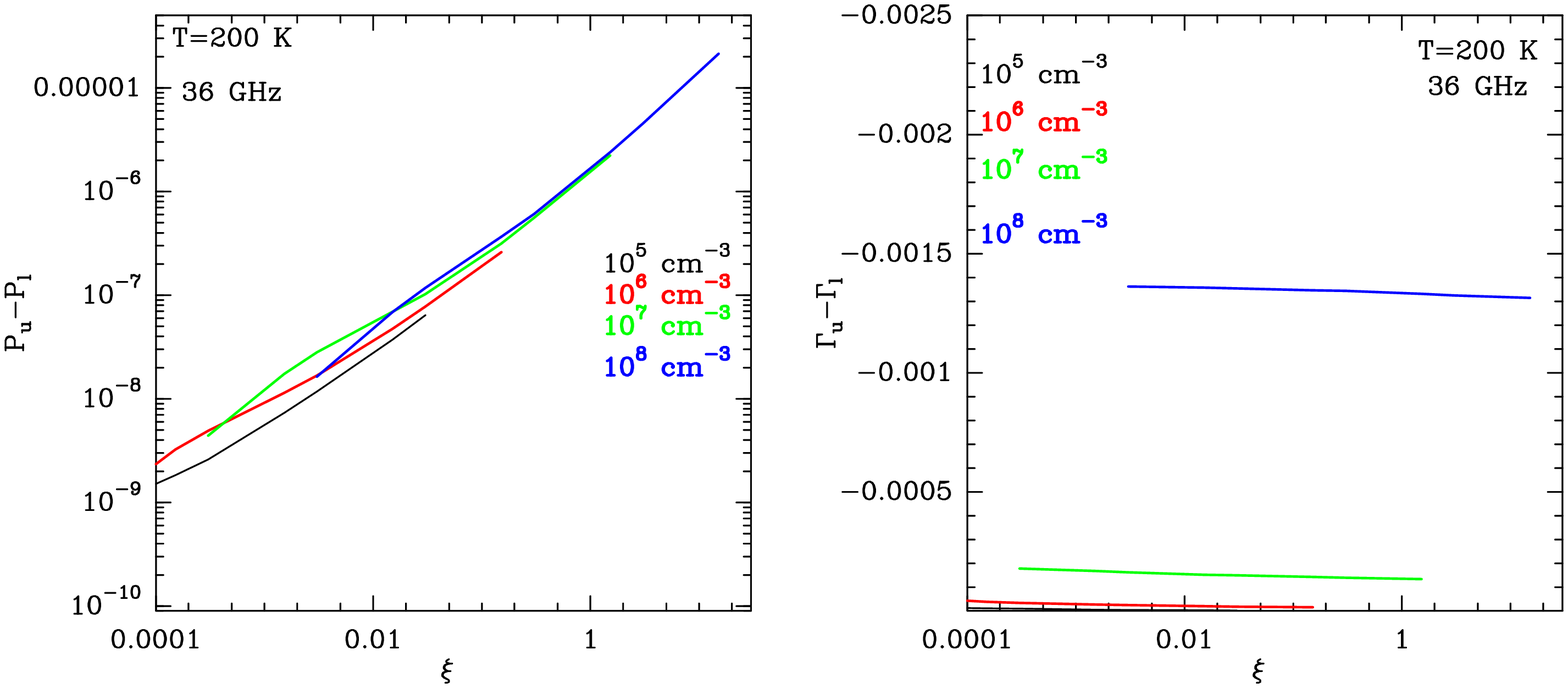}}
\subfigure{\includegraphics[bb=1 100 817 472 ,clip,width=0.85\textwidth]{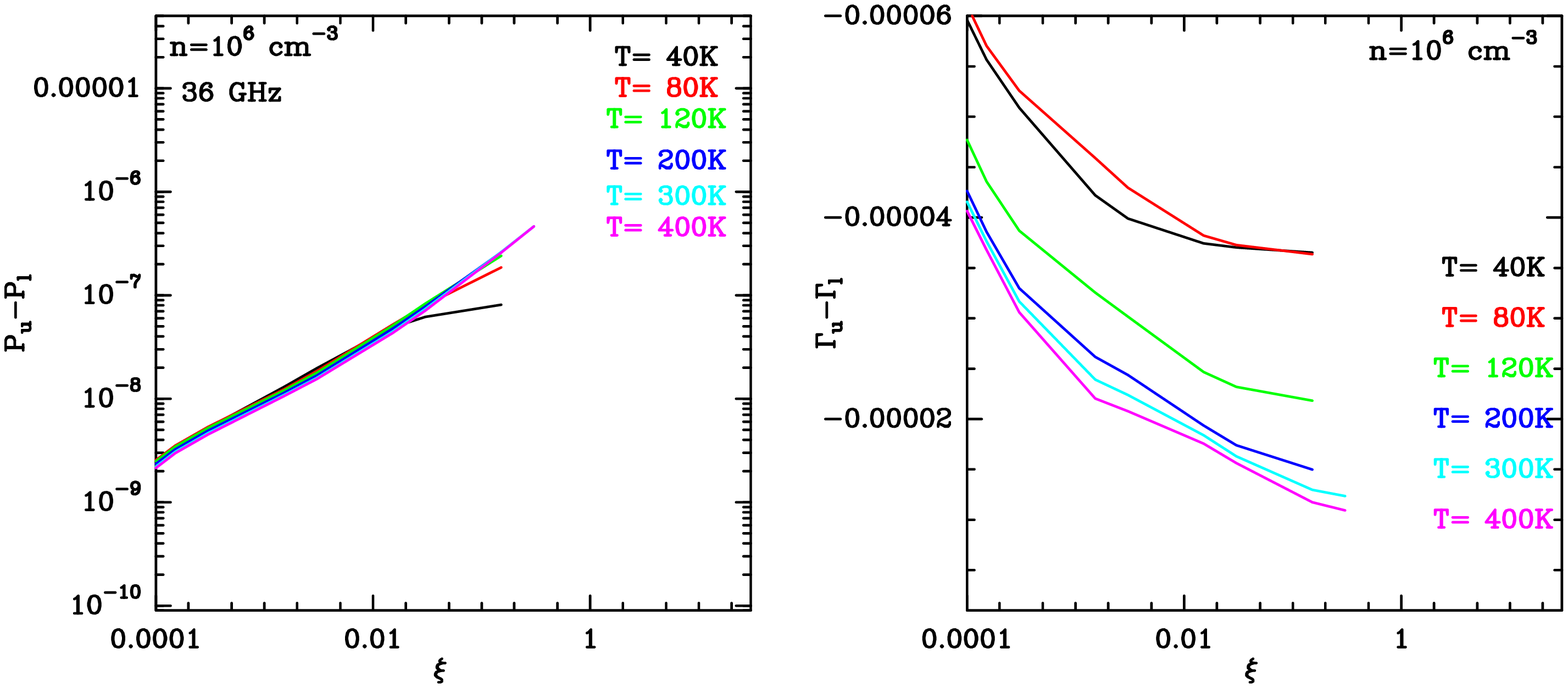}}
\caption{ Effective pump and loss rate for the 36\,GHz maser  at different densities and for a temperature of 200\,K (upper panels)
  and at different temperatures for $n=10^6$\,cm$^{-3}$ (lower panels).  It is possible to convert $\xi$ to methanol abundance using Eq.\,\ref{csi}.}\label{pl36}
\end{figure*}

To investigate the pumping mechanism of Class\,I methanol masers, we
define effective pump and loss rates of a given maser system as
$P=P_u-P_l$ and $L=\Gamma_u-\Gamma_l$, and study their behaviour as  functions of
$\xi$ for different temperatures and densities. As an example, $P$ and
$L$ for the 36\,GHz lines are represented in Fig.\,\ref{pl36}, which shows that both the relative population rates of the upper and lower
 maser levels, as well as their relative decay rates, play a role in
 causing population inversion.
This process does not depend significantly on
the temperature, while it does depend on the volume density of the gas, which 
emphasises the importance of collisions for Class\,I masers. 
The other Class\,I masers behave similarly.

\subsection{Conditions for population inversion}\label{inversion}

\begin{figure}
\centering
\subfigure{\includegraphics[width=0.35\textwidth]{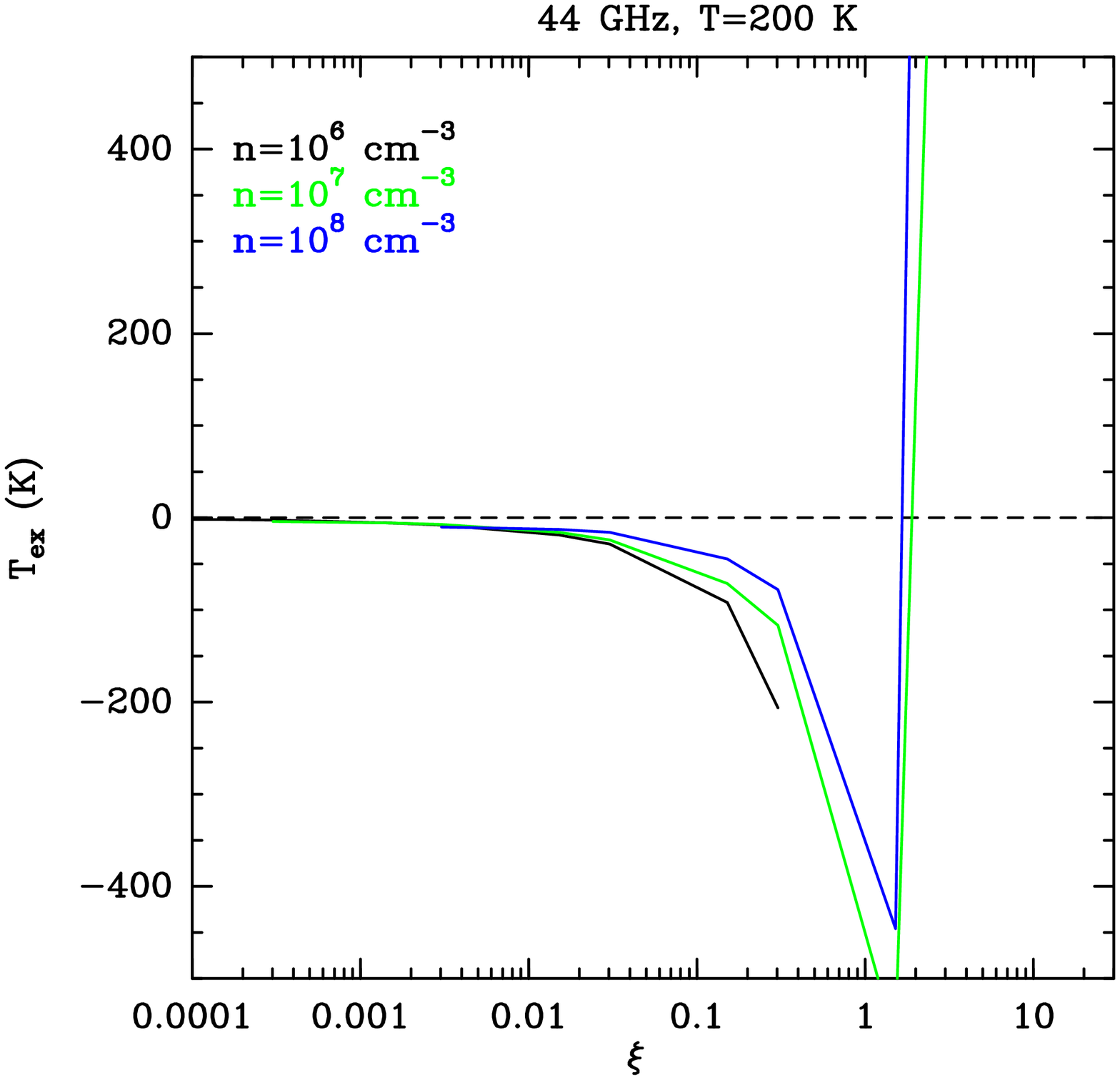}}
\subfigure{\includegraphics[width=0.35\textwidth]{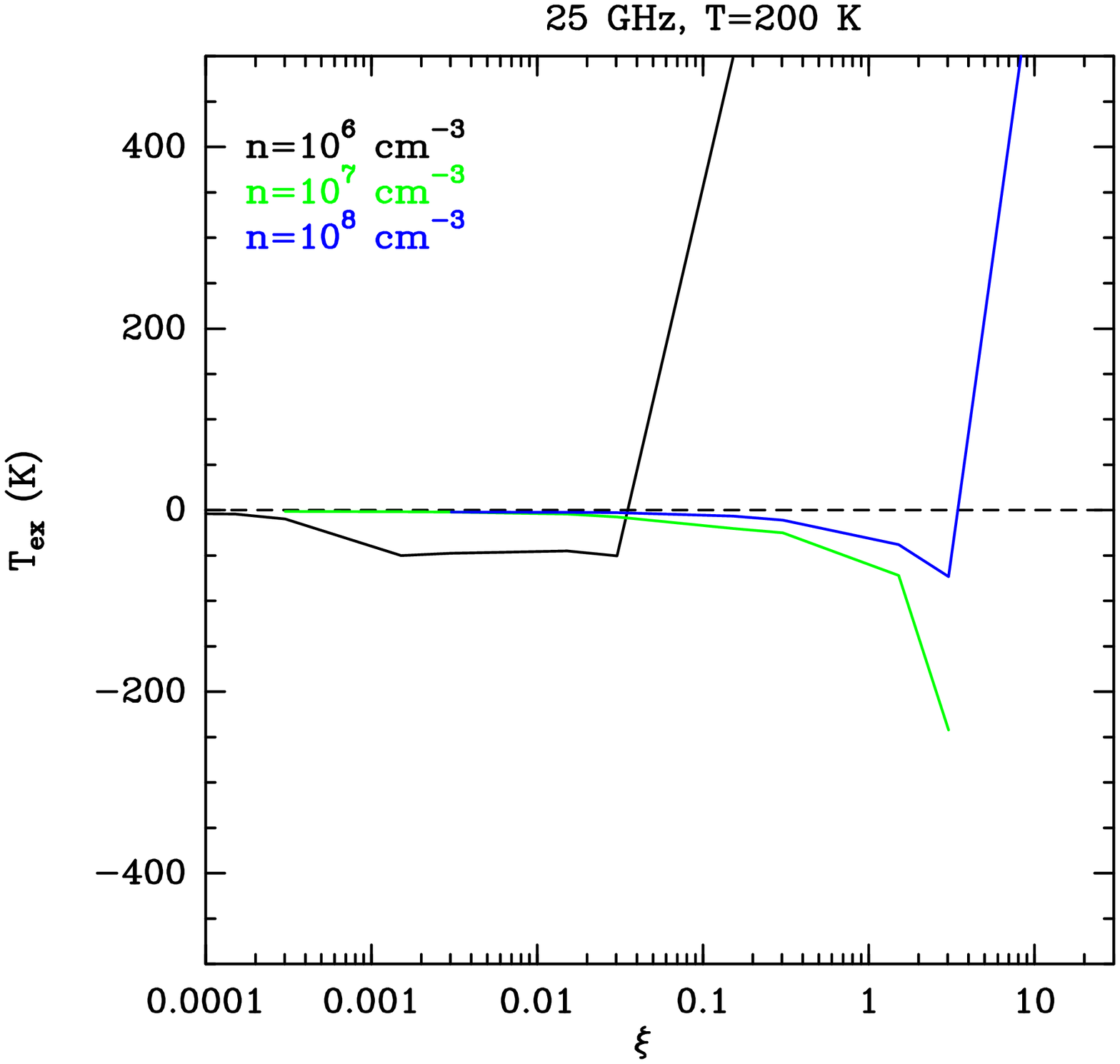}}

\caption{Excitation temperature, $T_{\rm ex}$, of the 44\,GHz (upper panel) 
and of the $6_2-6_1$-$E$ 25.018\,GHz line (lower panel)
at different densities and for a temperature of 200\,K. Curves are plotted until $T_{\rm ex}$ becomes positive.
It is possible to convert $\xi$ to methanol abundance using Eq.\,\ref{csi}.  We note that with increasing $\xi$, the excitation temperature 
 typically transits from -infinity to +infinity before thermalising. 
  The discrete nature of our grid in $\xi$ occasionally obscures this
 fact.}\label{tex4425}
\end{figure}

\begin{figure*}
\centering
\subfigure{\includegraphics[width=0.9\textwidth]{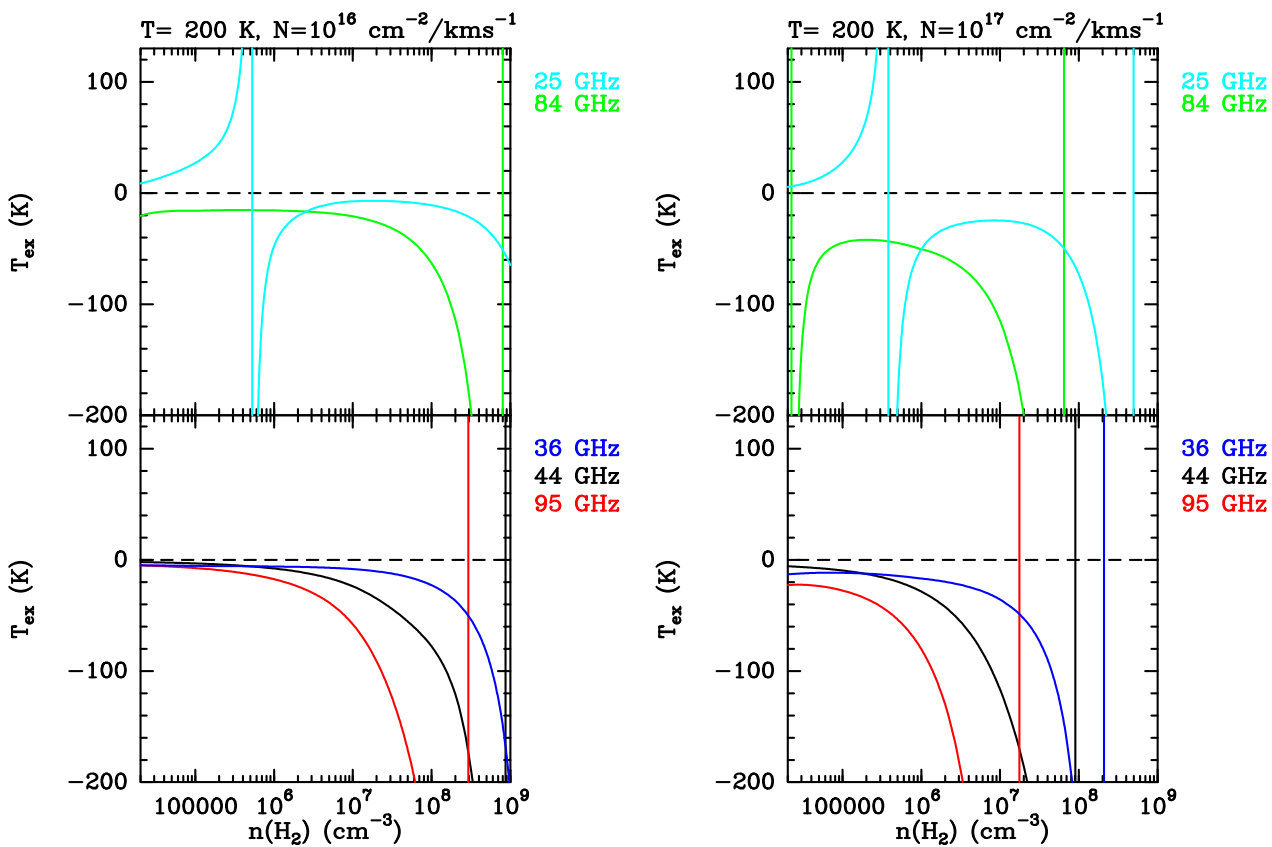}}
\caption{Excitation temperature of the 44\,GHz, 95\,GHz, and 36\,GHz lines (lower panels), and the  84\,GHz and $6_2-6_1$-$E$ masers (upper panels) for a temperature of 200\,K and a specific column density of $10^{16}$\,cm$^{-2}$\,km$^{-1}$\,s (left) and of  $10^{17}$\,cm$^{-2}$\,km$^{-1}$\,s (right) as functions of volume density. The dashed horizontal line separates maser from non-maser behaviour. Vertical lines mark the transitions from maser to non-maser behaviour.}\label{texden}
\end{figure*}

Figure\,\ref{tex4425} shows the excitation temperature of the 44\,GHz and of the $6_2-6_1$-$E$ 25.018\,GHz (representative of the 25\,GHz series) masers
 to illustrate the range of
$n$ and $\xi$ where the lines are inverted and where the pump and loss
rates are meaningful quantities for a temperature of 200\,K.  For completeness, similar plots for the 36\,GHz, 84\,GHz, and 95\,GHz lines are shown in Appendix\,\ref{online_c} (Fig.\,\ref{tex8495}).
Table\,\ref{quench} summarises the value of $\xi$, $\xi_q$,  at which population inversion is quenched for temperatures of 80\,K and of
200\,K for different Class\,I maser lines at different densities. From
Eq.\,\ref{csi2},  the specific methanol column density at
which each maser line quenches can be derived.  At a temperature of 200\,K, the
36\,GHz line thermalises around a few
$10^{17}$\,cm$^{-2}$\,km$^{-1}$\,s independently of density for
$n>10^6$\,cm$^{-3}$; a similar behaviour is shown by the 84\,GHz lines,
which also quenches at $\sim10^{17}$\,cm$^{-2}$\,km$^{-1}$\,s for all
densities. The same
is found for the other lines reported in Table\,\ref{quench}: they all
quench at a constant specific column density of approximately $10^{17}$\,cm$^{-2}$\,km$^{-1}$\,s independently of volume
density in the range $10^5-10^7$\,cm$^{-3}$, which  suggests that there is
a critical methanol column density at which quenching occurs. At a lower temperature of 80\,K, the specific column density at which the quenching occurs increases to approximately  $10^{18}$\,cm$^{-2}$\,km$^{-1}$\,s at a volume density of $10^8$\,cm$^{-3}$ except for the $6_2-6_1$-$E$ 25.018\,GHz line.
To give a complete view of the range of parameters where the Class\,I maser lines are inverted,
we also plot their excitation temperatures as functions of densities for a temperature of 200\,K and
a specific column density of $10^{17}$\,cm$^{-2}$\,km$^{-1}$\,s (Fig.\,\ref{texden}).
 Figure\,\ref{texden} clearly illustrates the effect that varying the volume density has on individual lines:
 the 25\,GHz series quenches at the highest volume density and it is  inverted
for $n>10^6$\,cm$^{-3}$; the 36\,GHz, 44\,GHz, and 95\,GHz lines are those inverted over the broadest
range of volume densities. Figure\,\ref{texden} also shows that the 84\,GHz may be not inverted at low densities (a few $10^4$\,cm$^{-3}$). However, this behaviour is found only at high specific column densities ($>10^{16}$\,cm$^{-2}$\,km$^{-1}$\,s) and in a narrow range of densities.

\begin{table*}
\begin{center}
\caption{$\xi_q$ for various Class\,I lines at different densities for $T=80$\,K and 200\,K.}\label{quench}
\begin{tabular}{lccccc}  
  \hline\hline
  \multicolumn{6}{c}{T=80\,K}\\
  \hline
Volume density & 36\,GHz& 84\,GHz& 44\,GHz& 95\,GHz & 25\,GHz$^a$\\
\hline                       
$10^6$\,cm$^{-3}$&$6 \times 10^{-2}$&$1 \times 10^{-1}$&$5 \times 10^{-2}$&$2 \times 10^{-2}$&$4 \times 10^{-2}$\\
$10^7$\,cm$^{-3}$&$1$              &$3 \times 10^{-1}$&$3 \times 10^{-1}$&$9 \times 10^{-2}$&$4 \times 10^{-1}$\\
$10^8$\,cm$^{-3}$&$5 \times 10^{-1}$&$2 \times 10^{-1}$&$2 \times 10^{-1}$&$7 \times 10^{-2}$&$3$\\
\hline
\multicolumn{6}{c}{T=200\,K}\\
  \hline
$10^6$\,cm$^{-3}$&$2 \times 10^{-1}$&$3 \times 10^{-2}$&--               &$2 \times 10^{-1}$&$3 \times 10^{-2}$\\
$10^7$\,cm$^{-3}$&$3 \times 10^{-1}$&$5 \times 10^{-1}$&$1.9$&$2$        &$8 \times 10^{-1}$\\
$10^8$\,cm$^{-3}$&$7$              &$2$               &$1.7$            &$5 \times 10^{-1}$&$3$\\
\hline
\end{tabular}
\tablefoot{\tablefoottext{a}{We refer here to $6_2-6_1$-$E$ transition.}}

\end{center}
\end{table*}

\subsection{Maser emission}

We follow the formalism developed by \citet{1989ApJ...346..983E} and \citet{2013ApJ...773...70H} for H$_2$O masers and describe the maser emission through the    pump rate coefficient ($q$), the  maser loss rate ($\Gamma$), and the inversion efficiency of the pumping scheme ($\eta$) as

\begin{equation}\label{q}
q = \frac{1}{2}\times\frac{P_u+P_l}{n^2\times (CH_3OH)}   
\end{equation}

\begin{equation}\label{gamma}
  \Gamma = \frac{g_u\Gamma_u+g_l\Gamma_l}{g_u+g_l}
\end{equation}

\begin{equation}\label{eta}
  \eta = \frac{P_u\Gamma_l-P_l\Gamma_u}{P_u\Gamma_l+P_l\Gamma_u}
.\end{equation}Here  
$g_u$ and $g_l$ are the statistical weights for the upper and lower level of a given maser system. We  note  that our definition of $\eta$ is different from that of  \citet{2013ApJ...773...70H} because
we do not assume that the upper and lower levels of a given maser system have the same decay rates.  

\begin{figure*}
\centering
\includegraphics[width=0.8\textwidth]{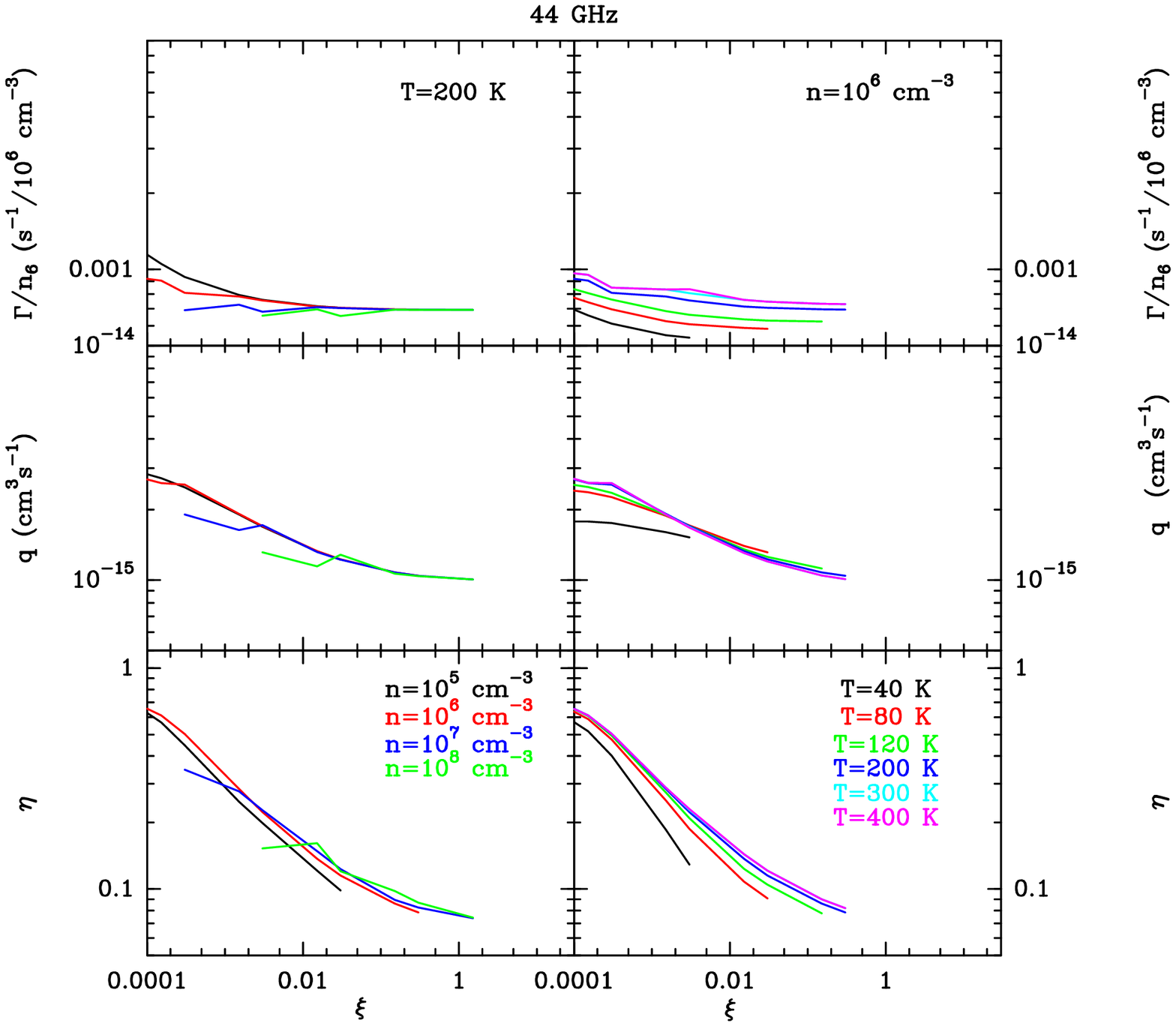}
\caption{Modelling results for the 44\,GHz maser pumping at T = 200\,K and various densities (left panel) and at $n=10^6$\,cm$^{-3}$ and several temperatures (right panel). Plotted are the  maser loss rate $\Gamma$ divided by $n_6$, the pump rate coefficient $q$, and the inversion efficiency $\eta$ as functions of $\xi$.  It is possible to convert $\xi$ to methanol abundance using Eq.\,\ref{csi}.}\label{hollenbach44}
\end{figure*}

\begin{figure*}
\centering
\includegraphics[width=0.8\textwidth]{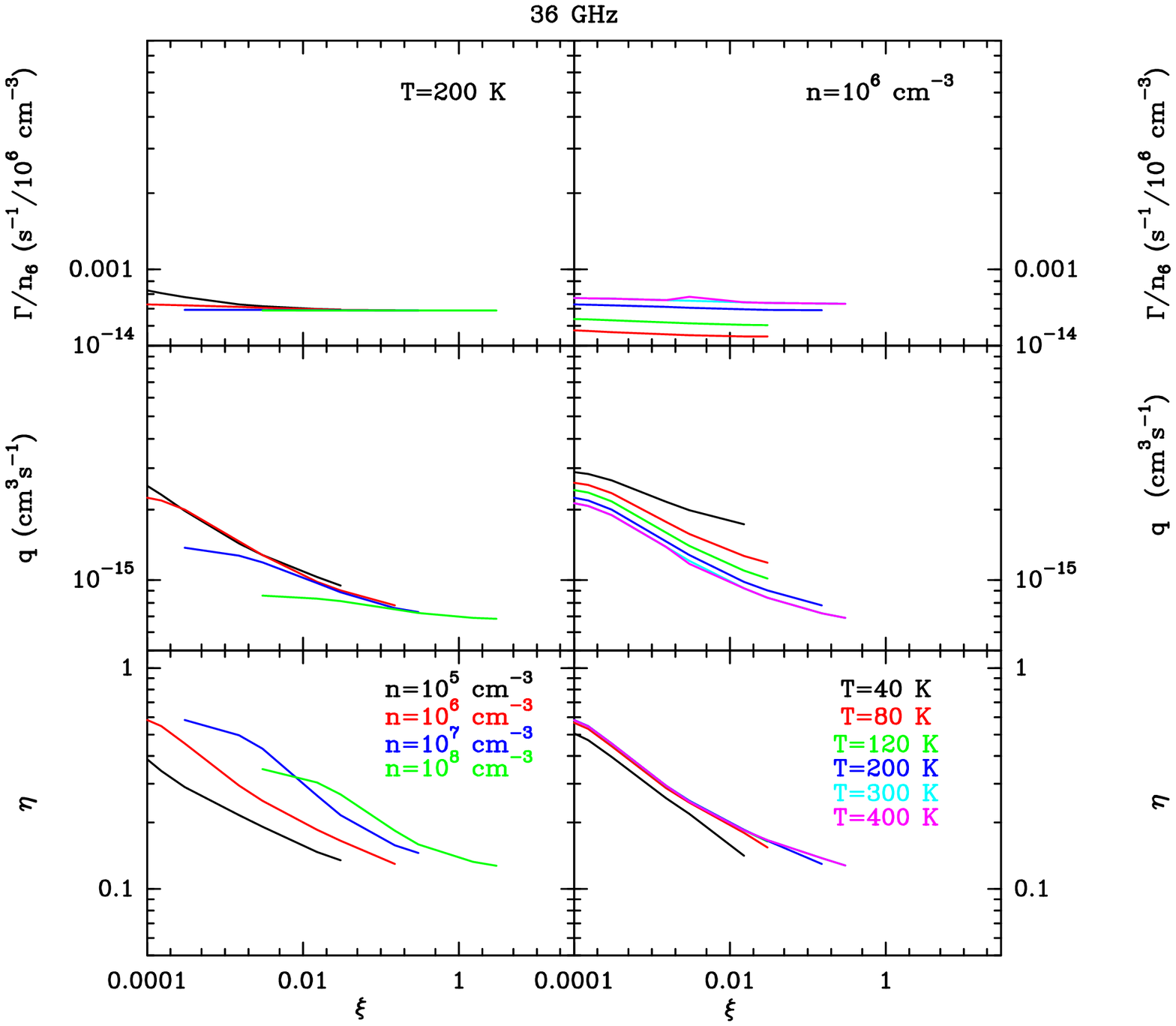}
\caption{Modelling results for the 36\,GHz maser pumping at T = 200\,K and various densities (left panel) and at $n=10^6$\,cm$^{-3}$ and several temperatures (right panel). Plotted are the  maser loss rate $\Gamma$ divided by $n_6$, the pump rate coefficient $q$, and the inversion efficiency $\eta$ as functions of $\xi$. It is possible to convert $\xi$ to methanol abundance using Eq.\,\ref{csi}.}\label{hollenbach36}
\end{figure*}

\begin{figure*}
\centering
\includegraphics[width=0.8\textwidth]{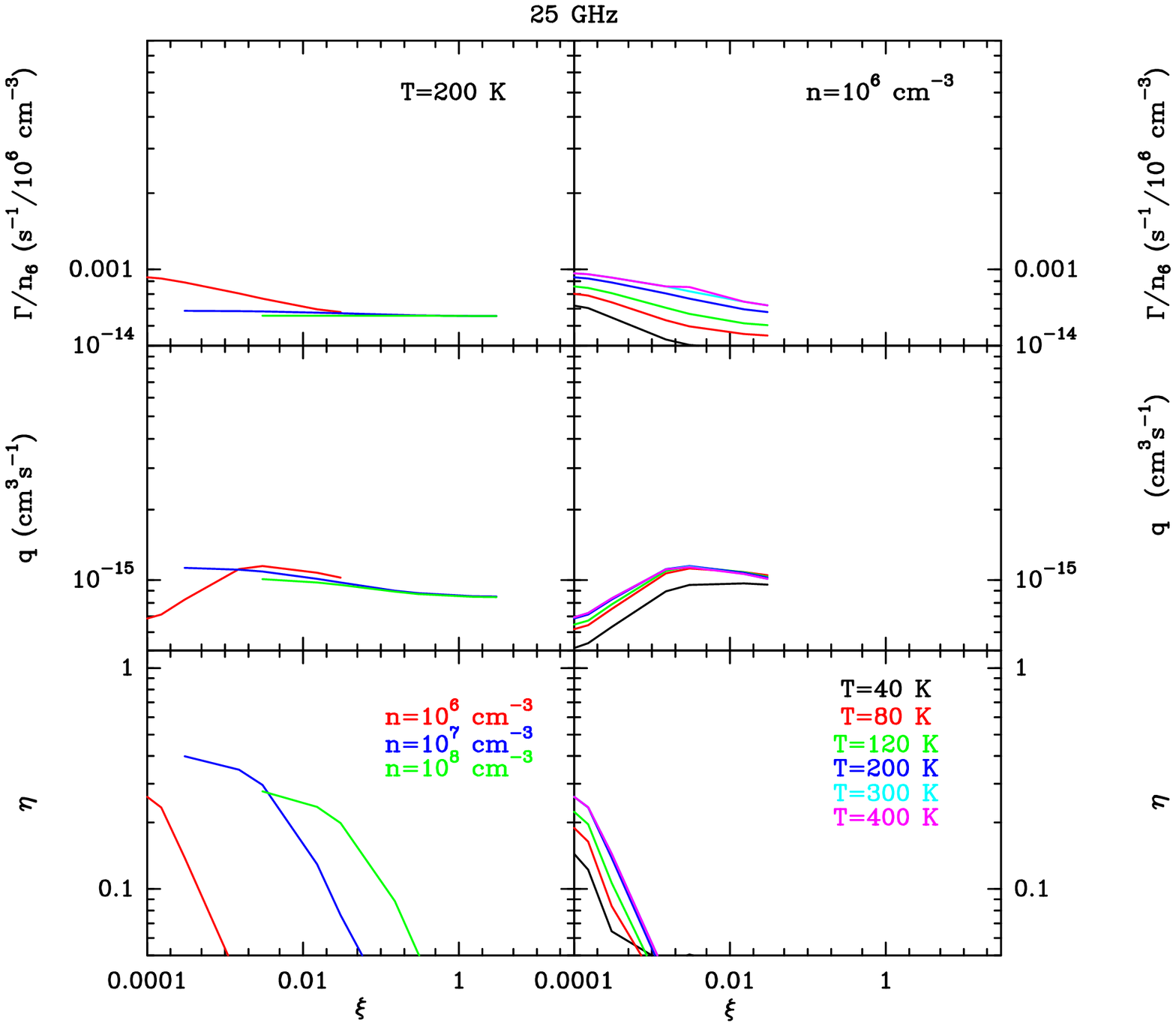}
\caption{Modelling results for the $6_2-6_1$-$E$ 25.018\,GHz maser pumping at T = 200\,K and various densities (left panel) and at $n=10^6$\,cm$^{-3}$ and several temperatures (right panel).  Plotted are the  maser loss rate $\Gamma$ divided by $n_6$, the pump rate coefficient $q$, and the inversion efficiency $\eta$ as functions of $\xi$. It is possible to convert  $\xi$ to methanol abundance using Eq.\,\ref{csi}.}\label{hollenbach25}
\end{figure*}

 Figures\,\ref{hollenbach44}, \ref{hollenbach36}, and \ref{hollenbach25} show our results for the 44\,GHz, the 36\,GHz, and the 25.018\,GHz masers for
$\eta$, $q$, and $\Gamma/n_6$ (the loss rate divided by 
density in units of $10^6$\,cm$^{-3}$) as functions of $\xi$ for a range of
densities and temperatures relevant for Class\,I masers. The parameters
$\Gamma$ and $q$  are the sum of a collisional part and a radiative
part: at low $\xi$, collisions are negligible and $\Gamma$ and $q$
are dominated by their radiative part, which decreases with increasing
values of $\xi$. Their collisional part is proportional to the H$_2$
density and has a weak dependence on temperature. Hence, $\Gamma/n_6$
is fairly flat as a function of $\xi$ when collisions dominate, while
$q$ decreases slowly with $\xi$ in this regime. Results for the other Class\,I maser lines are
shown in Appendix\,\ref{online_c}, Figs.\,\ref{hollenbach84} and \ref{hollenbach95}.
 For the 44\,GHz and the 95\,GHz lines, $\Gamma$, $q$, and $\eta$  have a very weak dependence on density and on temperature (above 80\,K) as function of $\xi$ (Figs.\,\ref{hollenbach44} and \ref{hollenbach95}). On the other hand, the 36\,GHz, 84\,GHz, and the 25\,GHz lines show a stronger dependency on density;  the 25\,GHz line is  the most sensitive to changes in the volume density of the gas (Figs.\,\ref{hollenbach36}, \ref{hollenbach25}, and \ref{hollenbach84}). All Class\,I masers analysed in this study do not show a strong dependence on temperature.

\subsection{Effects of the physical parameters on Class\,I masers}\label{physics}

 The parameters $\eta$ and $q$ discussed in the previous sections can be used  to
  characterise the maser emission
\citep[e.g., ][]{2013ApJ...773...70H}. 
The photon production rate $\Phi_m$ at line centre of a saturated
maser is directly linked to the observed flux ($F_m$) of a given maser
through  the geometry of the maser and can be defined as

\begin{eqnarray}\label{phi}
  \Phi_m = \frac{g_ug_l}{g_u+g_l}\times2n^2X(\rm{CH_3OH})\eta q  
.\end{eqnarray}

\begin{figure*}
\centering
\includegraphics[width=0.6\textwidth]{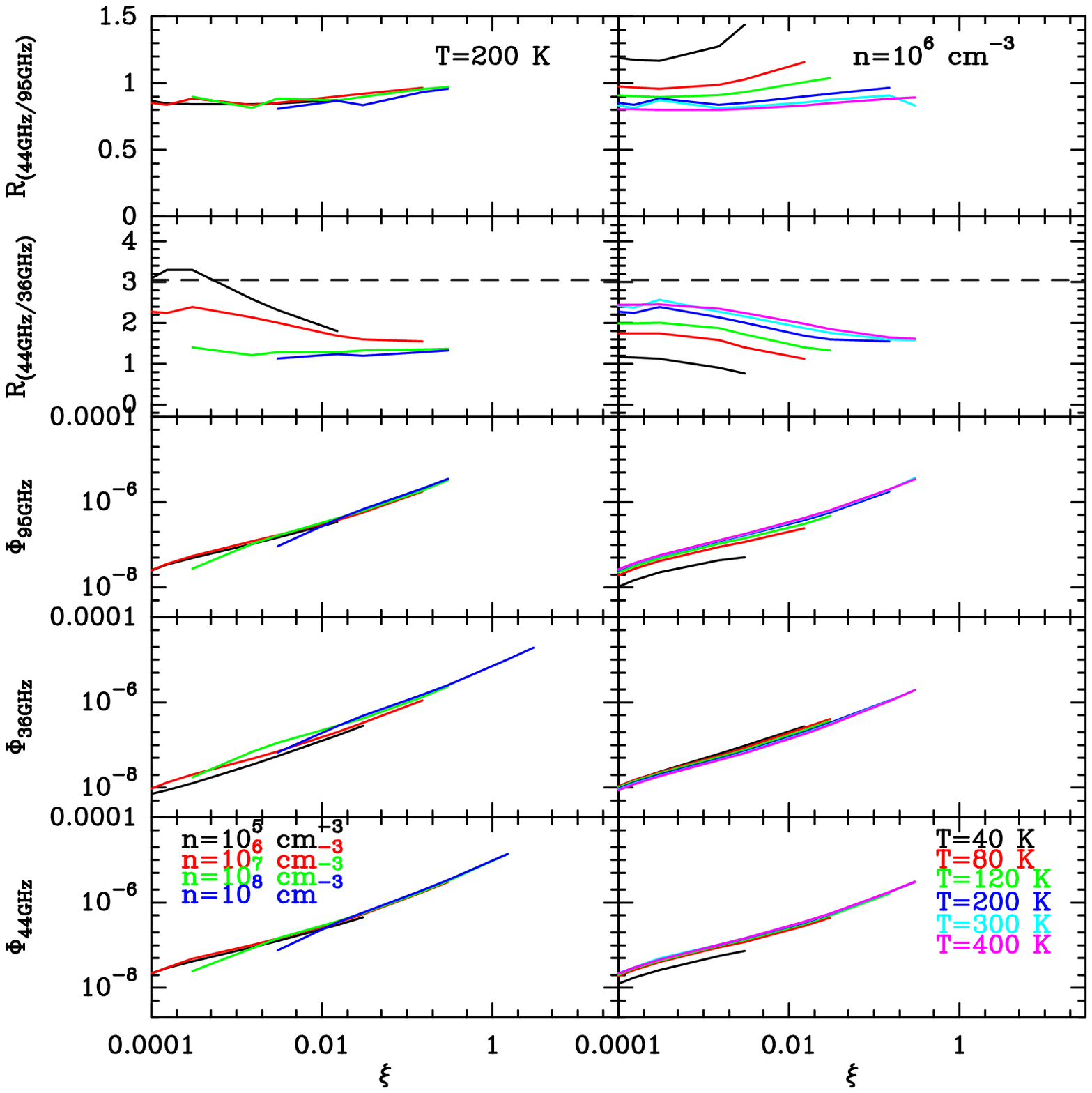}
\caption{Modelling results for the 44\,GHz, 36\,GHz, and 95\,GHz  maser photon production rates, and for their ratios   at 200\,K for various densities (left panel) and at $n=10^6$\,cm$^{-3}$ and several temperatures (right panel).
  The horizontal dashed line in the panels showing the
  44\,GHz to 36\,GHz ratio marks the peak value of the flux-density ratio distribution reported by \citet{2014MNRAS.439.2584V}.  It is possible to convert $\xi$ to methanol abundance using Eq.\,\ref{csi}.}\label{phi364495}
\end{figure*}

\begin{figure*}
\centering
\includegraphics[width=0.8\textwidth]{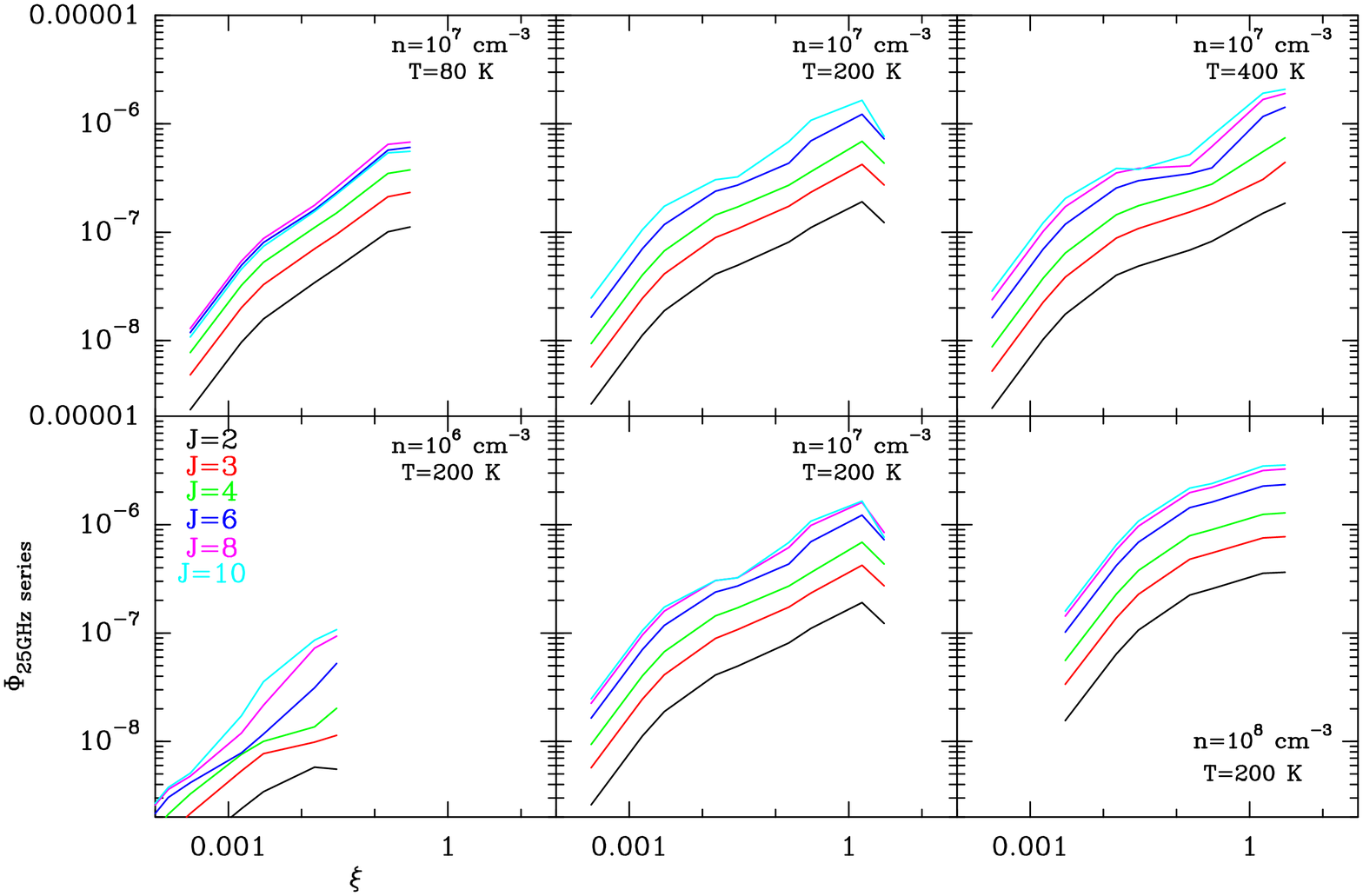}
\caption{Modelling results for the photon production rate of the $J_2-J_1$-$E$ ($J=2-10$) series at 25\,GHz  maser  at 200\,K for various densities (bottom panel) and at $n=10^7$\,cm$^{-3}$ and several temperatures (upper panel). It is possible to convert  $\xi$ to methanol abundance using Eq.\,\ref{csi}.}\label{phi25_all}
\end{figure*}

Figures\,\ref{phi364495}  and \ref{phi25_all} show the
dependence of $\Phi_m$ for several Class\,I methanol lines (36\,GHz, 44\,GHz, 95\,GHz, and the 25\,GHz series) as functions
of $\xi$ for different densities and temperatures for a range of
parameters for which the lines are inverted. For the the $J_2-J_1$-$E$ series we show  our model results for the photon production rate for lines between $J$=2 and 10.
The photon production rate of the 84\,GHz line is presented in Appendix\,\ref{online_c} (Fig.\,\ref{phi3684}).
For all lines, the photon
production rates follow almost a single curve as  functions  of $\xi$
independently of volume density. All lines are basically unaffected by $T$.
  For the 25\,GHz series, 
   we find that the 
 photon production rate  increases with $J$ with a peak between $J=8$ and 10
 (see discussion in Sect.\,\ref{comparison}).

\begin{figure*}
\centering
\includegraphics[width=0.85\textwidth]{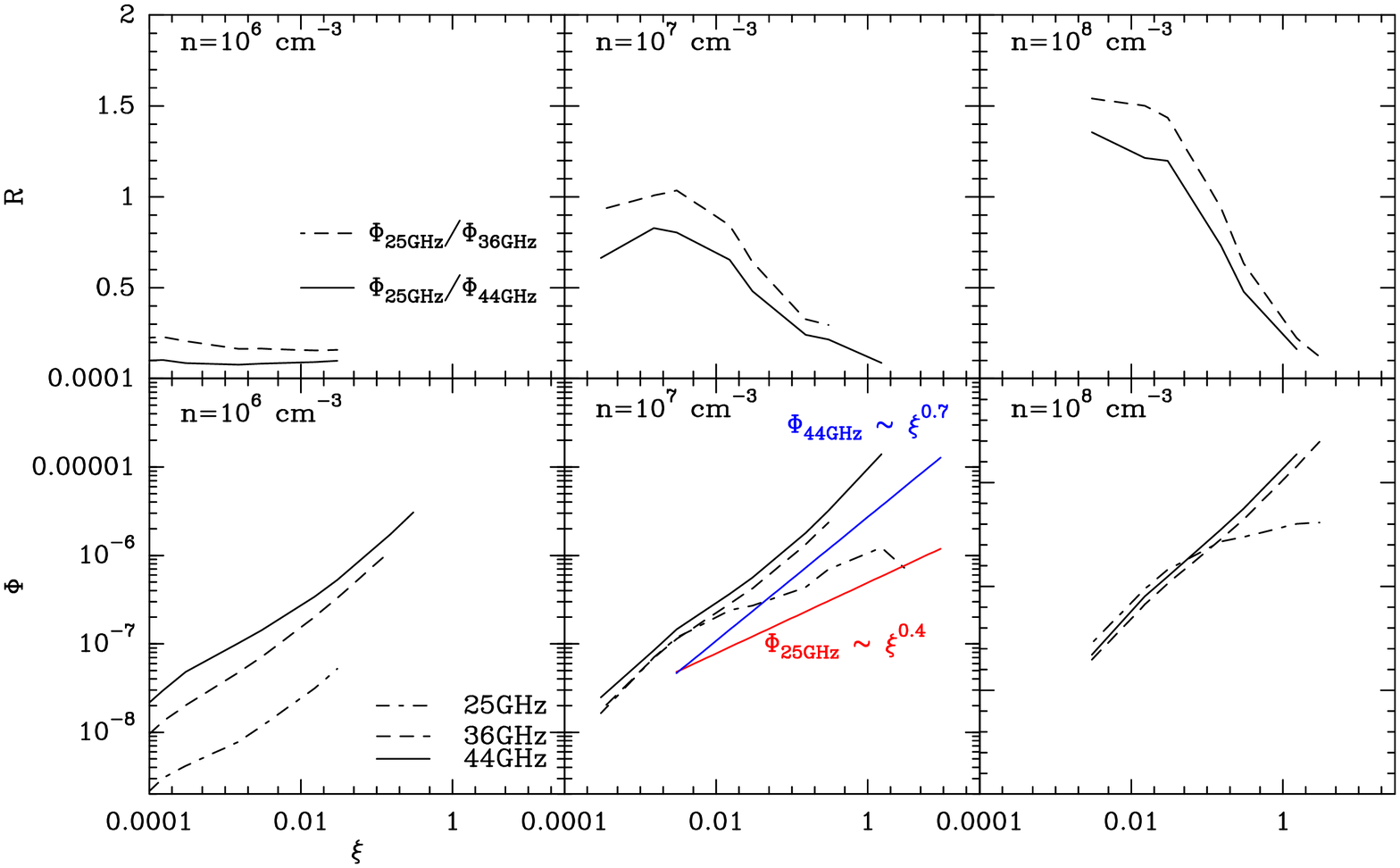}
\caption{{\bf Bottom panel:} Comparison  of  the maser photon production rates for the 44\,GHz, 36\,GHz, and 25\,GHz ($J=6$)
  for various densities at 200\,K. For example, the solid red and blue curves in the middle panel show a power law of $\xi^{0.4}$ and $\xi^{0.7}$, respectively. {\bf Top panel:} Modelling results  for the ratios of the $6_{2}-6_{1}$-$E$ maser  photon production rate to that of the 44\,GHz and 36\,GHz lines   for various densities  at 200\,K.
  It is possible to convert  $\xi$ to methanol abundance using Eq.\,\ref{csi}.}\label{phi_253644}
\end{figure*}

In Fig.\,\ref{phi_253644}, we compare the photon production rates for the $6_2\to6_1$ line with those of the 36\,GHz and 44\,GHz masers for densities of
 $10^6$\,cm$^{-3}$, $10^7$\,cm$^{-3}$, and $10^8$\,cm$^{-3}$. Clearly, the $6_2\to6_1$ line (and in general the whole $J_2 \to J_1$ series at 25\,GHz, see Fig.\,\ref{phi25_all}) behaves in a different fashion from the other maser transitions
discussed in this article  and is the Class\,I maser most sensitive to density.
This is not surprising. The 
  36\,GHz, 44\,GHz, 84\,GHz, and 95\,GHz masers are all directly connected
 to backbone ladders ($k=-1$ for $E$-type and $K=0$ for $A$-type)  and are
 inverted mainly because of the predominance of $\Delta k=0$ collisions.
 This is not the case for the $J_{2} \to J_{1}$ series whose inversion mechanism
 is more complex and which depends on $\Delta k \neq 0$ collisions to build
 up population in the $k=1$ and 2 $E$-type ladders.   As seen in Fig.\,\ref{texden},
 the $6_{2}-6_{1}$-$E$ transition is only  inverted at high densities
 above $10^6$\,cm$^{-3}$ in contrast to the other Class\,I masers treated
 here. Moreover, the density dependence of the photon production rate in
 $6_{2}-6_{1}$-$E$ is  steeper than in backbone-linked lines and the
 dependence on $\xi$ (emission measure) is flatter. 
This is clearly shown in Fig.\,\ref{phi_253644} at a density of
 $10^7$\,cm$^{-3}$ where  the 36 and 44\,GHz masers have photon
 production rates that are very similar and behave roughly as
  $\xi^{0.7}$. The photon production rate for the $6_{2}-6_{1}$-$E$ line  on the other hand is  close to the
 backbone-linked lines at low $\xi$, but has a much flatter $\xi$ dependence
 (roughly $\xi^{0.4}$) and consequently has a much lower photon output
 at $\xi$ values approaching unity.  It is
 clear from our models that at densities of the order of $10^6$ cm$^{-3}$ the photon
 output in the 25 GHz lines is much lower than at 36\,GHz or 44\,GHz. Indeed, observations suggest a much lower
 photon output at 25\,GHz than in backbone lines  \citep[e.g.,][]{1986A&A...157..318M,1990ApJ...354..556H,2006MNRAS.373..411V}, but  higher
 quality interferometric observations are needed to draw conclusions.

It is interesting to
note that despite the decrease in $\eta$ and $q$ with $\xi$ (see
Fig.\,\ref{hollenbach44}), the production rate increases with $\xi$
until thermalisation is reached, thus suggesting that the high $\xi$
before quenching (corresponding to specific column densities of
  $10^{16}-10^{17}$\,cm$^{-2}$\,km$^{-1}$\,s; see Table\,\ref{quench} and Sect.\,\ref{inversion}) are responsible for the observed  bright masers.  This
implies that high $\xi$ values are needed for high photon production rate and
that therefore masers with high brightness temperatures trace high
methanol emission measures.  These values are reached at high densities, high
temperatures, and high methanol column densities (Eq.\,\ref{csi}).  
In particular, the  36\,GHz line can reach higher photon production rates than the 84\,GHz, the 44\,GHz, and the 95\,GHz masers because it generally quenches at higher values of $\xi$ for $T<200$\,K.  A similar result has been
 obtained by \citet{2014ApJ...793..133M}  who find (at a temperature
 of 80\,K and a methanol abundance of $10^{-7}$)
 that optimal masing 
 conditions at 36\,GHz are at densities roughly an order of magnitude
 higher than at 44\,GHz.

Modelling intensities of saturated maser lines is not straightforward because the outcome depends on the geometry of the maser. In particular, the observed flux of a given maser transition depends on the geometry and on whether the maser is beamed towards us or not.
However, under the assumption that different maser lines of frequencies $\nu_1$ and $\nu_2$ are emitted by the same volume of gas and that they are both saturated, the observed ratios of their fluxes,  $F_1$ and $F_2$, is geometry independent and it can be written as a function of the ratio between their  photon production rates:

\begin{equation}\label{ratio}
  \frac{F_1}{F_2} = \frac{\nu_1}{\nu_2}\frac{\Phi_1}{\Phi_2}
.\end{equation}

Figure\,\ref{phi364495} shows the  ${\Phi_1}/{\Phi_2}$ ratio of the modelled maser photon production rates of the 36\,GHz, 44\, GHz, and 95\, GHz  maser lines for different densities and temperature as functions of $\xi$.  Figure\,\ref{phi_253644} shows the  photon production rate ratios of  $6_{2}-6_{1}$-$E$ transition relative to the 44\,GHz and 36\,GHz lines.
For completeness, the ratio between the 84\,GHz maser and the 36\,GHz line are shown in Fig.\,\ref{phi3684}. The ratio 
 ${\Phi_{\rm{44GHz}}}/{\Phi_{\rm{95GHz}}}$ is relatively flat and does not vary with density for a fixed temperature of 200\,K. On the other hand, it decreases with temperature for a fixed density of $10^6$\,cm$^{-3}$.
The ratio ${\Phi_{\rm{44GHz}}}/{\Phi_{\rm{36GHz}}}$ has a different behaviour as a function of temperature and density. At a fixed density of $10^6$\,cm$^{-3}$, the ratio increases with $T$, while at $T=200$\,K   it decreases with volume density. Finally,  ${\Phi_{\rm{36GHz}}}/{\Phi_{\rm{84GHz}}}$ decreases with density and temperature (Fig.\,\ref{phi3684}).

\section{Discussion}\label{dis}

\subsection{Maser luminosities}\label{lumi}

The luminosity of a given maser transition is defined as

\begin{equation}\label{lum}
L_{\rm{m}}\propto \Phi_{\rm{m}} \times V_{\rm{m}}
,\end{equation}where $V_{\rm{m}}$ is the maser volume. As discussed
 in Sect.\,\ref{size}, spot sizes for Class\,I masers are not well
 constrained. For a cylindrical geometry, a spot size of 100\,AU and
 an aspect ratio $\alpha=l_{los}/R=10$, where $l_{los}$ is the length oriented along the line of sight and $R$ the radius, we derive maser luminosities for the
  44\,GHz, 95\,GHz, 36\,GHz, and 84\,GHz lines, and the $6_2-6_1$-$E $ maser  (representative of the 25\,GHz series)   of
 $10^{41}$\,photons\,s$^{-1}$ for 80\,K and 200\,K for emission measures $\xi$ of 0.1 or 0.2.  The corresponding luminosities for a spot size of 50\,AU are  approximately $10^{39}$\,photons\,s$^{-1}$.
 The maser luminosity is linked to the observed isotropic luminosity, $L_{\rm{iso}}$, through the formula

\begin{equation}\label{lm}
L_{\rm{m}}=L_{\rm{iso}}\times\Omega_{b}/(4\pi)
,\end{equation}
where $\Omega_{b}$ is the beaming angle.
 For  a cylindrical
 maser spot with $l_{los}=\alpha R$ oriented along the line of sight, 
 $\Omega_{b} =\pi\times(R/l_{los})^{2}=\pi /\alpha^2$ making the
   assumption that  the beam solid angle is purely determined 
 by geometry (not necessarily the case; see \citealt{2013ApJ...773...70H} for a discussion). For an aspect 
 ratio $\alpha$ of 10  (corresponding to $\sim 10^{-3}$ steradian), observations of Class\,I masers ($L_{\rm{iso}}=10^{41}-10^{45}$ \,photons\,s$^{-1}$; see Table\,\ref{luminosity}) imply maser luminosities between 
 between $10^{38}$ and 
 $10^{42}$\,photons\,s$^{-1}$ in agreement with our models.

  For a collisionally excited maser, it is thus
 reasonable to expect  that  high density
 and a high  methanol abundance is required. This general expectation is confirmed
 by the results of our computations which show that the
 maser photon output per unit volume increases with the methanol
 emission measure $\xi$ defined by Eq.\,\ref{csi}.  
  From Eqs.\,\ref{lum} and \ref{lm}, 
 it can be seen that a small maser volume can be compensated for
 either by a larger $\Phi_{m}$ or by a small beaming angle. The former
 requires higher emission measures (or higher 
density and methanol abundance), while the latter becomes geometrically
 unlikely at a certain point (aspect ratios for the masing volume
 higher than 10 seem improbable). 
 We  can get some insights into the implications of this based
 on our computations of the photon production rates.  From Fig.\,\ref{phi364495},
 we note that in the most common lines,   a photon emissivity
 per unit volume of $10^{-6}$ photons s$^{-1}$ cm$^{-3}$ at  $\xi=0.1$ is achieved, while
 (see Fig.\,\ref{phi_253644}) the 25\,GHz emissivity is almost an order of magnitude smaller.
 
 From the methanol
 emission measure assuming velocity gradients of the order of 
 $5\, 10^4$ km s$^{-1}$ pc$^{-1}$ and a standard methanol abundance of 10$^{-6}$
 \citep[e.g.,][]{1988A&A...198..253M,2010A&A...521L..39K}, we find $n$(H$_2$)=$7\, 10^7$ cm$^{-3}$ consistent
 with a high density explanation of the masers. 

 Such high densities are consistent with the magnetic field
 measurement of \citet{2011ApJ...730L...5S} \citep[see also][]{2012AJ....144..189M}. They have detected a Zeeman pattern in the
 44 GHz maser towards OMC2 and estimate a line of sight
 field of 18 mG (with high uncertainty 
 owing to the poorly known Land\'{e} factor). If one uses the correlation between density and
 magnetic field of \citet{1999ApJ...520..706C}, this corresponds to a
 H$_2$ density of $10^7$ cm$^{-3}$, although with a very large uncertainty.
 We note here that Sarma and Momjian found a value ten times lower
 based on a different argument and we cannot exclude this. However,
 we conclude that the Zeeman result certainly favours high densities
 (upwards of $10^6$ cm$^{-3}$) in the masing gas.

\subsection{Maser saturation}\label{sat}

\begin{figure*}
  \centering
  \subfigure[]{\includegraphics[width=9cm]{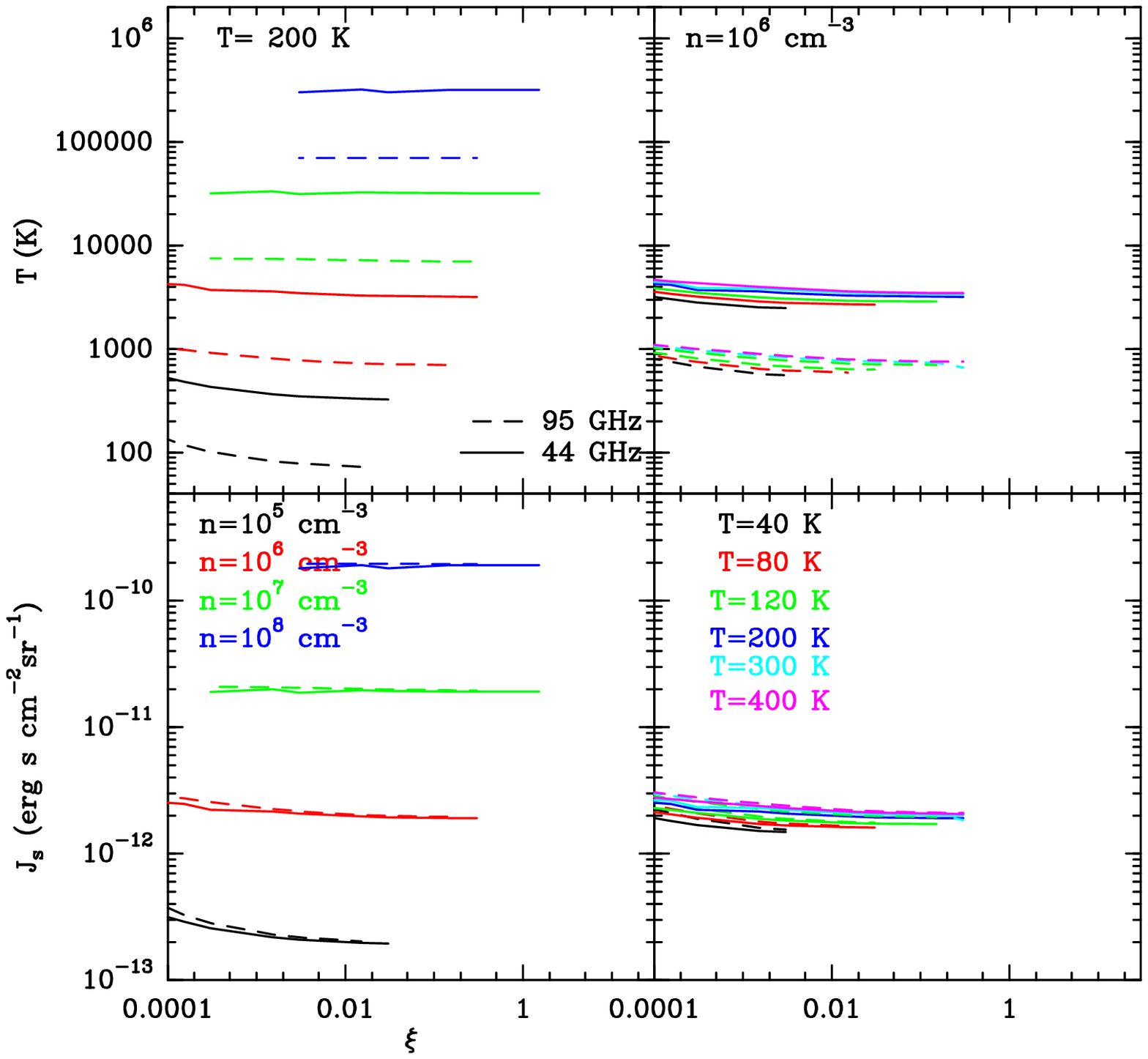}}
  \subfigure[]{\includegraphics[width=9cm]{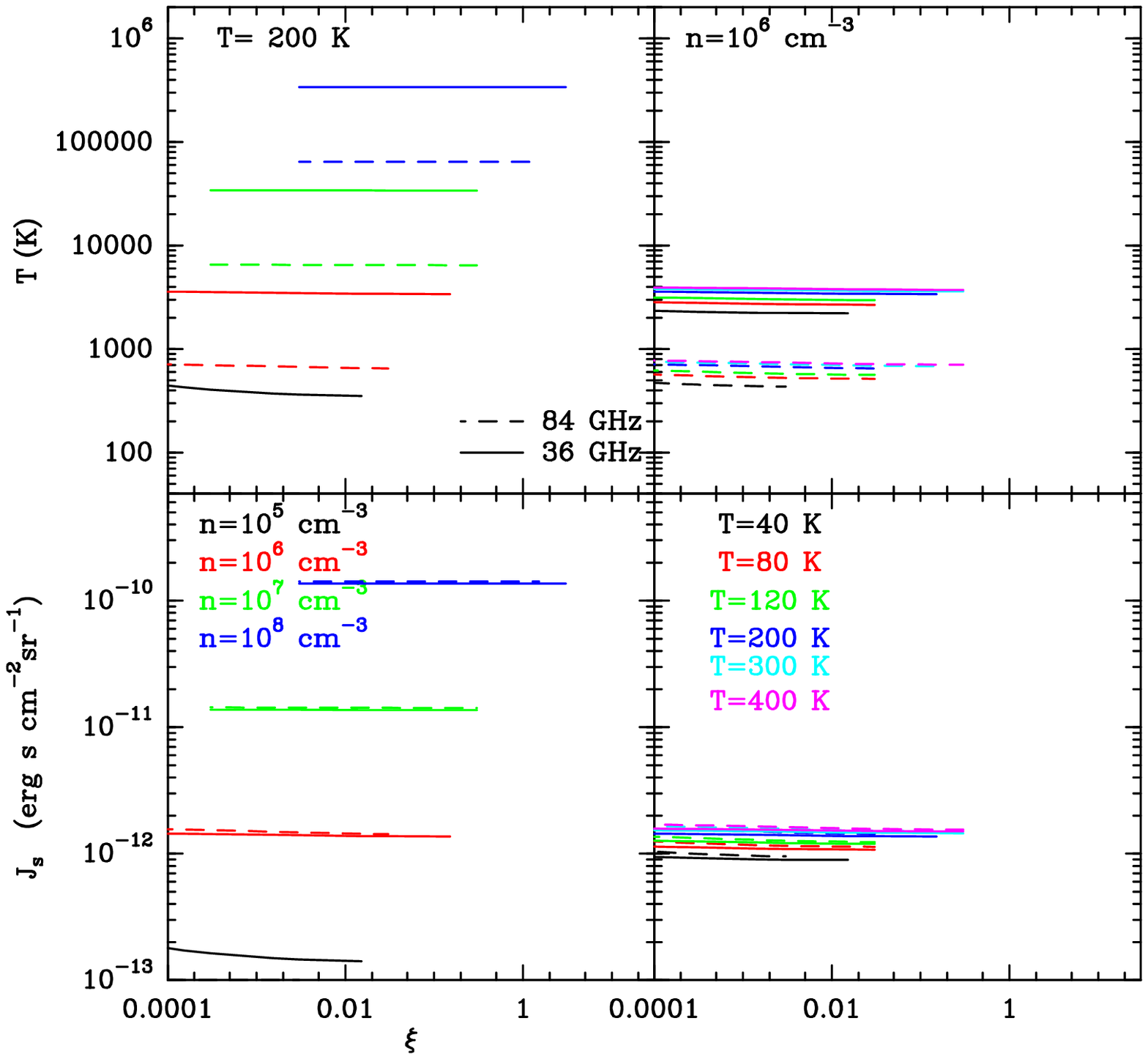}}
\caption{Plots of $J_s$ and $T_s$ as function of $\xi$ for different densities and temperatures for the 44\,GHz (solid line) and 95\,GHz (dashed lines) transitions ({\bf a}), and for the 36\,GHz (solid line) and 84\,GHz (dashed lines) masers ({\bf b}).  It is possible to convert $\xi$ to methanol abundance using Eq.\,\ref{csi}.}\label{jsts}
\end{figure*}

\begin{figure}
  \centering
  \includegraphics[width=9cm]{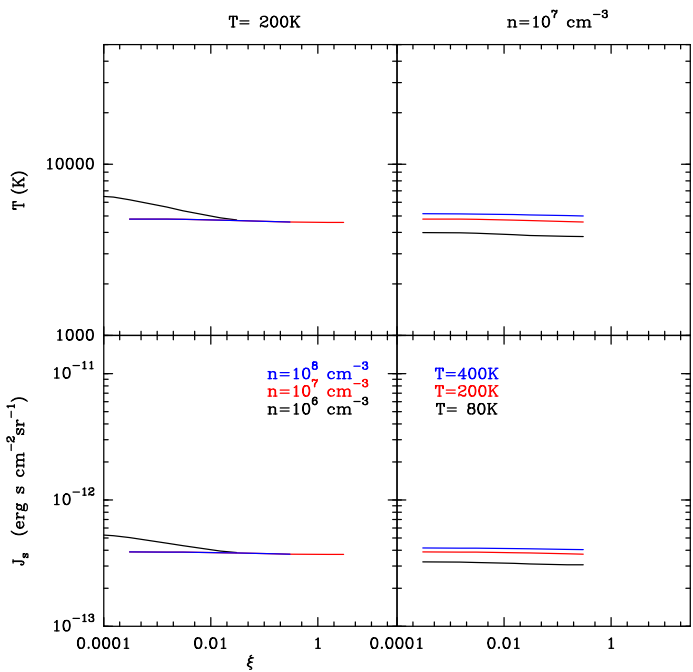}
\caption{Plots of $J_s$ and $T_s$ as function of $\xi$ for different densities and temperatures for the $6_2-6_1$-$E$ maser.  It is possible to  convert $\xi$ to methanol abundance using Eq.\,\ref{csi}.}\label{jsts25}
\end{figure}

The intensity of a maser reaches saturation at a value $J_s$ given by 

\begin{equation}\label{js}
J_s=\frac{g_l}{g_l+g_u} \frac{\Gamma}{B_{m}}
,\end{equation}
where $B_m$ is the Einstein coefficient for stimulated emission of the maser transition and $\Gamma$ the loss rate (Eq.\,\ref{gamma}). The corresponding brightness temperature $T_s$ of the line is 

\begin{equation}\label{ts}
T_s= J_{s}\times\frac{\lambda_m^2}{2k_{B}}= \frac{\Gamma}{A_m}\times\frac{h\nu_m}{k_{B}}\times\frac{g_l}{g_l+g_u}
,\end{equation}

where  $\lambda_m$ and $\nu_m$ are the wavelength and frequency of the maser line, $k_{B}$
  the Boltzmann constant, $h$ the Planck constant, and $A_m$ the Einstein coefficient for spontaneous emission of the maser transition. 
In this way, we do not correct for any given geometry to keep results
general.  Using $A_{44GHz}=2.0\times10^{-7}$\,s$^{-1}$,
$A_{95GHz}=2.1\times10^{-6}$\,s$^{-1}$,
$A_{36GHz}=1.5\times10^{-7}$\,s$^{-1}$,
$A_{84GHz}=2.0\times10^{-6}$\,s$^{-1}$, and $A_{25.018GHz}=8.6\times10^{-8}$\,s$^{-1}$ (from the Cologne database for
molecular spectroscopy,
\citealt{2001A&A...370L..49M, 2005JMoSt.742..215M}),
 we find brightness temperatures at saturation up to
approximately $10^5$\,K for the 44\,GHz, 95\,GHz, 36\,GHz, and 84\,GHz lines for
a temperature of 200\,K and a density of $10^8$\,cm$^{-3}$.  The $6_2-6_1$-$E$ has lower values of $T_s$ ($\sim 5000$\,K).
We show in Figs.\,\ref{jsts}  and \ref{jsts25} our computed
 values for $J_{s}$ and $T_{s}$ as a  function of $\xi$ for a range of
 temperatures and densities.
For the   44\,GHz, 95\,GHz, 36\,GHz, and 84\,GHz masers, we find similar behaviour; $T_s$ 
increases with the volume density of the gas. The 25\,GHz lines behave differently and show very little dependence on density.  All lines are basically
unaffected by changes in temperature. Moreover, the $T_s$ of the
masers scale with the density of the gas approximately as
$4\times10^3\,n_6$ for the 44\,GHz, and as $3\times10^3\,n_6$ for the
36\,GHz transition, and the $T_s$ of the 95\,GHz and 84\,GHz lines can
be obtained from the same relations corrected by a factor
$(44\,{\rm{GHz}}/95\,{\rm{GHz}})^2$ and
$(36\,{\rm{GHz}}/84\,{\rm{GHz}})^2$, respectively. From these
relations, it is possible to  derive constraints on the beaming angle $\Omega_b$
of a given maser: if brightness temperatures of $10^9$\,K are
observed, as in IRAS\,18151--1208 for the 44\,GHz line
\citep{2014ApJ...789L...1M}, then assuming that the density of the
gas is $\sim10^8$\,cm$^{-3}$, $\Omega_b$ is larger than
$5\times10^{-3}$ for a saturated maser. Similar values of $\Omega_b$ are estimated for the 25.018\,GHz line ($\Omega_b=0.03-0.003$ for $T_{\rm b}=10^6-10^7$\,K; see Table\,\ref{luminosity}).

\subsection{Comparison with observations}\label{comparison}
In Sect.\,\ref{saturation} we speculated that the Class\,I masers with the lowest excitation requirements
 are saturated, as suggested by their   slow variability (in general) and by the similar velocity profiles of different transitions found in several sources. Under this assumption, and assuming that different lines are emitted by the same volume of gas, we can  compare  observed line ratios to our model results to verify whether our calculations reproduce the observations. Recently, \citet{2014MNRAS.439.2584V} have reported observations  of a sample of 71 Class\,I maser methanol sources in the 36\,GHz and 44\,GHz lines with the ATCA array. These authors find that the 44\,GHz to 36\,GHz flux-density ratio  has a wide distribution with a peak at 2.5 (corresponding to ${\Phi_{\rm{44GHz}}}/{\Phi_{\rm{95GHz}}}\sim3$);
  only 14\% of the 292 maser groups discussed by \citet{2014MNRAS.439.2584V} were found to be stronger at 36\,GHz than at 44\,GHz (see their Fig.\,6). The authors also comment that owing to the complexity of spectral profiles, this is likely an upper limit to the number of groups where the 36\,GHz flux is stronger than  at 44\,GHz.
   These results seem to reflect  our model predictions, which show that the 44\,GHz line is always stronger than the 36\,GHz maser except at low temperatures and high densities (see Figs.\,\ref{phi364495} and \ref{phi_253644}). These conditions are probably met in
   regions with little  or no star formation activity.  Indeed \citet{2008AJ....135.1718P}  reported enhanced maser emission at 36\,GHz relative to the 44\,GHz maser line toward such regions. Moreover, 
    observations of supernovae remnants find 36\,GHz maser emission in the absence of a detectable 44\,GHz line \citep[e.g.,][]{2014AJ....147...73P}. Regarding other flux ratios, no large sample of sources observed at high angular resolution is available. \citet{2010A&A...517A..56F} detected  the 44\,GHz and the 95\,GHz masers in a sample of high-mass star forming regions with the Nobeyama telescope. Seven sources in the sample are detected in both lines in a similar range of velocities and with similar line widths. Their observed  ${\Phi_{\rm{44GHz}}}/{\Phi_{\rm{95GHz}}}$ ratio has an average value of $\sim5$ and ranges between 1 and 11, while the modelled ratio is never higher than 1. Clearly, our modelled ${\Phi_{\rm{44GHz}}}/{\Phi_{\rm{95GHz}}}$ fails to reproduce the observations.
However, a note of caution is needed in the case of single-dish observations because the maser locations can be significantly offset from the  pointing positions and because of the different beam sizes of the different lines (38\arcsec and 18\arcsec for the observations reported by \citealt{2010A&A...517A..56F}).

Concerning the relative intensities of  the $J_2 \to J_1$ lines at 25\,GHz,
it is noteworthy that our calculations predict
  that the ratio between the photon production rates of the $6_2 \to 6_1$ line to the $3_2 \to 3_1$ transition is close to 3 and slightly increases with density (Fig.\,\ref{phi25_all}), in reasonable agreement with observations \citep{1986A&A...157..318M}. These observations also show that the $J_2 \to J_1$ lines usually peak at $J=6,$ while our results show a peak around $J=9$. This may be the effect of the relatively high kinetic temperature we adopted for our calculations, 200\,K.

  Another implication of our results is that in sources where the 25\,GHz lines are seen in absorption \citep[e.g., W3OH, NGC7538][]{1986A&A...157..318M} the density must be lower than $10^6$\,cm$^{-3}$, and the 36\,GHz and 44\,GHz masers should be detected. This is indeed the case for both W3OH and NGC7538 \citep[e.g.,][]{1989ApJ...339..949H,1990ApJ...354..556H,2004ApJS..155..149K}, although the large beam of the single-dish observations does not allow verification of  the spatial association of the lines.

Our results show that interferometric 
 surveys  in the 25\,GHz lines and in the millimetre masers (at 84\,GHz and 95\,GHz) would be very useful. 
 Figure\,\ref{phi_253644} shows a comparison of our computed photon production rates
 at 200\,K for densities between $10^6$ and $10^8$\, cm$^{-3}$. 
 One interpretation of the small minority of objects with high
 44\,GHz to 36\,GHz flux-density ratio might be that they are high emission measure 
 (methanol abundance) objects with $\xi$ larger than 0.1
 and density above $10^7$\,cm$^{-3}$.  Certainly,  a good
 interferometric survey is needed  at 25 GHz and in the mm lines
 to make further progress.

 \section{Outlook}\label{outlook}

Refining  estimates of the beaming angle and related quantities
will require a proper resolution of
 the radiative transport problem for a plausible geometry
as well as a more consistent set
 of interferometric data.  This will allow a more reliable estimate
 of the expected brightness temperature and line profile.  To what
 extent does finite inclination of the shock propagation direction to
 the line of sight affect observed profiles? Is shock structure important
 and does this affect the observed maser structure?
 What angular size differences  are expected between different Class\,I
 masers? Answers to these questions will become important for the
 new generation of interferometric results from the VLA and ALMA, for example.

 As noted above, we conclude  that  bright Class\,I methanol 
 masers are mainly high-temperature high-density structures with methanol
maser emission measures $\xi$ close to the limits set by collisional
 quenching (roughly a hydrogen density of $10^8$ cm$^{-3}$).
 The basic physics is very similar to that discussed by \citet{2013ApJ...773...70H}
 for water masers and we find, for example, that in the case
 of methanol (as for water), the maser efficiency at high density is determined
 by the collisional selection rules rather than the relative 
 radiative decay rates from the upper and lower maser levels.
 The fall-off of maser efficiency $\eta \ $ with $\xi$ is similar
 to that found by \citet{2013ApJ...773...70H} for water masers, as is the increase of the  saturated
 maser photon production rate $\Phi $ with $\xi$. 
Thus, relatively high post shock densities and temperatures (though
 not sufficiently high to cause dissociation of H$_2$)
 are needed to account for most bright Class\,I masers.
  The results of \citet{2012AJ....144..189M} discussed above suggest
 that magnetic fields may be important in the dynamics of the
 structures that give rise to methanol masers, and we conclude
 that C-shocks (rather than J-shocks )
 give rise to Class\,I masers.
  The linear sizes observed (10-100\,AU, see Table\,\ref{luminosity})
  are  consistent with this.  \citet{1996ApJ...456..250K}
  have studied the structure of C-shocks in the parameter range
 of interest to us and find shock widths of roughly 10 AU 
 (dependent on pre-shock density, ionisation degree,  etc.) with
 compression factors (post shock/pre-shock density ratio) of the
 order of the Alfven Mach number $\varv_{sh}/\varv_{a}$ (where $\varv_{sh}$ is
 the shock velocity and $\varv_{a}$ the Alfven velocity).  The 
 post-shock temperatures are of the order of a few hundred K and the Alfven Mach number is
 typically a factor of 10. Thus for post-shock densities of the order of
 $10^7$\,cm$^{-3}$, we expect pre-shock densities of the order of $10^6$\,cm$^{-3}$. Hence,
 we  can conclude that saturated Class\,I masers are likely to occur
 in regions of high ambient density and pressure. High-mass star forming regions are thus a more likely locale for Class\,I
 methanol masers than lower pressure clouds close to the Sun. 

 In this context, calculations  based  on a C-shock geometry (analogous to those of \citealt{2013ApJ...773...70H}
and 
 \citealt{1996ApJ...456..250K})
 are clearly needed to produce  brightness
 temperature and line profiles for Class\,I masers that can be
 compared with observations. The computations of maser efficiency and
photo production rate in this article are a first step. A comprehensive set of interferometric data  which allow
 study of the spatial relationship between maser spots and their
 development in time is also  clearly
 needed.  In view of the importance of magnetic fields,
   calculations taking account of all Stokes
 parameters as well as more Zeeman measurements are needed as well. The essential
 first step here is to obtain reliable determinations of the
Land\'{e} factors for the different maser transitions.

 Another important question to be investigated is the possible influence
 of $o$-H$_2$ collisions on the properties  of Class\,I masers. It is 
 unclear what the ortho-to-para abundance ratio will be in the
 shocks  we postulate, but it is clear that $o$-H$_2$ collisions
 have different selection rules than $p$-H$_2$ collisions \citep{2010MNRAS.406...95R}  and that, if the $o/p$ ratio is approximately equal to  or greater than 1, this  will change our results appreciably. It could
 conceivably switch off the maser. Thus, calculations assuming
 different $o/p$ ratios are needed. 
 
 One important consequence of such studies is obtaining  direct
 observational information about C-shock structure. Masers allow
 this to be studied both in time and space. It seems likely that
 it will be possible to measure proper motions in Class\,I masers 
 and also effects due to inhomogeneities in the pre--shock 
 medium. There are presumably gradients in temperature and density 
 along the direction of shock propagation and differences in the
 spatial distribution of Class\,I masers may provide a guide to this.
 Understanding C-shock structure on the other hand is a valuable
 input to studies of the interaction of protostellar outflows
with their surroundings.  We conclude that further work along these
 lines is needed.

\section{Summary and conclusions}\label{summary}

Using new collisional rates for collisions of CH$_3$OH with
para-H$_2$, we computed the pump and decay rates as well as the pump
efficiency for the most commonly observed Class\,I methanol maser
transitions  with the aim of investigating the physical conditions leading
  to bright Class\,I masers in regions of massive star formation.
   This is, to our knowledge, the first time that calculations of this
 type have been carried out for methanol masers. We hope that our
 results can be applied to an appropriate model of  
 the geometry of
 saturated Class\,I methanol
 masers in order to compute a realistic brightness temperature 
 distribution  and beaming solid angle.  Ultimately, one hopes
 that this may allow insight into the structure and velocity
 field of the shocks responsible for Class\,I maser emission. 
 Our analysis of the photon production rates has allowed a few
 tentative conclusions on the properties of the high-gain 
 masers such as those observed towards IRAS\,18151-1208 by
 \citet{2014ApJ...789L...1M}. They can be summarised as follows:

\begin{enumerate}
\item We predict inversion in all transitions where maser emission is observed.
\item The production rates of all Class\,I masers increase with $\xi$
  until thermalisation is reached. This implies that masers with high
  brightness temperature trace high methanol emission measure reached
  at high densities ($n({\rm{H_2}})\sim 10^7-10^8$\,cm$^{-3}$), high temperatures ($>100$\,K), and high methanol column
  densities  with methanol maser emission measures close to the limits set by collisional quenching.
\item Class\,I masers can reasonably be separated into the
 three families: the $(J+1)_{-1}-J_{0}$-$E$ type series, the 
 $(J+1)_0-J_1$-$A$ type, and the $J_{2}-J_{1}$-$E$ series at 25\,GHz.
\item The 25\,GHz lines behave in a different way from the other maser transitions likely because they are inverted through  $\Delta k \neq 0$ collisions. In particular, they are only inverted at high densities
  above $10^6$\,cm$^{-3}$ in contrast to other Class\,I masers, which mase already at densities of approximately $10^4$\,cm$^{-3}$.

\item  Under the assumption that Class\,I masers are saturated, our calculations  reproduce reasonably well most of the observed properties of Class\,I methanol masers  in star forming regions. In particular, they predict that the 44\,GHz line is almost always stronger than the 36\,GHz maser, except at low temperatures and high densities and that
the luminosity of the  25\,GHz series is lower than that of the other Class\,I
masers. The observed constant ratio between different lines in the 25\,GHz series is also reproduced.
\item Maser emission in the 25\,GHz lines is not expected  at low densities. Therefore, detection
  of all three families in the same maser feature 
 is a clear indication of high densities ($>10^6$\,cm$^{-3}$).  On the other hand,
 detection of  maser emission at 36\,GHz and 44\,GHz with no corresponding 25\,GHz masers
indicates  densities  lower than $10^6$\,cm$^{-3}$
and probably larger size scales than the spot
 sizes of the order of 50 AU observed in some sources.
\item The maser luminosities derived from our model predictions for the photon production rates are in agreement with observed isotropic luminosities for maser beam angles of the order of $10^{-3}$ steradian.

\end{enumerate}  

\begin{acknowledgements}
We  thank  the  anonymous  referee
for  helpful  comments  that  have  improved  the  content  and
overall clarity of this manuscript. C.M.W. would like to acknowledge travel
support from the Science Foundation Ireland (Grant 13/ERC/I12907).
C.M.W. acknowledges
 the hospitality of the MPIfR during the course of this work.
\end{acknowledgements}

\begin{appendix}

\section{Additional figures}\label{online_c}

\begin{figure*}
  \centering
  \subfigure{\includegraphics[width=0.45\textwidth]{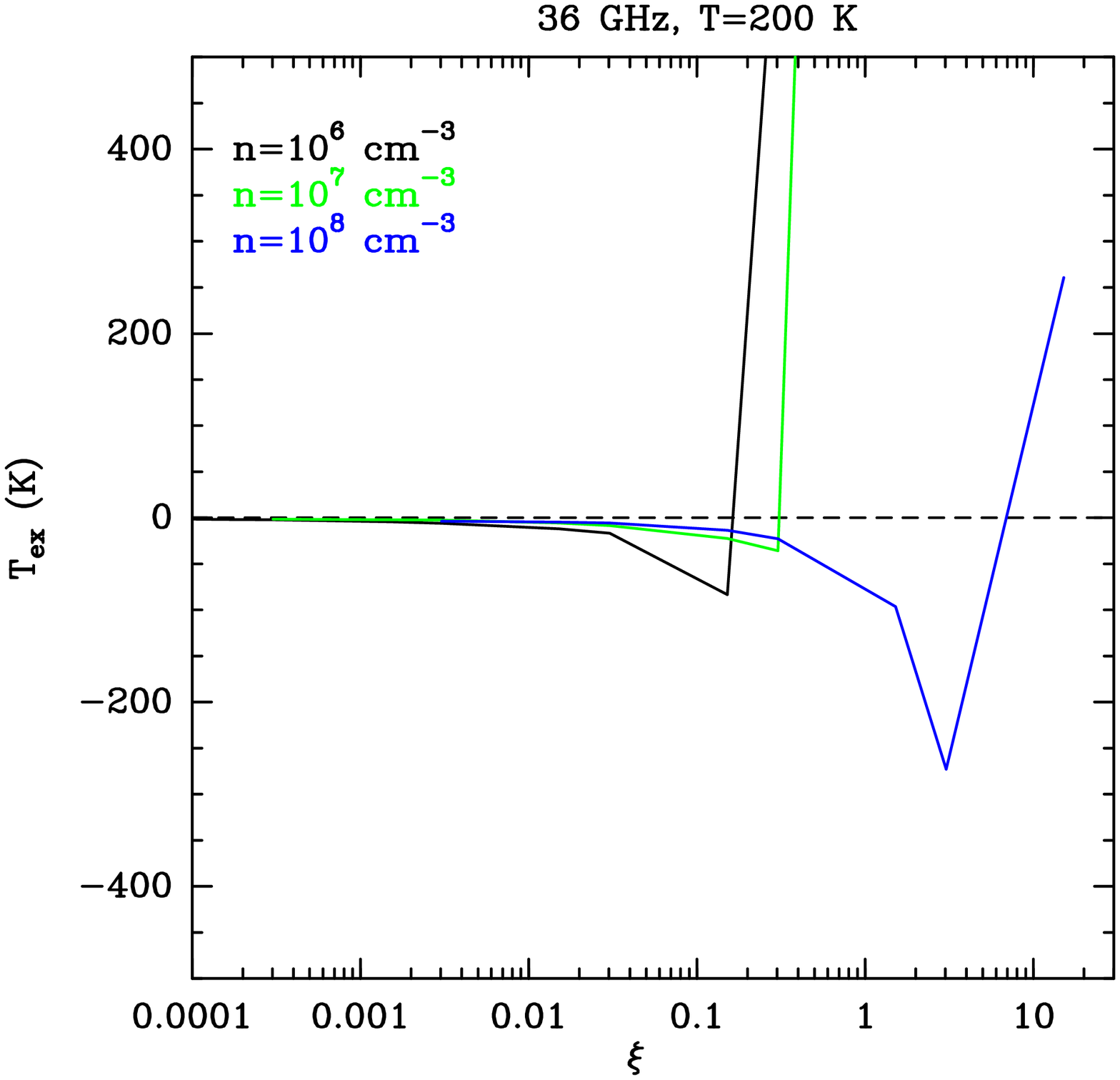}}
\subfigure{  \includegraphics[width=0.45\textwidth]{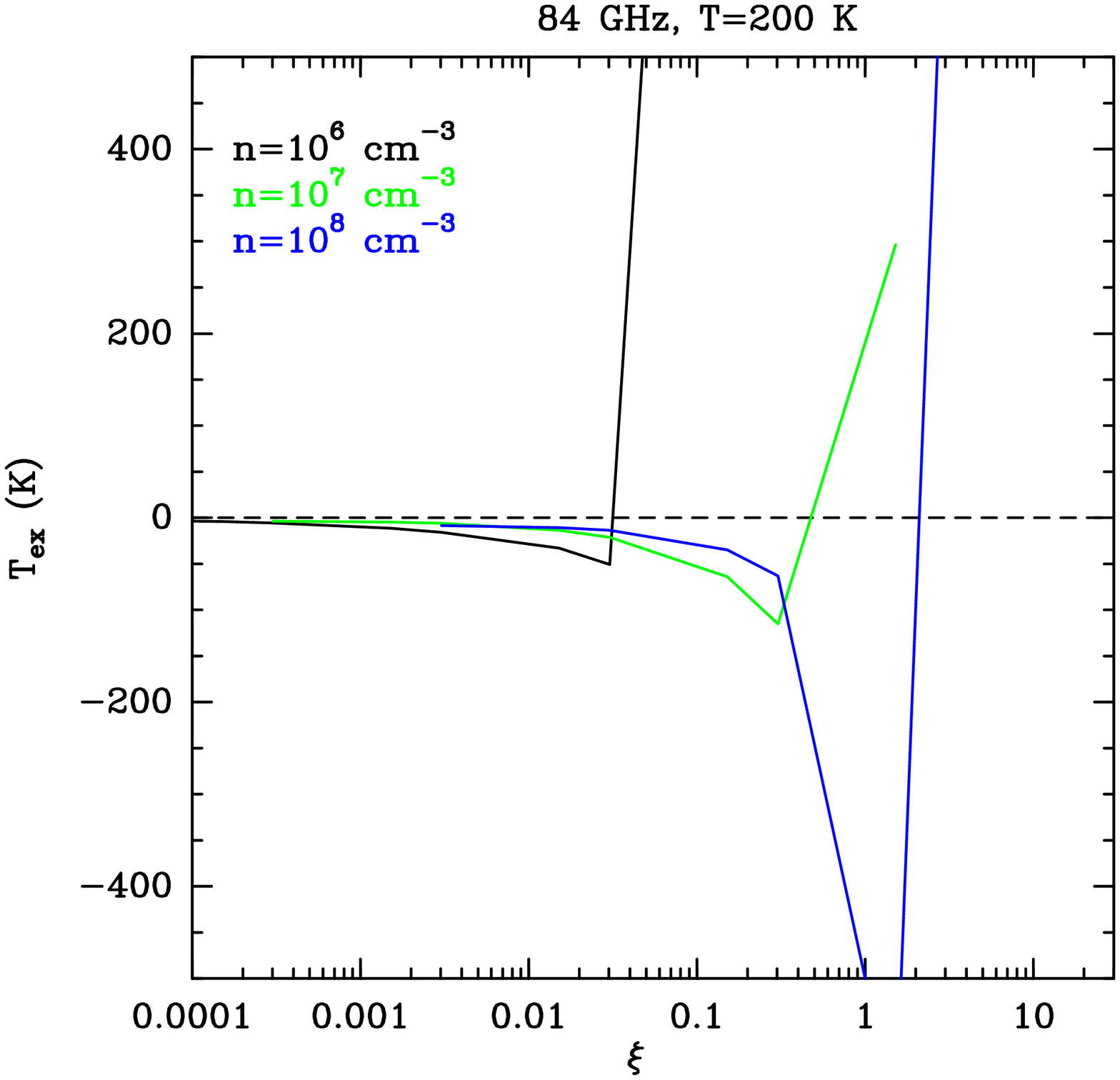}}
\subfigure{  \includegraphics[width=0.45\textwidth]{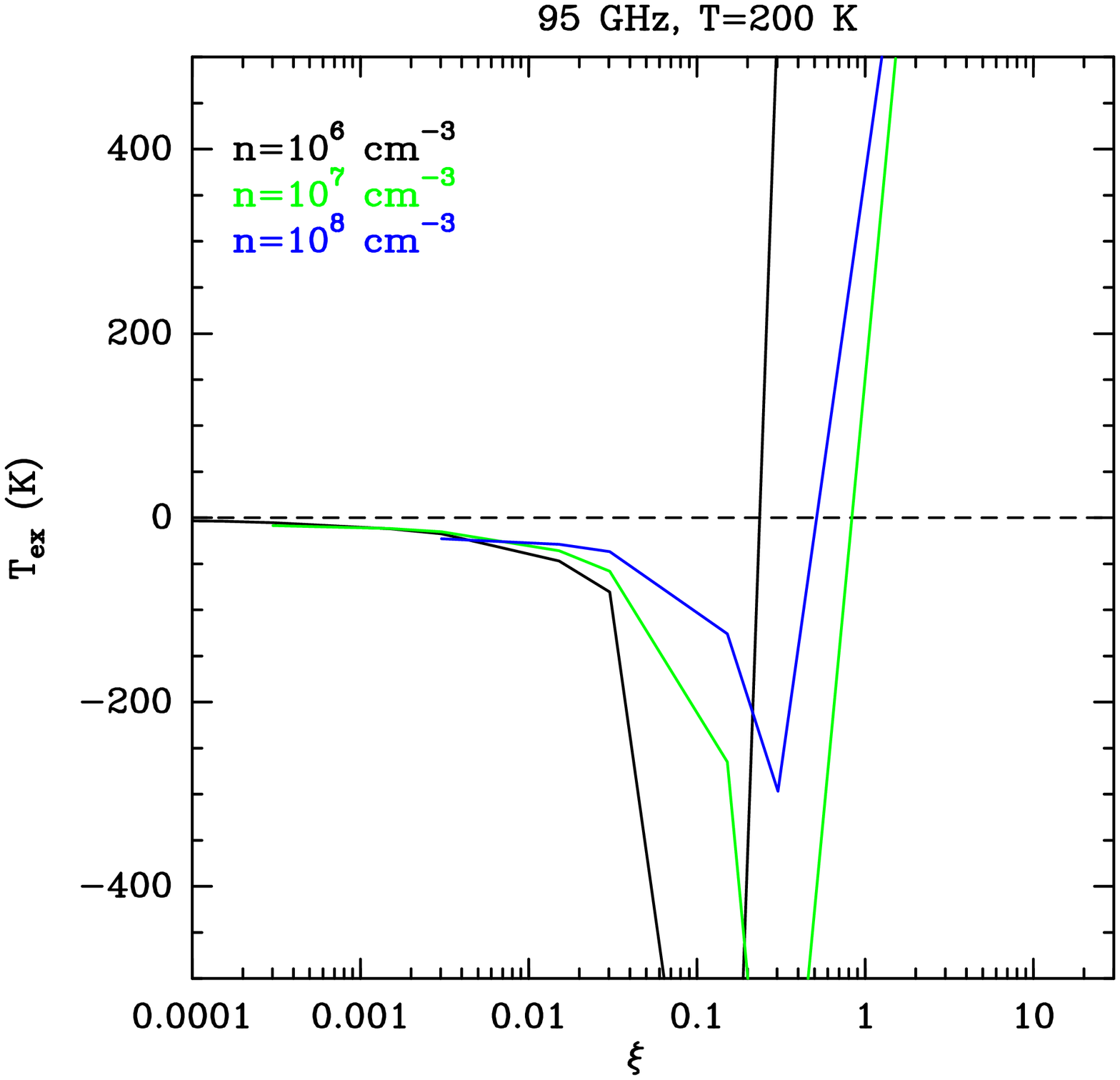}}
\caption{Excitation temperature of the 36\,GHz (upper panel), 84\,GHz (middle panel),
and 95\,GHz lines (lower panel)
  at different densities and for a temperature of 200\,K.   Curves are plotted until $T_{\rm ex}$ becomes positive. It is possible to  convert $\xi$ to methanol abundance using Eq.\,\ref{csi}.}\label{tex8495}
\end{figure*}

\begin{figure*}
\centering
\includegraphics[width=0.8\textwidth]{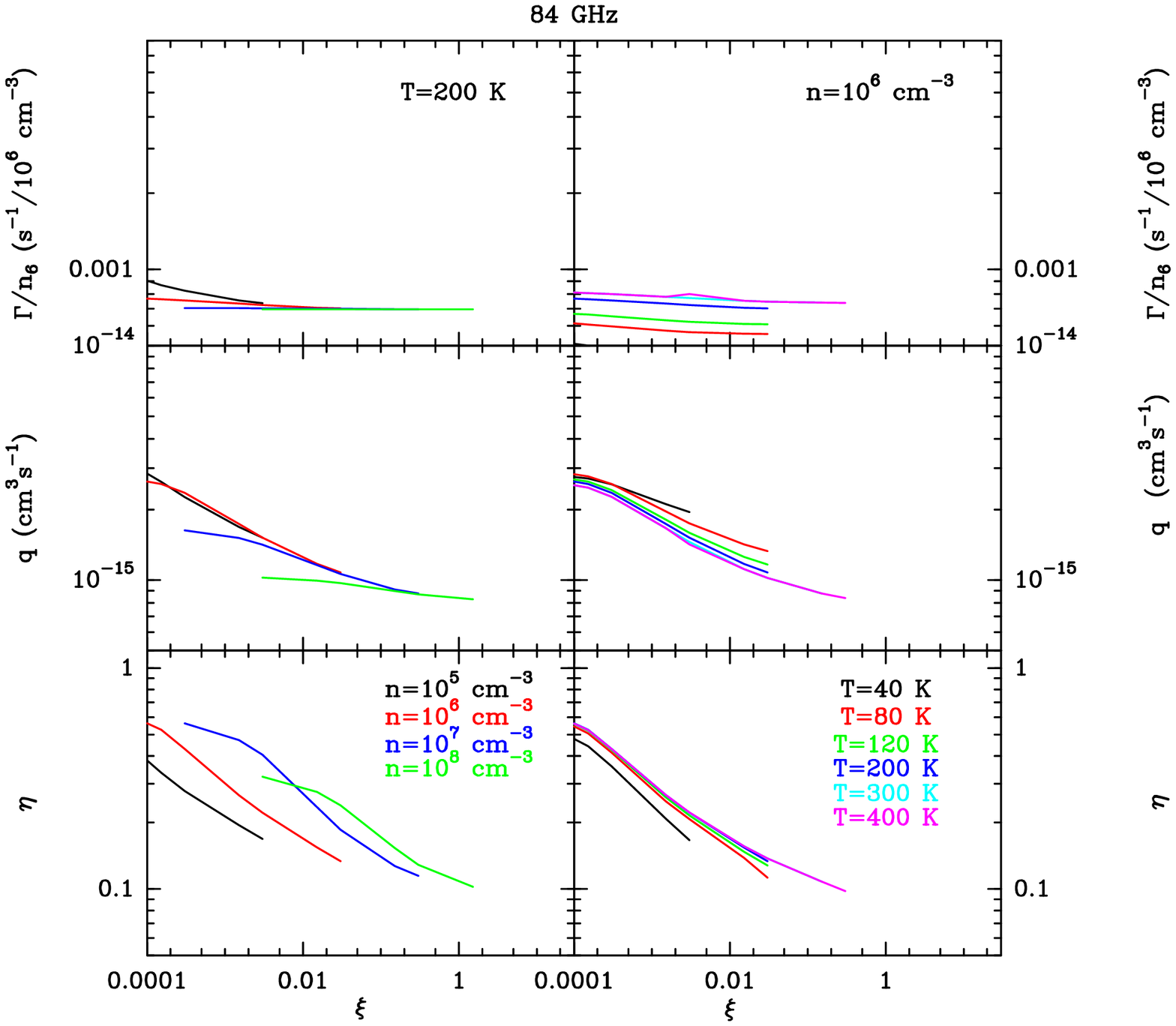}
\caption{Modelling results for the 84\,GHz maser pumping at T = 200\,K and various densities (left panel) and at $n=10^6$\,cm$^{-3}$ and several temperatures (right panel).  Plotted are the  maser loss rate $\Gamma$ divided by $n_6$, the pump rate coefficient $q$, and the inversion efficiency $\eta$ as functions of $\xi$. It is possible to convert $\xi$ to methanol abundance using Eq.\,\ref{csi}.}\label{hollenbach84}
\end{figure*}

\begin{figure*}
\centering
\includegraphics[width=0.8\textwidth]{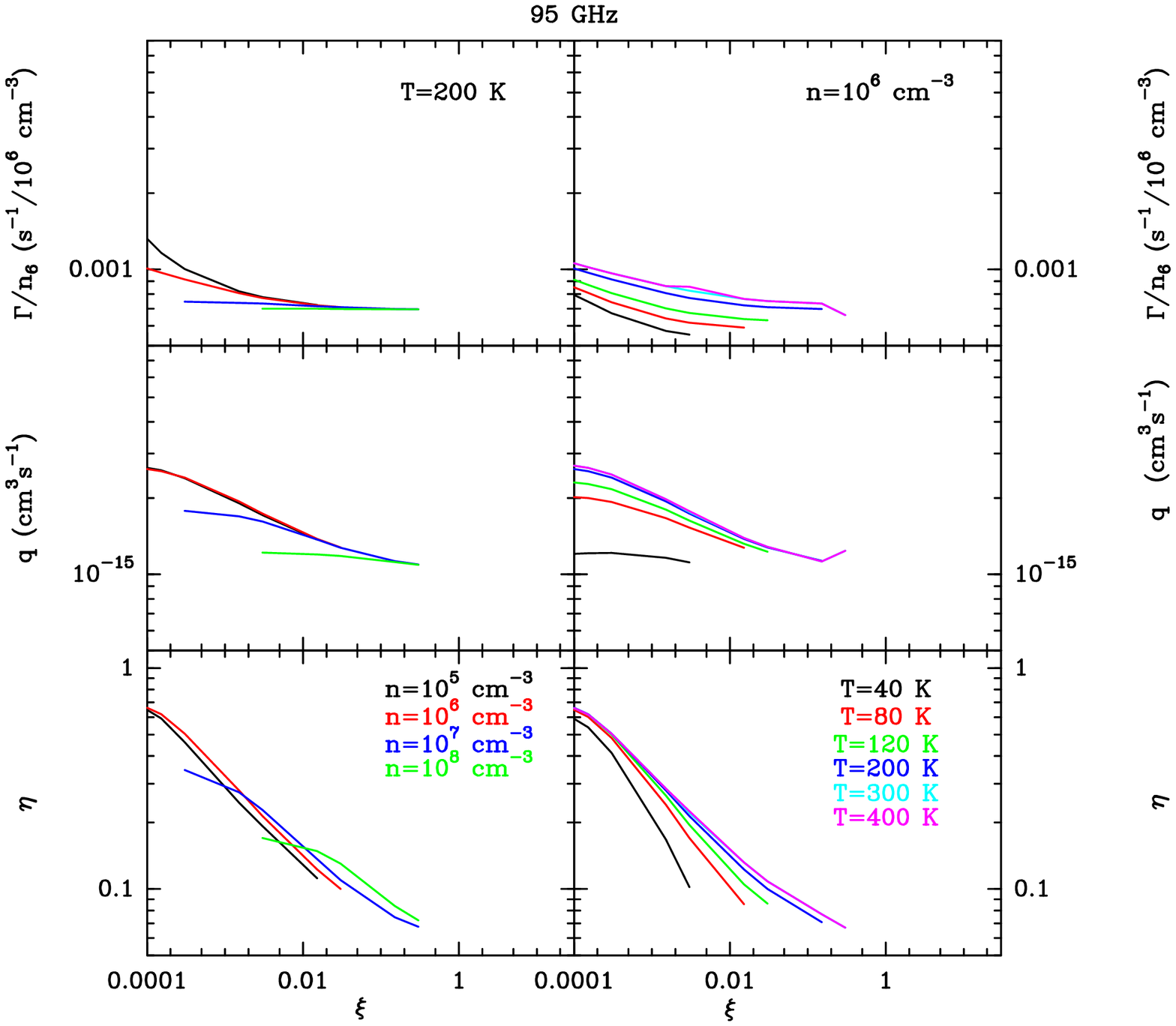}
\caption{Modelling results for the 95\,GHz maser pumping at T = 200\,K and various densities (left panel) and at $n=10^6$\,cm$^{-3}$ and several temperatures (right panel). Plotted are the  maser loss rate $\Gamma$ divided by $n_6$, the pump rate coefficient $q$, and the inversion efficiency $\eta$ as functions of $\xi$. It is possible to convert $\xi$ to methanol abundance using Eq.\,\ref{csi}.}\label{hollenbach95}
\end{figure*}

\begin{figure*}
\centering
\includegraphics[width=0.7\textwidth]{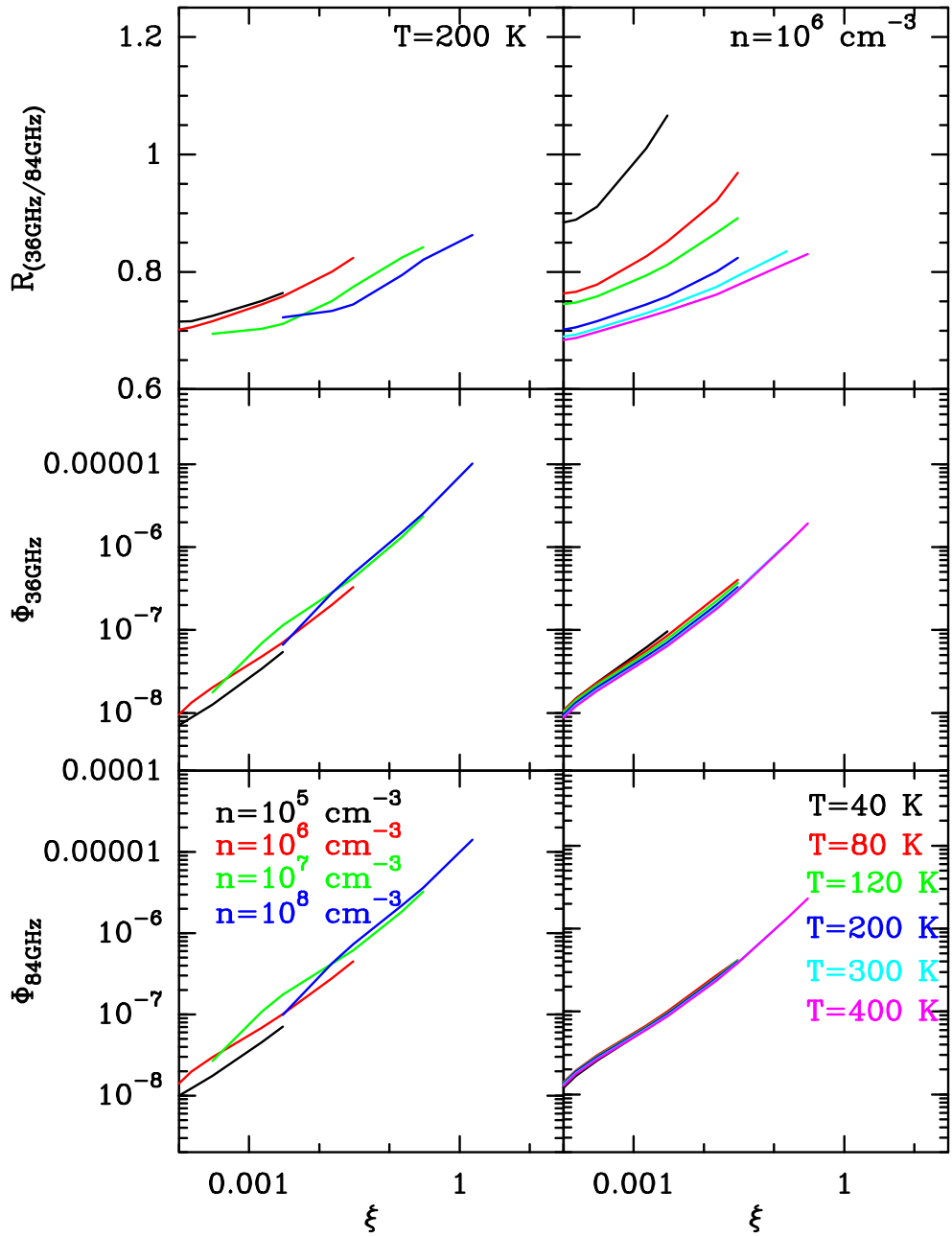}
\caption{Modelling results for the 36\,GHz (bottom) and 84\,GHz (middle) maser photon production rates, and for their ratios (top panel)    at 200\,K for various densities (left panel) and at $n=10^6$\,cm$^{-3}$ and several temperatures (right panel). It is possible to convert $\xi$ to methanol abundance using Eq.\,\ref{csi}.}\label{phi3684}
\end{figure*}

Here we present the additional figures discussed
in Sect.\,\ref{stateq}. We first present the excitation temperatures of the 36\,GHz, 
84\,GHz, and 95\,GHz masers as functions of $\xi$ for 200\,K and for different
densities to complement the plots for the other lines presented in the main text (Fig.\,\ref{tex8495}).
Figures\,\ref{hollenbach84}--\ref{hollenbach95} illustrate the maser loss rate $\Gamma$, the  pump rate coefficient $q$, and the inversion efficiency $\eta$ as functions of $\xi$
for the  84\,GHz and 95\,GHz  masers at various densities and temperatures.
Finally, in Fig.\,\ref{phi3684}  we present the photon production rates of the 36\,GHz and 84\,GHz lines  as functions of the methanol maser emission measures for different physical parameters.

\end{appendix}


\begin{thebibliography}{123}
\expandafter\ifx\csname natexlab\endcsname\relax\def\natexlab#1{#1}\fi

\bibitem[{{Araya} {et~al.}(2009){Araya}, {Kurtz}, {Hofner}, \&
  {Linz}}]{2009ApJ...698.1321A}
{Araya}, E.~D., {Kurtz}, S., {Hofner}, P., \& {Linz}, H. 2009, \apj, 698, 1321

\bibitem[{{Bachiller} {et~al.}(1990){Bachiller}, {Gomez-Gonzalez}, {Barcia}, \&
  {Menten}}]{1990A&A...240..116B}
{Bachiller}, R., {Gomez-Gonzalez}, J., {Barcia}, A., \& {Menten}, K.~M. 1990,
  \aap, 240, 116

\bibitem[{{Barrett} {et~al.}(1975){Barrett}, {Ho}, \&
  {Martin}}]{1975ApJ...198L.119B}
{Barrett}, A.~H., {Ho}, P., \& {Martin}, R.~N. 1975, \apjl, 198, L119

\bibitem[{{Barrett} {et~al.}(1976){Barrett}, {Hot}, {Martin}, {Bologna},
  {Johnston}, {Schwartz}, {Cheung}, {Chui}, {Matsakis}, \&
  {Moran}}]{1976ApL....18...13B}
{Barrett}, A.~H., {Hot}, P.~T.~P., {Martin}, R.~N., {et~al.} 1976, \aplett, 18,
  13

\bibitem[{{Barrett} {et~al.}(1971){Barrett}, {Schwartz}, \&
  {Waters}}]{1971ApJ...168L.101B}
{Barrett}, A.~H., {Schwartz}, P.~R., \& {Waters}, J.~W. 1971, \apjl, 168, L101

\bibitem[{{Batrla} {et~al.}(1987){Batrla}, {Matthews}, {Menten}, \&
  {Walmsley}}]{1987Natur.326...49B}
{Batrla}, W., {Matthews}, H.~E., {Menten}, K.~M., \& {Walmsley}, C.~M. 1987,
  \nat, 326, 49

\bibitem[{{Batrla} \& {Menten}(1988)}]{1988ApJ...329L.117B}
{Batrla}, W. \& {Menten}, K.~M. 1988, \apjl, 329, L117

\bibitem[{{Breckenridge} \& {Kukolich}(1995)}]{1995ApJ...438..504B}
{Breckenridge}, S.~M. \& {Kukolich}, S.~G. 1995, \apj, 438, 504

\bibitem[{{Britton} \& {Voronkov}(2012)}]{2012IAUS..287..282B}
{Britton}, T.~R. \& {Voronkov}, M.~A. 2012, in IAU Symposium, Vol. 287, IAU
  Symposium, ed. R.~S. {Booth}, W.~H.~T. {Vlemmings}, \& E.~M.~L. {Humphreys},
  282--283

\bibitem[{{Buxton} {et~al.}(1977){Buxton}, {Barrett}, {Ho}, \&
  {Schneps}}]{1977AJ.....82..985B}
{Buxton}, R.~B., {Barrett}, A.~H., {Ho}, P.~T.~P., \& {Schneps}, M.~H. 1977,
  \aj, 82, 985

\bibitem[{{Cesaroni} \& {Walmsley}(1991)}]{1991A&A...241..537C}
{Cesaroni}, R. \& {Walmsley}, C.~M. 1991, \aap, 241, 537

\bibitem[{{Chen} {et~al.}(2011){Chen}, {Ellingsen}, {Shen}, {Titmarsh}, \&
  {Gan}}]{2011ApJS..196....9C}
{Chen}, X., {Ellingsen}, S.~P., {Shen}, Z.-Q., {Titmarsh}, A., \& {Gan}, C.-G.
  2011, \apjs, 196, 9

\bibitem[{{Chui} {et~al.}(1974){Chui}, {Cheung}, {Matsakis}, {Townes}, \&
  {Cardiasmenos}}]{1974ApJ...187L..19C}
{Chui}, M.~F., {Cheung}, A.~C., {Matsakis}, D., {Townes}, C.~H., \&
  {Cardiasmenos}, A.~G. 1974, \apjl, 187, L19

\bibitem[{{Cragg} {et~al.}(1992){Cragg}, {Johns}, {Godfrey}, \&
  {Brown}}]{1992MNRAS.259..203C}
{Cragg}, D.~M., {Johns}, K.~P., {Godfrey}, P.~D., \& {Brown}, R.~D. 1992,
  \mnras, 259, 203

\bibitem[{{Crutcher}(1999)}]{1999ApJ...520..706C}
{Crutcher}, R.~M. 1999, \apj, 520, 706

\bibitem[{{Cyganowski} {et~al.}(2009){Cyganowski}, {Brogan}, {Hunter}, \&
  {Churchwell}}]{2009ApJ...702.1615C}
{Cyganowski}, C.~J., {Brogan}, C.~L., {Hunter}, T.~R., \& {Churchwell}, E.
  2009, \apj, 702, 1615

\bibitem[{{Davis} {et~al.}(2007){Davis}, {Kumar}, {Sandell}, {Froebrich},
  {Smith}, \& {Currie}}]{2007MNRAS.374...29D}
{Davis}, C.~J., {Kumar}, M.~S.~N., {Sandell}, G., {et~al.} 2007, \mnras, 374,
  29

\bibitem[{{Elitzur}(1992)}]{1992ASSL..170.....E}
{Elitzur}, M. 1992, Astronomical masers, ed. D.~Kluwer

\bibitem[{{Elitzur} {et~al.}(1989){Elitzur}, {Hollenbach}, \&
  {McKee}}]{1989ApJ...346..983E}
{Elitzur}, M., {Hollenbach}, D.~J., \& {McKee}, C.~F. 1989, \apj, 346, 983

\bibitem[{{Ellingsen}(2005)}]{2005MNRAS.359.1498E}
{Ellingsen}, S.~P. 2005, \mnras, 359, 1498

\bibitem[{{Ellingsen} {et~al.}(2010){Ellingsen}, {Breen}, {Caswell}, {Quinn},
  \& {Fuller}}]{2010MNRAS.404..779E}
{Ellingsen}, S.~P., {Breen}, S.~L., {Caswell}, J.~L., {Quinn}, L.~J., \&
  {Fuller}, G.~A. 2010, \mnras, 404, 779

\bibitem[{{Ellingsen} {et~al.}(2014){Ellingsen}, {Chen}, {Qiao}, {Baan}, {An},
  {Li}, \& {Breen}}]{2014ApJ...790L..28E}
{Ellingsen}, S.~P., {Chen}, X., {Qiao}, H.-H., {et~al.} 2014, \apjl, 790, L28

\bibitem[{{Fish} {et~al.}(2011){Fish}, {Muehlbrad}, {Pratap}, {Sjouwerman},
  {Strelnitski}, {Pihlstr{\"o}m}, \& {Bourke}}]{2011ApJ...729...14F}
{Fish}, V.~L., {Muehlbrad}, T.~C., {Pratap}, P., {et~al.} 2011, \apj, 729, 14

\bibitem[{{Fontani} {et~al.}(2010){Fontani}, {Cesaroni}, \&
  {Furuya}}]{2010A&A...517A..56F}
{Fontani}, F., {Cesaroni}, R., \& {Furuya}, R.~S. 2010, \aap, 517, A56

\bibitem[{{Frail}(2011)}]{2011MmSAI..82..703F}
{Frail}, D.~A. 2011, \memsai, 82, 703

\bibitem[{{Gaines} {et~al.}(1974){Gaines}, {Casleton}, \&
  {Kukolich}}]{1974ApJ...191L..99G}
{Gaines}, L., {Casleton}, K.~H., \& {Kukolich}, S.~G. 1974, \apjl, 191, L99

\bibitem[{{Gan} {et~al.}(2013){Gan}, {Chen}, {Shen}, {Xu}, \&
  {Ju}}]{2013ApJ...763....2G}
{Gan}, C.-G., {Chen}, X., {Shen}, Z.-Q., {Xu}, Y., \& {Ju}, B.-G. 2013, \apj,
  763, 2

\bibitem[{{Goldreich} \& {Kwan}(1974)}]{1974ApJ...189..441G}
{Goldreich}, P. \& {Kwan}, J. 1974, \apj, 189, 441

\bibitem[{{Green} {et~al.}(2009){Green}, {Caswell}, {Fuller}, {Avison},
  {Breen}, {Brooks}, {Burton}, {Chrysostomou}, {Cox}, {Diamond}, {Ellingsen},
  {Gray}, {Hoare}, {Masheder}, {McClure-Griffiths}, {Pestalozzi}, {Phillips},
  {Quinn}, {Thompson}, {Voronkov}, {Walsh}, {Ward-Thompson}, {Wong-McSweeney},
  {Yates}, \& {Cohen}}]{2009MNRAS.392..783G}
{Green}, J.~A., {Caswell}, J.~L., {Fuller}, G.~A., {et~al.} 2009, \mnras, 392,
  783

\bibitem[{{Green} {et~al.}(2008){Green}, {Caswell}, {Fuller}, {Breen},
  {Brooks}, {Burton}, {Chrysostomou}, {Cox}, {Diamond}, {Ellingsen}, {Gray},
  {Hoare}, {Masheder}, {McClure-Griffiths}, {Pestalozzi}, {Phillips}, {Quinn},
  {Thompson}, {Voronkov}, {Walsh}, {Ward-Thompson}, {Wong-McSweeney}, {Yates},
  \& {Cohen}}]{2008MNRAS.385..948G}
{Green}, J.~A., {Caswell}, J.~L., {Fuller}, G.~A., {et~al.} 2008, \mnras, 385,
  948

\bibitem[{{Gusdorf} {et~al.}(2008){Gusdorf}, {Cabrit}, {Flower}, \& {Pineau Des
  For{\^e}ts}}]{2008A&A...482..809G}
{Gusdorf}, A., {Cabrit}, S., {Flower}, D.~R., \& {Pineau Des For{\^e}ts}, G.
  2008, \aap, 482, 809

\bibitem[{{Haschick} \& {Baan}(1989)}]{1989ApJ...339..949H}
{Haschick}, A.~D. \& {Baan}, W.~A. 1989, \apj, 339, 949

\bibitem[{{Haschick} {et~al.}(1990){Haschick}, {Menten}, \&
  {Baan}}]{1990ApJ...354..556H}
{Haschick}, A.~D., {Menten}, K.~M., \& {Baan}, W.~A. 1990, \apj, 354, 556

\bibitem[{{Hills} {et~al.}(1975){Hills}, {Pankonin}, \&
  {Landecker}}]{1975A&A....39..149H}
{Hills}, R., {Pankonin}, V., \& {Landecker}, T.~L. 1975, \aap, 39, 149

\bibitem[{{Hollenbach} {et~al.}(2013){Hollenbach}, {Elitzur}, \&
  {McKee}}]{2013ApJ...773...70H}
{Hollenbach}, D., {Elitzur}, M., \& {McKee}, C.~F. 2013, \apj, 773, 70

\bibitem[{{Hunter} {et~al.}(2014){Hunter}, {Brogan}, {Cyganowski}, \&
  {Young}}]{2014ApJ...788..187H}
{Hunter}, T.~R., {Brogan}, C.~L., {Cyganowski}, C.~J., \& {Young}, K.~H. 2014,
  \apj, 788, 187

\bibitem[{{Johnston} {et~al.}(1992){Johnston}, {Gaume}, {Stolovy}, {Wilson},
  {Walmsley}, \& {Menten}}]{1992ApJ...385..232J}
{Johnston}, K.~J., {Gaume}, R., {Stolovy}, S., {et~al.} 1992, \apj, 385, 232

\bibitem[{{Johnston} {et~al.}(1997){Johnston}, {Gaume}, {Wilson}, {Nguyen}, \&
  {Nedoluha}}]{1997ApJ...490..758J}
{Johnston}, K.~J., {Gaume}, R.~A., {Wilson}, T.~L., {Nguyen}, H.~A., \&
  {Nedoluha}, G.~E. 1997, \apj, 490, 758

\bibitem[{{Kalenskii} {et~al.}(1992){Kalenskii}, {Bachiller}, {Berulis},
  {Val'tts}, {Gomez-Gonzalez}, {Martin-Pintado}, {Rodriguez-Franco}, \&
  {Slysh}}]{1992AZh....69.1002K}
{Kalenskii}, S.~V., {Bachiller}, R., {Berulis}, I.~I., {et~al.} 1992, \azh, 69,
  1002

\bibitem[{{Kalenskii} {et~al.}(2010{\natexlab{a}}){Kalenskii}, {Johansson},
  {Bergman}, {Kurtz}, {Hofner}, {Walmsley}, \& {Slysh}}]{2010MNRAS.405..613K}
{Kalenskii}, S.~V., {Johansson}, L.~E.~B., {Bergman}, P., {et~al.}
  2010{\natexlab{a}}, \mnras, 405, 613

\bibitem[{{Kalenskii} {et~al.}(2010{\natexlab{b}}){Kalenskii}, {Kurtz},
  {Slysh}, {Hofner}, {Walmsley}, {Johansson}, \&
  {Bergman}}]{2010ARep...54..932K}
{Kalenskii}, S.~V., {Kurtz}, S., {Slysh}, V.~I., {et~al.} 2010{\natexlab{b}},
  Astronomy Reports, 54, 932

\bibitem[{{Kalenskii} {et~al.}(1994){Kalenskii}, {Liljestroem}, {Val'tts},
  {Vasil'kov}, {Slysh}, \& {Urpo}}]{1994A&AS..103..129K}
{Kalenskii}, S.~V., {Liljestroem}, T., {Val'tts}, I.~E., {et~al.} 1994, \aaps,
  103, 129

\bibitem[{{Kalenskii} {et~al.}(2001){Kalenskii}, {Slysh}, {Val'Tts},
  {Winnberg}, \& {Johansson}}]{2001ARep...45...26K}
{Kalenskii}, S.~V., {Slysh}, V.~I., {Val'Tts}, I.~E., {Winnberg}, A., \&
  {Johansson}, L.~E. 2001, Astronomy Reports, 45

\bibitem[{{Kama} {et~al.}(2010){Kama}, {Dominik}, {Maret}, {van der Tak},
  {Caux}, {Ceccarelli}, {Fuente}, {Crimier}, {Lord}, {Bacmann}, {Baudry},
  {Bell}, {Benedettini}, {Bergin}, {Blake}, {Boogert}, {Bottinelli}, {Cabrit},
  {Caselli}, {Castets}, {Cernicharo}, {Codella}, {Comito}, {Coutens}, {Demyk},
  {Encrenaz}, {Falgarone}, {Gerin}, {Goldsmith}, {Helmich}, {Hennebelle},
  {Henning}, {Herbst}, {Hily-Blant}, {Jacq}, {Kahane}, {Klotz}, {Langer},
  {Lefloch}, {Lis}, {Lorenzani}, {Melnick}, {Nisini}, {Pacheco}, {Pagani},
  {Parise}, {Pearson}, {Phillips}, {Salez}, {Saraceno}, {Schilke}, {Schuster},
  {Tielens}, {van der Wiel}, {Vastel}, {Viti}, {Wakelam}, {Walters},
  {Wyrowski}, {Yorke}, {Cais}, {G{\"u}sten}, {Philipp}, {Klein}, \&
  {Helmich}}]{2010A&A...521L..39K}
{Kama}, M., {Dominik}, C., {Maret}, S., {et~al.} 2010, \aap, 521, L39

\bibitem[{{Kaufman} \& {Neufeld}(1996)}]{1996ApJ...456..250K}
{Kaufman}, M.~J. \& {Neufeld}, D.~A. 1996, \apj, 456, 250

\bibitem[{{Kogan} \& {Slysh}(1998)}]{1998ApJ...497..800K}
{Kogan}, L. \& {Slysh}, V. 1998, \apj, 497, 800

\bibitem[{{Kurtz} {et~al.}(2004){Kurtz}, {Hofner}, \&
  {{\'A}lvarez}}]{2004ApJS..155..149K}
{Kurtz}, S., {Hofner}, P., \& {{\'A}lvarez}, C.~V. 2004, \apjs, 155, 149

\bibitem[{{Lees}(1973)}]{1973ApJ...184..763L}
{Lees}, R.~M. 1973, \apj, 184, 763

\bibitem[{{Lees} \& {Haque}(1974)}]{lees1974}
{Lees}, R.~M. \& {Haque}, S. 1974, {Canadian J.Phys.}, 52, 2250

\bibitem[{{Lees} \& {Oka}(1969)}]{lees1969}
{Lees}, R.~M. \& {Oka}, T. 1969, {J. Chem. Phys.}, 51, 3027

\bibitem[{{Leurini} {et~al.}(2004){Leurini}, {Schilke}, {Menten}, {Flower},
  {Pottage}, \& {Xu}}]{2004A&A...422..573L}
{Leurini}, S., {Schilke}, P., {Menten}, K.~M., {et~al.} 2004, \aap, 422, 573

\bibitem[{{Liechti} \& {Wilson}(1996)}]{1996A&A...314..615L}
{Liechti}, S. \& {Wilson}, T.~L. 1996, \aap, 314, 615

\bibitem[{{Lonsdale} {et~al.}(1998){Lonsdale}, {Doeleman}, {Liechti}, {Slysh},
  {Alcolea}, {Bujarrabal}, {Colomer}, {Goluber}, {Kalenski}, {Phillips},
  {Pratap}, {Predmore}, {Val'tts}, \& {Devicente}}]{1998AAS...193.7101L}
{Lonsdale}, C.~J., {Doeleman}, S.~S., {Liechti}, S., {et~al.} 1998, Bulletin of
  the American Astronomical Society, 30, 1355

\bibitem[{{M{\" u}ller} {et~al.}(2001){M{\" u}ller}, {Thorwirth}, {Roth}, \&
  {Winnewisser}}]{2001A&A...370L..49M}
{M{\" u}ller}, H.~S.~P., {Thorwirth}, S., {Roth}, D.~A., \& {Winnewisser}, G.
  2001, \aap, 370, L49

\bibitem[{{Matsakis} {et~al.}(1980){Matsakis}, {Wright}, {Townes}, {Welch},
  {Cheung}, \& {Askne}}]{1980ApJ...236..481M}
{Matsakis}, D.~N., {Wright}, M.~C.~H., {Townes}, C.~H., {et~al.} 1980, \apj,
  236, 481

\bibitem[{{Matsumoto} {et~al.}(2014){Matsumoto}, {Hirota}, {Sugiyama}, {Kim},
  {Kim}, {Byun}, {Jung}, {Chibueze}, {Honma}, {Kameya}, {Kim}, {Lyo}, {Motogi},
  {Oh}, {Shino}, {Sunada}, {Bae}, {Chung}, {Chung}, {Cho}, {Han}, {Han},
  {Hwang}, {Je}, {Jike}, {Jung}, {Jung}, {Kang}, {Kang}, {Kang}, {Kan-ya},
  {Kawaguchi}, {Kim}, {Kim}, {Ryoung Kim}, {Kim}, {Kobayashi}, {Kono},
  {Kurayama}, {Lee}, {Lee}, {Lee}, {Lee}, {Lee}, {Lee}, {Minh}, {Miyazaki},
  {Oh}, {Oyama}, {Park}, {Roh}, {Sasao}, {Sawada-Satoh}, {Shibata}, {Sohn},
  {Song}, {Tamura}, {Wajima}, {Wi}, {Yeom}, \& {Yun}}]{2014ApJ...789L...1M}
{Matsumoto}, N., {Hirota}, T., {Sugiyama}, K., {et~al.} 2014, \apjl, 789, L1

\bibitem[{{McEwen} {et~al.}(2014){McEwen}, {Pihlstr{\"o}m}, \&
  {Sjouwerman}}]{2014ApJ...793..133M}
{McEwen}, B.~C., {Pihlstr{\"o}m}, Y.~M., \& {Sjouwerman}, L.~O. 2014, \apj,
  793, 133

\bibitem[{{Mehringer} \& {Menten}(1997)}]{1997ApJ...474..346M}
{Mehringer}, D.~M. \& {Menten}, K.~M. 1997, \apj, 474, 346

\bibitem[{{Mehrotra} {et~al.}(1985){Mehrotra}, {Dreizler}, \&
  {Mäder}}]{mehrotra}
{Mehrotra}, S.~C., {Dreizler}, H., \& {Mäder}, H. 1985, Naturforschung, 40a,
  683

\bibitem[{{Menten}(1991{\natexlab{a}})}]{1991aimn.conf..119M}
{Menten}, K. 1991{\natexlab{a}}, in ASP Conf. Ser. 16: Atoms, Ions and
  Molecules: New Results in Spectral Line Astrophysics, 119

\bibitem[{{Menten}(1991{\natexlab{b}})}]{1991ApJ...380L..75M}
{Menten}, K.~M. 1991{\natexlab{b}}, \apjl, 380, L75

\bibitem[{{Menten} \& {Batrla}(1989)}]{1989ApJ...341..839M}
{Menten}, K.~M. \& {Batrla}, W. 1989, \apj, 341, 839

\bibitem[{{Menten} {et~al.}(1988{\natexlab{a}}){Menten}, {Johnston}, {Wadiak},
  {Walmsley}, \& {Wilson}}]{1988ApJ...331L..41M}
{Menten}, K.~M., {Johnston}, K.~J., {Wadiak}, E.~J., {Walmsley}, C.~M., \&
  {Wilson}, T.~L. 1988{\natexlab{a}}, \apjl, 331, L41

\bibitem[{{Menten} {et~al.}(1988{\natexlab{b}}){Menten}, {Reid}, {Moran},
  {Wilson}, {Johnston}, \& {Batrla}}]{1988ApJ...333L..83M}
{Menten}, K.~M., {Reid}, M.~J., {Moran}, J.~M., {et~al.} 1988{\natexlab{b}},
  \apjl, 333, L83

\bibitem[{{Menten} {et~al.}(1992){Menten}, {Reid}, {Pratap}, {Moran}, \&
  {Wilson}}]{1992ApJ...401L..39M}
{Menten}, K.~M., {Reid}, M.~J., {Pratap}, P., {Moran}, J.~M., \& {Wilson},
  T.~L. 1992, \apjl, 401, L39

\bibitem[{{Menten} {et~al.}(1986){Menten}, {Walmsley}, {Henkel}, \&
  {Wilson}}]{1986A&A...157..318M}
{Menten}, K.~M., {Walmsley}, C.~M., {Henkel}, C., \& {Wilson}, T.~L. 1986,
  \aap, 157, 318

\bibitem[{{Menten} {et~al.}(1988{\natexlab{c}}){Menten}, {Walmsley}, {Henkel},
  \& {Wilson}}]{1988A&A...198..267M}
{Menten}, K.~M., {Walmsley}, C.~M., {Henkel}, C., \& {Wilson}, T.~L.
  1988{\natexlab{c}}, \aap, 198, 267

\bibitem[{{Menten} {et~al.}(1988{\natexlab{d}}){Menten}, {Walmsley}, {Henkel},
  \& {Wilson}}]{1988A&A...198..253M}
{Menten}, K.~M., {Walmsley}, C.~M., {Henkel}, C., \& {Wilson}, T.~L.
  1988{\natexlab{d}}, \aap, 198, 253

\bibitem[{{Momjian} \& {Sarma}(2012)}]{2012AJ....144..189M}
{Momjian}, E. \& {Sarma}, A.~P. 2012, \aj, 144, 189

\bibitem[{{Morimoto} {et~al.}(1985){Morimoto}, {Kanzawa}, \&
  {Ohishi}}]{1985ApJ...288L..11M}
{Morimoto}, M., {Kanzawa}, T., \& {Ohishi}, M. 1985, \apjl, 288, L11

\bibitem[{{Moscadelli} {et~al.}(2007){Moscadelli}, {Goddi}, {Cesaroni},
  {Beltr{\'a}n}, \& {Furuya}}]{2007A&A...472..867M}
{Moscadelli}, L., {Goddi}, C., {Cesaroni}, R., {Beltr{\'a}n}, M.~T., \&
  {Furuya}, R.~S. 2007, \aap, 472, 867

\bibitem[{{Moscadelli} {et~al.}(1999){Moscadelli}, {Menten}, {Walmsley}, \&
  {Reid}}]{1999ApJ...519..244M}
{Moscadelli}, L., {Menten}, K.~M., {Walmsley}, C.~M., \& {Reid}, M.~J. 1999,
  \apj, 519, 244

\bibitem[{{M{\"u}ller} {et~al.}(2004){M{\"u}ller}, {Menten}, \&
  {M{\"a}der}}]{2004A&A...428.1019M}
{M{\"u}ller}, H.~S.~P., {Menten}, K.~M., \& {M{\"a}der}, H. 2004, \aap, 428,
  1019

\bibitem[{{M{\"u}ller} {et~al.}(2005){M{\"u}ller}, {Schl{\"o}der}, {Stutzki},
  \& {Winnewisser}}]{2005JMoSt.742..215M}
{M{\"u}ller}, H.~S.~P., {Schl{\"o}der}, F., {Stutzki}, J., \& {Winnewisser}, G.
  2005, Journal of Molecular Structure, 742, 215

\bibitem[{{Nakano} \& {Yoshida}(1986)}]{1986PASJ...38..531N}
{Nakano}, M. \& {Yoshida}, S. 1986, \pasj, 38, 531

\bibitem[{{Nesterenok}(2016)}]{2016MNRAS.455.3978N}
{Nesterenok}, A.~V. 2016, \mnras, 455, 3978

\bibitem[{{Pelling}(1975)}]{1975MNRAS.172...41P}
{Pelling}, M. 1975, \mnras, 172, 41

\bibitem[{{Pihlstr{\"o}m} {et~al.}(2011){Pihlstr{\"o}m}, {Sjouwerman}, \&
  {Fish}}]{2011ApJ...739L..21P}
{Pihlstr{\"o}m}, Y.~M., {Sjouwerman}, L.~O., \& {Fish}, V.~L. 2011, \apjl, 739,
  L21

\bibitem[{{Pihlstr{\"o}m} {et~al.}(2014){Pihlstr{\"o}m}, {Sjouwerman}, {Frail},
  {Claussen}, {Mesler}, \& {McEwen}}]{2014AJ....147...73P}
{Pihlstr{\"o}m}, Y.~M., {Sjouwerman}, L.~O., {Frail}, D.~A., {et~al.} 2014,
  \aj, 147, 73

\bibitem[{{Plambeck} \& {Menten}(1990)}]{1990ApJ...364..555P}
{Plambeck}, R.~L. \& {Menten}, K.~M. 1990, \apj, 364, 555

\bibitem[{{Plambeck} \& {Wright}(1988)}]{1988ApJ...330L..61P}
{Plambeck}, R.~L. \& {Wright}, M.~C.~H. 1988, \apjl, 330, L61

\bibitem[{{Pratap} {et~al.}(2008){Pratap}, {Shute}, {Keane}, {Battersby}, \&
  {Sterling}}]{2008AJ....135.1718P}
{Pratap}, P., {Shute}, P.~A., {Keane}, T.~C., {Battersby}, C., \& {Sterling},
  S. 2008, \aj, 135, 1718

\bibitem[{{Rabli} \& {Flower}(2010)}]{2010MNRAS.406...95R}
{Rabli}, D. \& {Flower}, D.~R. 2010, \mnras, 406, 95

\bibitem[{{Sanna} {et~al.}(2010){Sanna}, {Moscadelli}, {Cesaroni}, {Tarchi},
  {Furuya}, \& {Goddi}}]{2010A&A...517A..71S}
{Sanna}, A., {Moscadelli}, L., {Cesaroni}, R., {et~al.} 2010, \aap, 517, A71

\bibitem[{{Sarma} \& {Momjian}(2009)}]{2009ApJ...705L.176S}
{Sarma}, A.~P. \& {Momjian}, E. 2009, \apjl, 705, L176

\bibitem[{{Sarma} \& {Momjian}(2011)}]{2011ApJ...730L...5S}
{Sarma}, A.~P. \& {Momjian}, E. 2011, \apjl, 730, L5

\bibitem[{{Sastry} {et~al.}(1984){Sastry}, {Lees}, \& {de
  Lucia}}]{1984JMoSp.103..486S}
{Sastry}, K.~V.~L.~N., {Lees}, R.~M., \& {de Lucia}, F.~C. 1984, Journal of
  Molecular Spectroscopy, 103, 486

\bibitem[{{Schilke} {et~al.}(1997){Schilke}, {Walmsley}, {Pineau des Forets},
  \& {Flower}}]{1997A&A...321..293S}
{Schilke}, P., {Walmsley}, C.~M., {Pineau des Forets}, G., \& {Flower}, D.~R.
  1997, \aap, 321, 293

\bibitem[{{Sjouwerman} {et~al.}(2010{\natexlab{a}}){Sjouwerman}, {Murray},
  {Pihlstr{\"o}m}, {Fish}, \& {Araya}}]{2010ApJ...724L.158S}
{Sjouwerman}, L.~O., {Murray}, C.~E., {Pihlstr{\"o}m}, Y.~M., {Fish}, V.~L., \&
  {Araya}, E.~D. 2010{\natexlab{a}}, \apjl, 724, L158

\bibitem[{{Sjouwerman} {et~al.}(2010{\natexlab{b}}){Sjouwerman},
  {Pihlstr{\"o}m}, \& {Fish}}]{2010ApJ...710L.111S}
{Sjouwerman}, L.~O., {Pihlstr{\"o}m}, Y.~M., \& {Fish}, V.~L.
  2010{\natexlab{b}}, \apjl, 710, L111

\bibitem[{{Slysh} \& {Kalenskii}(2009)}]{2009ARep...53..519S}
{Slysh}, V.~I. \& {Kalenskii}, S.~V. 2009, Astronomy Reports, 53, 519

\bibitem[{{Slysh} {et~al.}(1997){Slysh}, {Kalenskii}, {Val'tts}, \&
  {Golubev}}]{1997ApJ...478L..37S}
{Slysh}, V.~I., {Kalenskii}, S.~V., {Val'tts}, I.~E., \& {Golubev}, V.~V. 1997,
  \apjl, 478, L37

\bibitem[{{Slysh} {et~al.}(1999){Slysh}, {Kalenskii}, {Val'tts}, {Golubev}, \&
  {Mead}}]{1999ApJS..123..515S}
{Slysh}, V.~I., {Kalenskii}, S.~V., {Val'tts}, I.~E., {Golubev}, V.~V., \&
  {Mead}, K. 1999, \apjs, 123, 515

\bibitem[{{Slysh} {et~al.}(1994){Slysh}, {Kalenskii}, {Valtts}, \&
  {Otrupcek}}]{1994MNRAS.268..464S}
{Slysh}, V.~I., {Kalenskii}, S.~V., {Valtts}, I.~E., \& {Otrupcek}, R. 1994,
  \mnras, 268, 464

\bibitem[{{Slysh} {et~al.}(1993){Slysh}, {Kalenskij}, \&
  {Val'tts}}]{1993ApJ...413L.133S}
{Slysh}, V.~I., {Kalenskij}, S.~V., \& {Val'tts}, I.~E. 1993, \apjl, 413, L133

\bibitem[{{Slysh} {et~al.}(2002){Slysh}, {Kalenski{\u i}}, \&
  {Val'Tts}}]{2002ARep...46...49S}
{Slysh}, V.~I., {Kalenski{\u i}}, S.~V., \& {Val'Tts}, I.~E. 2002, Astronomy
  Reports, 46, 49

\bibitem[{{Sobolev}(1992)}]{1992SvA....36..590S}
{Sobolev}, A.~M. 1992, \sovast, 36, 590

\bibitem[{{Sobolev} {et~al.}(1997){Sobolev}, {Cragg}, \&
  {Godfrey}}]{1997A&A...324..211S}
{Sobolev}, A.~M., {Cragg}, D.~M., \& {Godfrey}, P.~D. 1997, \aap, 324, 211

\bibitem[{{Sobolev} \& {Deguchi}(1994{\natexlab{a}})}]{1994ApJ...433..719S}
{Sobolev}, A.~M. \& {Deguchi}, S. 1994{\natexlab{a}}, \apj, 433, 719

\bibitem[{{Sobolev} \& {Deguchi}(1994{\natexlab{b}})}]{1994A&A...291..569S}
{Sobolev}, A.~M. \& {Deguchi}, S. 1994{\natexlab{b}}, \aap, 291, 569

\bibitem[{{Sobolev} \& {Strelnitskii}(1983)}]{1983SvAL....9...12S}
{Sobolev}, A.~M. \& {Strelnitskii}, V.~S. 1983, Soviet Astronomy Letters, 9, 12

\bibitem[{{Sobolev} {et~al.}(1998){Sobolev}, {Wallin}, \&
  {Watson}}]{1998ApJ...498..763S}
{Sobolev}, A.~M., {Wallin}, B.~K., \& {Watson}, W.~D. 1998, \apj, 498, 763

\bibitem[{{Szczepanski} {et~al.}(1989){Szczepanski}, {Ho}, {Haschick}, \&
  {Baan}}]{1989IAUS..136..383S}
{Szczepanski}, J.~C., {Ho}, P.~T.~P., {Haschick}, A.~D., \& {Baan}, W.~A. 1989,
  in IAU Symposium, Vol. 136, The Center of the Galaxy, ed. {M.~Morris}, 383

\bibitem[{{Tsunekawa} {et~al.}(1995){Tsunekawa}, {Ukai}, {Toyama}, \&
  {Takagi}}]{tsunekawa}
{Tsunekawa}, S., {Ukai}, T., {Toyama}, A., \& {Takagi}, K. 1995, Department of
  Physics, Toyama Univerity, Japan

\bibitem[{{Urquhart} {et~al.}(2015){Urquhart}, {Moore}, {Menten}, {K{\"o}nig},
  {Wyrowski}, {Thompson}, {Csengeri}, {Leurini}, \&
  {Eden}}]{2015MNRAS.446.3461U}
{Urquhart}, J.~S., {Moore}, T.~J.~T., {Menten}, K.~M., {et~al.} 2015, \mnras,
  446, 3461

\bibitem[{{Val'tts} {et~al.}(1995){Val'tts}, {Dzyura}, {Kalenskii}, {Slysh},
  {Bus}, \& {Vinnberg}}]{1995ARep...39...18V}
{Val'tts}, I.~E., {Dzyura}, A.~M., {Kalenskii}, S.~V., {et~al.} 1995, Astronomy
  Reports, 39, 18

\bibitem[{{Val'tts} {et~al.}(2000){Val'tts}, {Ellingsen}, {Slysh}, {Kalenskii},
  {Otrupcek}, \& {Larionov}}]{2000MNRAS.317..315V}
{Val'tts}, I.~E., {Ellingsen}, S.~P., {Slysh}, V.~I., {et~al.} 2000, \mnras,
  317, 315

\bibitem[{{van der Tak} {et~al.}(2007){van der Tak}, {Black}, {Sch{\"o}ier},
  {Jansen}, \& {van Dishoeck}}]{2007A&A...468..627V}
{van der Tak}, F.~F.~S., {Black}, J.~H., {Sch{\"o}ier}, F.~L., {Jansen}, D.~J.,
  \& {van Dishoeck}, E.~F. 2007, \aap, 468, 627

\bibitem[{{Voronkov} {et~al.}(2005){Voronkov}, {Sobolev}, {Ellingsen},
  {Ostrovskii}, \& {Alakoz}}]{2005Ap&SS.295..217V}
{Voronkov}, M., {Sobolev}, A., {Ellingsen}, S., {Ostrovskii}, A., \& {Alakoz},
  A. 2005, \apss, 295, 217

\bibitem[{{Voronkov} {et~al.}(2006){Voronkov}, {Brooks}, {Sobolev},
  {Ellingsen}, {Ostrovskii}, \& {Caswell}}]{2006MNRAS.373..411V}
{Voronkov}, M.~A., {Brooks}, K.~J., {Sobolev}, A.~M., {et~al.} 2006, \mnras,
  373, 411

\bibitem[{{Voronkov} {et~al.}(2007){Voronkov}, {Brooks}, {Sobolev},
  {Ellingsen}, {Ostrovskii}, \& {Caswell}}]{2007IAUS..242..182V}
{Voronkov}, M.~A., {Brooks}, K.~J., {Sobolev}, A.~M., {et~al.} 2007, in IAU
  Symposium, Vol. 242, IAU Symposium, ed. J.~M. {Chapman} \& W.~A. {Baan},
  182--183

\bibitem[{{Voronkov} {et~al.}(2010{\natexlab{a}}){Voronkov}, {Caswell},
  {Britton}, {Green}, {Sobolev}, \& {Ellingsen}}]{2010MNRAS.408..133V}
{Voronkov}, M.~A., {Caswell}, J.~L., {Britton}, T.~R., {et~al.}
  2010{\natexlab{a}}, \mnras, 408, 133

\bibitem[{{Voronkov} {et~al.}(2012){Voronkov}, {Caswell}, {Ellingsen}, {Breen},
  {Britton}, {Green}, {Sobolev}, \& {Walsh}}]{2012IAUS..287..433V}
{Voronkov}, M.~A., {Caswell}, J.~L., {Ellingsen}, S.~P., {et~al.} 2012, in IAU
  Symposium, Vol. 287, IAU Symposium, ed. R.~S. {Booth}, W.~H.~T. {Vlemmings},
  \& E.~M.~L. {Humphreys}, 433--440

\bibitem[{{Voronkov} {et~al.}(2014){Voronkov}, {Caswell}, {Ellingsen}, {Green},
  \& {Breen}}]{2014MNRAS.439.2584V}
{Voronkov}, M.~A., {Caswell}, J.~L., {Ellingsen}, S.~P., {Green}, J.~A., \&
  {Breen}, S.~L. 2014, \mnras, 439, 2584

\bibitem[{{Voronkov} {et~al.}(2010{\natexlab{b}}){Voronkov}, {Caswell},
  {Ellingsen}, \& {Sobolev}}]{2010MNRAS.405.2471V}
{Voronkov}, M.~A., {Caswell}, J.~L., {Ellingsen}, S.~P., \& {Sobolev}, A.~M.
  2010{\natexlab{b}}, \mnras, 405, 2471

\bibitem[{{Voronkov} {et~al.}(2011){Voronkov}, {Walsh}, {Caswell}, {Ellingsen},
  {Breen}, {Longmore}, {Purcell}, \& {Urquhart}}]{2011MNRAS.413.2339V}
{Voronkov}, M.~A., {Walsh}, A.~J., {Caswell}, J.~L., {et~al.} 2011, \mnras,
  413, 2339

\bibitem[{{Walmsley} {et~al.}(1988){Walmsley}, {Menten}, {Batrla}, \&
  {Matthews}}]{1988A&A...197..271W}
{Walmsley}, C.~M., {Menten}, K.~M., {Batrla}, W., \& {Matthews}, H.~E. 1988,
  \aap, 197, 271

\bibitem[{{Wilgenbus} {et~al.}(2000){Wilgenbus}, {Cabrit}, {Pineau des
  For{\^e}ts}, \& {Flower}}]{2000A&A...356.1010W}
{Wilgenbus}, D., {Cabrit}, S., {Pineau des For{\^e}ts}, G., \& {Flower}, D.~R.
  2000, \aap, 356, 1010

\bibitem[{{Wilson} {et~al.}(1996){Wilson}, {Zeng}, {Huettemeister}, \&
  {Dahmen}}]{1996A&A...307..209W}
{Wilson}, T.~L., {Zeng}, Q., {Huettemeister}, S., \& {Dahmen}, G. 1996, \aap,
  307, 209

\bibitem[{{Yanagida} {et~al.}(2014){Yanagida}, {Sakai}, {Hirota}, {Sakai},
  {Foster}, {Sanhueza}, {Jackson}, {Furuya}, {Aikawa}, \&
  {Yamamoto}}]{2014ApJ...794L..10Y}
{Yanagida}, T., {Sakai}, T., {Hirota}, T., {et~al.} 2014, \apjl, 794, L10

\bibitem[{{Yusef-Zadeh} {et~al.}(2013){Yusef-Zadeh}, {Cotton}, {Viti},
  {Wardle}, \& {Royster}}]{2013ApJ...764L..19Y}
{Yusef-Zadeh}, F., {Cotton}, W., {Viti}, S., {Wardle}, M., \& {Royster}, M.
  2013, \apjl, 764, L19

\bibitem[{{Zapata} {et~al.}(2012){Zapata}, {Loinard}, {Su}, {Rodr{\'{\i}}guez},
  {Menten}, {Patel}, \& {Galv{\'a}n-Madrid}}]{2012ApJ...744...86Z}
{Zapata}, L.~A., {Loinard}, L., {Su}, Y.-N., {et~al.} 2012, \apj, 744, 86

\bibitem[{{Zeng} {et~al.}(1987){Zeng}, {Lou}, \& {Li}}]{1987Ap&SS.132..263Z}
{Zeng}, Q., {Lou}, G.~F., \& {Li}, S.~Z. 1987, \apss, 132, 263

\end{thebibliography}
\end{document}